\documentclass[aip,jcp,amsmath,amssymb,reprint, citeautoscript, longbibliography, floatfix]{revtex4-2}

\usepackage{graphicx}
\graphicspath{{figures/}}
\usepackage{dcolumn}
\usepackage{bm}
\usepackage[caption=false]{subfig}
\usepackage{tikz}
\usepackage{textcomp}
\usepackage{booktabs}
\usepackage{hhline}
\usepackage{braket}
\usepackage{textgreek}
\usepackage{xr}
\usepackage{float}
\usepackage{natbib}
\usepackage{url}
\begin{document}
\title[]{Computing linear optical spectra in the presence of nonadiabatic effects on Graphics Processing Units using molecular dynamics and tensor-network approaches}

\author{Evan Lambertson}
 \affiliation{Department of Chemistry, Oregon State University, Corvallis, Oregon 97331, USA}

\author{Dayana Bashirova}
 \affiliation{Department of Chemistry, Oregon State University, Corvallis, Oregon 97331, USA}

\author{Kye E. Hunter}
 \affiliation{Department of Chemistry, Oregon State University, Corvallis, Oregon 97331, USA}
 \affiliation{Present address: CRG (Barcelona Collaboratorium for Modelling and Predictive Biology),  Dr. Aiguader 88, Barcelona 08003, Spain}

 \author{Benhardt Hansen}
 \affiliation{Department of Chemistry, Oregon State University, Corvallis, Oregon 97331, USA}

\author{Tim J. Zuehlsdorff}
\email{tim.zuehlsdorff@oregonstate.edu}
 \affiliation{Department of Chemistry, Oregon State University, Corvallis, Oregon 97331, USA}

\begin{abstract}
We compare two recently developed strategies, implemented in open source software packages, for computing linear optical spectra in condensed phase environments in the presence of nonadiabatic effects. Both approaches rely on computing excitation energy and transition dipole fluctuations along molecular dynamics (MD) trajectories, treating molecular and environmental degrees of freedom on the same footing. Spectra are then generated in two ways: In the recently developed Gaussian Non-Condon Theory (GNCT), the linear response functions are computed in terms of independent \emph{adiabatic} excited states, with non-Condon effects described through spectral densities of transition dipole fluctuations. For strongly coupled excited states, we instead parameterize a linear vibronic coupling (LVC) Hamiltonian directly from spectral densities of energy fluctuations and \emph{diabatic} couplings computed along the MD trajectory. The optical spectrum is then calculated using powerful, numerically exact tensor-network approaches. Both the electronic structure calculations to sample system fluctuations and the quantum dynamics simulations using tensor-network methods are carried out on graphics processing units (GPUs), enabling rapid calculations on complex condensed phase systems. We assess the performance of the approaches using model systems in the presence of a conical intersection (CI), and the pyrazine molecule in different solvent environments. 
\end{abstract}

\maketitle

\section{Introduction}

Steady-state absorption and emission spectra provide important insights into the underlying electronic structure of molecules and their interactions with complex condensed phase environments, especially when paired with computational modeling to facilitate interpretation of experimentally observed features. However, computationally efficient methodologies that can accurately model optical lineshapes in complex systems are still an area of open research\cite{Loco2018b,Loco_2018,Zuehlsdorff_IJQC_2019, Zuehlsdorff2021}. In general, these methods need to account for the coupling of the electronic states to nuclear degrees of freedom to describe vibronic effects, and explicitly capture chromophore-environment interactions such as hydrogen bonding and slow collective solute-solvent motion, as well as environmental polarization effects. Additionally, in many situations, multiple electronic excited states contribute to the spectral lineshape, and their complex interactions, in the form of intensity borrowing between electronic states\cite{Orlandi1973}, and nonadiabatic effects in the form of \emph{conical intersections} (CIs)\cite{Curchod2018} are particularly challenging to model from first principles. 

Traditional approaches\cite{Santoro_2008,Baiardi2013, deSouza2018} to modeling the optical spectra, such as the Franck-Condon Herzberg-Teller (FCHT) scheme, often rely on the \emph{adiabatic}, or Born-Oppenheimer approximation\cite{Born1927}, treating electronic and nuclear dynamics as separate by neglecting their couplings. Non-Condon effects, such as intensity borrowing of dipole-forbidden states, are then described through a first-order Taylor expansion of the transition dipole moment (Herzberg-Teller coupling)\cite{Santoro_2008}. Additionally, to make computations feasible, the ground and excited state potential-energy surfaces (PESs) have to be approximated as harmonic around their respective minima. This approximation, while allowing for a straightforward implementation of the computation of spectral lineshapes in a range of electronic structure packages, can break down for large semi-flexible molecules\cite{Ferrer2014, Cerezo2024} and often limits the inclusion of solvent effects to approximate treatments through polarizable continuum models (PCMs)\cite{Cammi_2005}. A range of approaches to account for interactions with the condensed phase environment exist, from deriving effective environmental broadening of spectral lineshapes from molecular-dynamics (MD) sampling of solvent configurations\cite{Cerezo2015}, to averaging over vibronic lineshapes computed in frozen solvent environments\cite{Zuehlsdorff_IJQC_2019,Cerezo2019,Cerezo2024}. However, these approaches generally invoke some form of timescale separation between ``environmental" and ``chromophore'' degrees of freedom, which might not be justified in cases of strong solvent coupling or large-amplitude collective motion.   

Additionally, \emph{nonadiabatic} effects\cite{Curchod2018} on optical spectra can often be significant, with their influence typically most appreciable when nuclear dynamics takes place in the vicinity of a CI where adiabatic potential energy surfaces meet. In these scenarios, the optical spectra cannot be modeled in terms of independent \emph{adiabatic} excitations, and describing the full quantum dynamics of the system generally poses a significant challenge, especially in the context of a vast number of degrees of freedom. Methods such as the multi-configuration time-dependent Hartree\cite{Meyer1990, Beck2000, Meyer2009, Wang2015, Worth2020} (MCTDH) approach can provide numerically exact solutions to the dynamics, but are often applied to a limited number of degrees of freedom to remain tractable, and require a parameterization of the PESs, leading to additional approximations based on timescale separation arguments when explicitly including environmental interactions\cite{Green2021,Segalina2022,Cerezo2023}. 

In this work, we compare and contrast two recently developed methodologies, \cite{Dunnett2021,Wiethorn2023,Hunter2024} implemented in open-source software packages,
that integrate MD sampling of system and environment interactions in equilibrium to construct spectral densities with efficient algorithms to compute optical lineshapes, both in the presence of weakly and strongly coupled excited states. Spectra are either computed within the recently introduced Gaussian Non-Condon Theory (GNCT) for describing non-Condon effects in weakly interacting \emph{adiabatic} states\cite{Wiethorn2023}, or tensor network-based approaches for describing strongly interacting \emph{diabatic states}\cite{Prior2010, Schroder2016,Schroder2019,Dunnett2021,Hunter2024}, and are compared to the well-known cumulant method\cite{mukamel_book} that relies on the Condon\cite{Condon1926,Condon1928} approximation. Crucially, the MD-based sampling of the system retains a full coupling of the electronic states to environmental degrees of freedom, without the need to invoke frozen solvent environments. The approaches leverage graphics processing units (GPUs) at all stages, and we demonstrate a massive speed-up in computations, especially for the tensor-network simulations involving explicitly coupled excited states. 

We test the tensor-network approach on a minimal (2-mode) model systems of CIs, before contrasting all methods using the the pyrazine molecule in different solvent environments, a well-studied example\cite{Raab1999,Burghardt2008} featuring a CI between its two lowest-lying excited electronic states. By directly computing excitation energies and dipole fluctuations along MD trajectories, the three methods applied in this work serve as successively more sophisticated approaches to determination of the absorption spectrum of pyrazine. The resulting spectra are compared to experimental data\cite{Halverson1951_pyrazine_solution,Samir2020_pyrazine_vac}, demonstrating the strengths the different approaches in capturing both nonadiabatic and environmental coupling effects.

\section{Theoretical background and computational methods}
\label{sec:theory_and_methods}

\begin{figure*}
\centering
\includegraphics[width=0.85\textwidth]{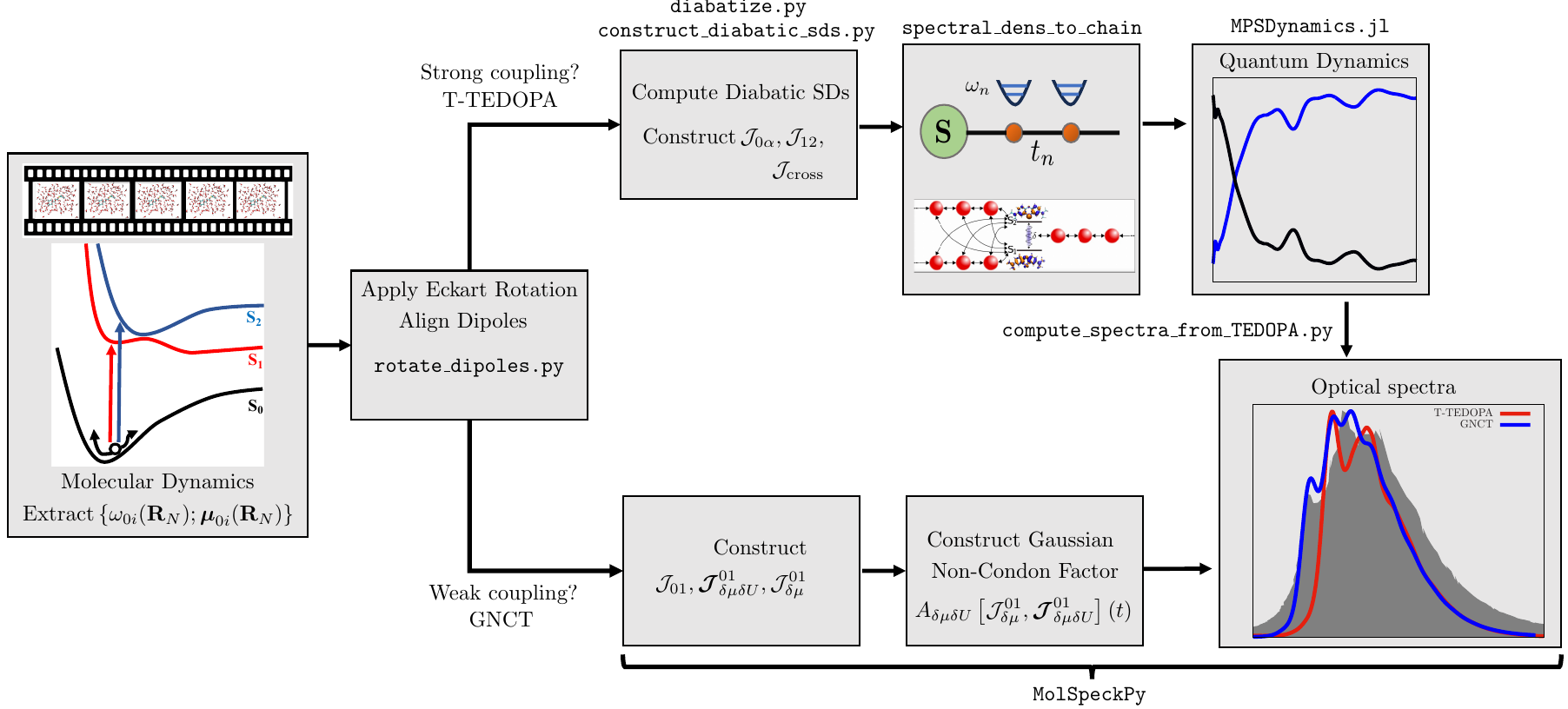}
\caption{Overview of the computational strategy employed in this work, detailing two methods to compute linear optical spectra in complex systems of multiple excited states: The Gaussian Non-Condon Theory (GNCT); and the thermalized time-evolving density operator with orthogonal polynomials algorithm (T-TEDOPA). Python scripts handling the data processing are provided in the Data Availability section.}
\label{fig:fig_1_overview}
\end{figure*}

For the purpose of this work, we focus on the computational modeling of linear spectroscopy in complex condensed phase environments, namely absorption and steady-state fluorescence spectra. The absorption and emission lineshapes can be expressed in terms of Fourier transforms of quantum correlation functions of the transition dipole operator: 

\begin{eqnarray} \nonumber
\sigma_\textrm{abs}(\omega)&=&\alpha_\textrm{abs}(\omega) \mathcal{FT}\left[\left \langle  \hat{\mu}^-(t) \hat{\mu}^+(0)\right\rangle_{\rho_0}\right] \\
\sigma_\textrm{emi}(\omega)&=&\alpha_\textrm{emi}(\omega) \mathcal{FT}\left[\left \langle \hat{\mu}^+(t) \hat{\mu}^-(0) \right \rangle_{\rho_\textrm{ex}}\right].
\label{eqn:abs_emi}
\end{eqnarray}

Here, $\hat{\mu}^-=\sum_{i=1}^{{N}_\textrm{ex}} \mu_{0 i}|S_{0}\rangle \langle S_i |$ and $\hat{\mu}^+=\left(\hat{\mu}^-\right)^\dagger$ are the transition dipole operators causing transitions between the electronic ground state $|S_0\rangle $ and $N_{\textrm{ex}}$ excited states $\{ |S_i \rangle \}$, and $\mu_{0i}$ denotes the electronic transition dipole moment between the ground state and state $i$. The quantities $\rho_0$ and $\rho_\textrm{ex}$ denote equilibrium density matrices in the electronic ground and the excited states respectively. The factors $\alpha_\textrm{abs}(\omega)$ and $\alpha_\textrm{emi}(\omega)$ are frequency-dependent prefactors necessary when comparing directly to experiment\cite{Baiardi2013}. For simplicity, both prefactors are set to 1 for the purpose of this work. 

A direct evaluation of Eqn.~\ref{eqn:abs_emi} is generally highly challenging, especially if several excited states couple strongly, for example due to the presence of a CI near the Condon region. Popular approaches, such as the FCHT method\cite{Baiardi2013,deSouza2018}, rely on reducing the complexity of the PES by invoking the harmonic approximation, and in general assume that the coupling between multiple excited states is weak. Other approaches, such as the Multi-Configuration Time-dependent Hartree (MCTDH) method\cite{Beck2000, Meyer2009}, and its multi-layer variant\cite{Wang2015,Worth2020}, allow for numerically exact evaluations of the relevant dipole correlation functions in Eqn.~\ref{eqn:abs_emi} even in the limit of strongly coupled excited states, but require a parameterization of the PESs that makes the treatment of complex solvated systems challenging. Extensions of the MCTDH approach to the condensed phase often rely on an effective time-scale separation of solute and solvent nuclear degrees of freedom\cite{Green2021,Segalina2022,Cerezo2023}, which might not be valid for strongly interacting environments and collective chromophore-environment motion. Below we summarize two approaches, both in the limit of weakly interacting (Sec.~\ref{sec:Theory_GNCT}) and strongly interacting (Sec.~\ref{sec:Theory_tensor_networks}) excited states, that are based on open-quantum system approaches and can naturally account for chromophore and environmental degrees of freedom on the same footing (see Fig.~\ref{fig:fig_1_overview} for a full overview of the approaches). 

\subsection{Weak nonadiabatic effects using open quantum system approaches: The Gaussian Non-Condon Theory (GNCT)}
\label{sec:Theory_GNCT}
For sufficiently well-separated and weakly interacting excited states, optical excitations can be treated as independent. The absorption lineshape can then be approximated as the sum of absorption lineshapes computed individually for each excited state, whereas the emission lineshape is due to a transition from the lowest electronic excited state to the ground state, following Kasha's rule\cite{Kasha1950}. Emission and absorption processes can thus be treated in terms of a two-level system or a collection of independent two-level systems, providing a significant simplification to the direct evaluation of Eqn.~\ref{eqn:abs_emi}. 

 The expression can be further simplified by invoking the Condon approximation\cite{Condon1926,Condon1928}, which states that the electronic transition dipole moment is approximately constant with respect to nuclear degrees of freedom. Further assuming that the fluctuations of the energy-gap operator defined as the difference between the ground and excited state PES obey Gaussian statistics, the dipole-dipole response functions can be evaluated through the well-known second-order cumulant approximation\cite{Mukamel-book,Mukamel04, Zuehlsdorff2019b}:
\begin{equation}
    \left \langle  \hat{\mu}^-(t) \hat{\mu}^+(0)\right\rangle_{\rho_0}\approx \chi_\textrm{GCT}^{01}(t)=|\boldsymbol{\mu}_{01}|^2 e^{-\textrm{i}\omega_{01}^\textrm{av}t-g_2\left[\mathcal{J}_{01}\right](t)}
\end{equation}
The appropriate response function for emission spectra can be formulated in an analogous way. Here, GCT denotes Gaussian Condon theory, $\boldsymbol{\mu}_{01}$ is the (constant) transition dipole moment between electronic ground and the first excited state, and $\omega_{01}^\textrm{av}$ is the quantum mechanical average the excitation energy between the ground- and the first excited state in thermal equilibrium. The lineshape function $g_2(t)$ in the second-order cumulant approximation is given by:
\begin{eqnarray} \nonumber 
g_2[\mathcal{J}](t)&=&\frac{1}{\pi}\int_0^\infty \textrm{d}\omega\,\frac{\mathcal{J}(\omega)}{\omega^2} \left[\textrm{coth}\left(\frac{\beta \omega}{2}\right)\left[1-\cos(\omega t)\right] \right. \\ \label{eqn:cumulant}
&&+\textrm{i}\left[\sin(\omega t)-\omega t\right]\bigg]\\ \nonumber
 \mathcal{J}_{01}(\omega) &=& \textrm{i}\theta(\omega) \int \textrm{d}t\ e^{\textrm{i} \omega t} \ \mathrm{Im}\  \langle \delta U_{01}(\hat{\mathbf{q}}, t) \delta U_{01}(\hat{\mathbf{q}}, 0) \rangle_{\rho_{0}}
\end{eqnarray}
where $\beta=1/k_\textrm{B}T$, $\delta U_{01}=\hat{H}_1-\hat{H}_0-\omega_{01}^\textrm{av}$, $\hat{\mathbf{q}}$ denotes nuclear degrees of freedom, $\theta(\omega)$ is the Heaviside step function and $C_{\delta U}^{01}(t)=\langle \delta U_{01}(\hat{\mathbf{q}}, t) \delta U_{01}(\hat{\mathbf{q}}, 0) \rangle_{\rho_{0}}$ is the equilibrium quantum autocorrelation function of energy gap fluctuations. The quantity $\mathcal{J}(\omega)$ is known as the \emph{spectral density} of system-bath coupling, encoding the coupling of nuclear degrees of freedom to the energy gap. While the second order cumulant approach is only truly valid for Gaussian energy gap fluctuations, approximate higher order correction terms can be formulated in terms of increasing orders of energy gap quantum correlation functions\cite{Fidler2013,Zuehlsdorff2019b,Allan2024}. 

Eqn.~\ref{eqn:cumulant} is only valid within the Condon approximation, and as such the cumulant approach cannot account for weakly nonadiabatic coupling effects between excited states, such as intensity-borrowing of dark states from nearby transitions (Herzberg-Teller effects\cite{Orlandi1973,Santoro_2008}). In a recent collaborative work, we have introduced a Gaussian Non-Condon Theory (GNCT) that approximately accounts for non-Condon effects, while still retaining a fully decoupled treatment of multiple excited states\cite{Wiethorn2023}. In the GNCT approach, the dipole response function is expressed as:
\begin{equation}
\chi^{01}_\textrm{GNCT}(t)=A_{\delta \mu \delta U}\left[\mathcal{J}^{01}_{\delta \mu},\boldsymbol{\mathcal{J}}^{01}_{\delta \mu\delta U}\right](t)e^{-\textrm{i}\omega_{01}^\textrm{av}t-g_2\left[\mathcal{J}_{01}\right](t)}.
\label{eqn:GNCT_expression}
\end{equation}
Here, the prefactor $A_{\delta \mu \delta U}(t)$ is explicitly dependent on two additional spectral defined through quantum correlation functions involving fluctuations of the transition dipole moment $\delta\boldsymbol{\mu}_{01}=\boldsymbol{\mu}_{01}-\boldsymbol{\mu}_{01}^\textrm{av}$:
\begin{eqnarray}
C_{\delta \mu}(t)&=&\langle \delta \boldsymbol{\mu}_{01}(\hat{\mathbf{q}}, t) \cdot \delta\boldsymbol{\mu}_{01}(\hat{\mathbf{q}}, 0) \rangle_{\rho_{0}} \\
\mathbf{C}_{\delta\mu\delta U}(t)&=&\langle \delta \boldsymbol{\mu}_{01}(\hat{\mathbf{q}}, t) \delta U_{01}(\hat{\mathbf{q}}, 0) \rangle_{\rho_{0}}
\end{eqnarray}
The full functional form of $A_{\delta \mu \delta U}$ can be found in Ref.~\onlinecite{Wiethorn2023} and SI Sec.~III. The GNCT approach has been shown to yield comparable results to the commonly used FCHT theory for gas-phase systems, but has the advantage that non-Condon terms are defined in terms of fluctuations of the transition dipole moment\cite{Wiethorn2023}. This means in contrast to FCHT, no Taylor expansion of the transition dipole moment around some fixed reference geometry has to be invoked, making the approach ideally suited for modeling optical properties of condensed phase systems. 

\subsubsection{Spectral densities from MD: Eckart rotations}
The quantum correlation functions required to define spectral densities in the cumulant and GNCT approach (Eqns.~\ref{eqn:cumulant} and \ref{eqn:GNCT_expression}) are in general inaccessible in realistic condensed phase systems, where a large number of nuclear degrees of freedom cause fluctuations in the respective operators. Instead, quantum correlation functions are approximately reconstructed from \emph{classical} correlation functions using quantum correction factors\cite{Egorov1999,Craig2004,Ramirez2004}, such that, for a given spectral density, 
\begin{equation}
\mathcal{J}(\omega)\approx \theta(\omega) \frac{\beta \omega }{2} \int \textrm{d}t\ e^{\textrm{i} \omega t} \ C^{\mathrm{cl}}(t)
\end{equation}
where the classical correlation function of energy gap fluctuations, transition dipole moments, or the cross correlation of dipole and energy fluctuations can be evaluated from an ab-initio molecular dynamics (AIMD) simulation\cite{Valleau2012, Shim2012,Lee2016,Lee2016b,Loco2018b,Zuehlsdorff2019b,Cignoni2022, Wiethorn2023} on the ground state (for absorption) or an excited state (for emission) PES. For the computation of emission spectra, it is again assumed that excited states are sufficiently well-separated such that the system can be propagated on a well-defined adiabatic excited state PES. Calculating optical excitation energies and dipole moments $\left \{\omega_{0i}(\textbf{R}_N); \boldsymbol{\mu}_{0i}(\textbf{R}_N) \right\}$ for every snapshot $\textbf{R}_N$ along the trajectory, for example using time-dependent density-functional theory (TDDFT)\cite{Runge1984,Casida1995}, then yields all information required to construct the response function in the GNCT approach. 

In MD simulations, the molecule is free to rotate, causing rotational contributions to any correlation function of the transition dipole moment, which are undesirable. Additionally, its sign is not expected to be consistent between different MD snapshots, leading to ill-behaved classical correlation functions. As such, some care has to be taken when directly computing spectral densities from the transition dipole moment. Moreover, along a given MD trajectory, the ordering of closely-lying excited states with respect to excitation energy may change, and tracing a consistent \emph{adiabatic} state is essential for obtaining physically meaningful spectral densities.
These issues can be overcome using two strategies (see Fig.~\ref{fig:fig_1_overview} and SI Sec.~II for a detailed description). First, snapshots along the trajectory are rotated into a consistent Eckart frame\cite{Eckart1935,Krasnoshchekov2014}, removing any contributions from rotational motion. Consistent adiabatic states are selected by maximizing the overlap with reference states computed for the ground state optimized geometry of the molecule in vacuum. By aligning the rotated transition dipole moments with the reference dipole moments, the sign problem can also be alleviated, leading to well-behaved transition dipole response functions. 

Constructing spectral densities directly from MD comes with two main advantages. First, in the condensed phase the spectral densities are continuous functions, and describe the coupling of the system to an infinite bath of nuclear degrees of freedom. Thus, solvent relaxation, spectral broadening and polarization effects are accounted for, as is the effect of direct solute solvent coupling for example through hydrogen bonding interactions. Additionally, the direct sampling of the (generally anharmonic) PES encodes some anharmonic effects in the spectral densities, such as shifts in frequency of prominent vibronic peaks\cite{Zuehlsdorff2019b, Zuehlsdorff2020}. The formalism is thus ideally suited for studying molecules embedded in condensed phase environments such as solvents or proteins. 

All calculations of spectra in the GCT and GNCT formalisms presented in this work are performed using an open-source spectroscopy \texttt{Python} package developed in our group that is freely available on GitHub\cite{Spectroscopy_python_code}. The code takes as an input a text file with a list of excitation energies and corresponding oscillator strengths and transition dipole moments in the Eckart frame, as well as a separate input file controlling calculation parameters. A \texttt{Python} script performing the Eckart rotation and subsequent alignment of transition dipole moments relative to a set of reference states is provided in the data repository accompanying this publication (see data availability section). 

\subsection{Strong nonadiabatic effects: Quantum dynamics with tensor networks}
\label{sec:Theory_tensor_networks}

The approach outlined in Sec.~\ref{sec:Theory_GNCT} is suitable for incorporating nonadiabatic effects beyond the Condon approximation in well-separated excited states, i.e. where it is still meaningful to describe the linear response function in terms of non-interacting states. However, for systems with several close-lying excited states in the Condon region, strong nonadiabatic effects due to \emph{conical intersections}\cite{Curchod2018} are expected to strongly influence excited state relaxation dynamics and optical properties\cite{Domcke2012} and the GNCT describing non-Condon effects purely due to transition dipole fluctuations is no longer valid. In this case, explicit simulation of the quantum dynamics involving multiple excited states is necessary to obtain the dipole response functions in Eqn.~\ref{eqn:abs_emi}. 

In collaboration with others\cite{Dunnett2021, Hunter2024}, we have recently shown that finite-temperature absorption and emission lineshapes in the presence of a conical intersection can be computed by combining an MD-based sampling approach of spectral densities in complex environments with the thermalized time-evolving density operator with orthogonal polynomials algorithm (T-TEDOPA)\cite{Prior2010, Schroder2016,Schroder2019, tamascelli2019efficient} for numerically exact quantum dynamics simulations. Here, we briefly summarize the main features of the formalism, focusing specifically on computational tools and algorithms developed by us to apply the approach to a wide range of complex condensed-phase systems. 

\subsubsection{Three-level systems with linear vibronic coupling}
We consider the simplest example of a system with strong nonadiabatic effects, namely a three-level system with an electronic ground- and two electronic excited states, where the two excited states are explicitly coupled due to $N$ vibrational modes. Approximating all couplings as linear with respect to nuclear degrees of freedom yields the well-known linear vibronic coupling (LVC) Hamiltonian\cite{Koppel1984,Domcke2012}:

\begin{eqnarray} \nonumber
\hat{H}_\textrm{LVC}&=&\hat{H}_\textrm{BOM}+\hat{H}_\textrm{c} 
\\
&=& \begin{pmatrix}
    {H}_0 & \mu_{01} & \mu_{02} \\
    \mu_{10} & H_1 & 0 \\
    \mu_{20} & 0 & H_2 
    \end{pmatrix}+\sum_j^N
    \begin{pmatrix}
    0 & 0 & 0 \\
    0 & 0 & \Lambda_j \hat{q}_j  \\
    0 & \Lambda_j \hat{q}_j  & 0  \\
    \end{pmatrix}.
    \label{eqn:LVChamiltonian}
\end{eqnarray}
Here, $\hat{H}_\textrm{BOM}$ is the Spin-Boson or Brownian Oscillator Model Hamiltonian\cite{Zuehlsdorff2020b}, and $\mu_{0i}$ are taken to be transition dipole moments \emph{independent} of nuclear degrees of freedom. For $\hat{H}_\textrm{BOM}$, transitions between the electronic ground and the two excited states can only be induced by the electronic transition dipole operator. The coupling term $\hat{H}_c$ explicitly couples the two excited states S$_1$ and S$_2$ along nuclear degrees of freedom $\{\hat{q}_j\}$ and is thus responsible for nonadiabatic effects. ${H}_0$, $H_1$ and $H_2$ are taken to be displaced harmonic oscillator Hamiltonians, such that:
\begin{eqnarray}
H_0&=&\sum_j^N  \left(\frac{\hat{p}_j^2}{2}+\frac{1}{2}\omega^2_j \hat{q}^2_j \right) \\
H_1&=&\sum_j^N  \left(\frac{\hat{p}_j^2}{2}+\frac{1}{2}\omega^2_j \left(\hat{q}_j-K_j^{\{1\}}\right)^2 \right)+\Delta_{01} \\
H_2&=&\sum_j^N  \left(\frac{\hat{p}_j^2}{2}+\frac{1}{2}\omega^2_j \left(\hat{q}_j-K_j^{\{2\}}\right)^2 \right) +\Delta_{02},
\end{eqnarray}
where $\Delta_{0i}$ denotes the adiabatic energy gap between the ground state and the $i$-th excited state. The Hamiltonian $\hat{H}_\textrm{LVC}$ written in this way contains diagonal vibronic couplings and off-diagonal couplings to a common set of $N$ vibrational normal modes.

The LVC Hamiltonian in Eqn.~\ref{eqn:LVChamiltonian} is widely used in the study of nonadiabatic quantum dynamics in the presence of conical intersections\cite{Domcke1981,Worth1996,Capano2014,Papai2016, Neville2018,Aranda2021,Zobel2021,Green2021,Segalina2022,Segatta2023}, but is generally formulated in terms of a finite number of (chromophore) degrees of freedom. Despite its simple form, the off-diagonal couplings contained in $\hat{H}_\textrm{c}$ make numerically exact solutions, for example using MCTDH, computationally challenging for large numbers of vibrational modes, although approaches (such as multi-layer MCTDH) exist that can accurately treat systems with a few hundreds of degrees of freedom\cite{Wang2015}. To retain a full coupling to an infinite bath of condensed phase interactions, however, it is desirable to parameterize the LVC Hamiltonian through \emph{spectral densities} sampled from molecular dynamics simulations. We have demonstrated recently\cite{Dunnett2021,Hunter2024}, that such a parameterization can be achieved for a three-level system of two coupled excited states by defining four spectral densities: 

 \begin{eqnarray}\label{eq:spectral-density-from-classical-md} \nonumber
    \mathcal{J}_{0\alpha}(\omega) &\approx& \theta(\omega) \frac{\beta \omega }{2} \int \textrm{d}t\ e^{\textrm{i} \omega t} \ C^{\mathrm{cl}}_{0\alpha}(t) \\ \nonumber
    \mathcal{J}_\textrm{cross}(\omega) &\approx& \theta(\omega) \frac{\beta \omega }{2} \int \textrm{d}t\ e^{\textrm{i} \omega t} \ C^{\mathrm{cl}}_\textrm{cross}(t) \\
    \mathcal{J}_{12}(\omega) &\approx& \theta(\omega) \frac{\beta \omega }{2} \int \textrm{d}t\ e^{\textrm{i} \omega t} \ C^{\mathrm{cl}}_{12}(t).
\end{eqnarray}
Here, $\alpha=1,2$ labels electronic exited states 1 and 2, $C^\textrm{cl}_{0\alpha}=\langle E_{0\alpha}(t) E_{0\alpha}(0) \rangle_\textrm{cl}$, $C^\textrm{cl}_\textrm{cross}=\langle E_{01}(t) E_{02}(0) \rangle_\textrm{cl}$, and $C^\textrm{cl}_{12}=\langle \delta_{12}(t) \delta_{12}(0) \rangle_\textrm{cl}$. $E_{01}(t)$, $E_{02}(t)$, and $\delta(t)$ are energy fluctuations and couplings of \emph{diabatic} states that can be constructed from their adiabatic counterparts computed along an MD trajectory via a suitable diabatization procedure (see Sec.~\ref{sec:Theory_diabatization}). The system is sampled on the \emph{ground state} PES, thus avoiding explicit propagation on the (strongly coupled) adiabatic excited states surfaces. Such parameterization of the LVC Hamiltonian is expected to be accurate as long as excited state dynamics studied remain sufficiently close to the Condon region\cite{Penfold2022}. 

The spectral densities $\mathcal{J}_{0\alpha}(\omega)$ describe the fluctuations of diabatic states due to so-called \emph{tuning modes}, whereas $\mathcal{J}_{12}(\omega)$ contains nonadiabatic effects due to \emph{coupling modes} of the two diabatic electronic states. The cross-correlation spectral density $\mathcal{J}_\textrm{cross}(\omega)$ encodes the degree to which fluctuations due to tuning modes in the two diabatic states are \emph{correlated} or \emph{anti-correlated}. The inclusion of cross-correlations in a spectral density based formalism is necessary, as the $\mathcal{J}_{0\alpha}(\omega)$ spectral densities are positive semi-definite functions, i.e. they encode the strength of the couplings of energy gap fluctuations to nuclear degrees of freedom but not their relative signs (see also SI Sec.~VI). 

\subsubsection{Spectral densities from MD: Diabatization}
\label{sec:Theory_diabatization}

Computing vertical excitation energies along an MD trajectory, such as the ones needed as input for the GCT and GNCT formalism, yields \emph{adiabatic} excited states, whereas the LVC Hamiltonian is formulated in terms of \emph{diabatic} states. Constructing truly diabatic states is generally impossible in complex molecular systems and a wide variety of schemes exist for constructing \emph{quasi}-diabatic states\cite{Subotnik2015}. Assuming that only two excited states of interest, well separated from the manifold of other states, mix in the Condon region, a straightforward diabatization scheme can be formulated based on minimizing the overlap of the transition dipole moments of the adiabatic states\cite{Medders2017}. 

Defining the matrix of transition dipole moments $\textbf{D}$
 \begin{equation}
     \textbf{D}(t)=
     \begin{pmatrix}
     \boldsymbol{\mu}_{01} (t)\cdot \boldsymbol{\mu}_{01}(t) & \boldsymbol{\mu}_{01} (t)\cdot \boldsymbol{\mu}_{02}(t) \\
     \boldsymbol{\mu}_{01}(t)\cdot \boldsymbol{\mu}_{02}(t) & \boldsymbol{\mu}_{02}(t)\cdot \boldsymbol{\mu}_{02}(t)
     \end{pmatrix} 
 \end{equation}
 along the MD trajectory, diagonalizing $\textbf{D}$ for every time step, the eigenvectors can then be used to transform the diagonal matrix of \emph{adiabatic} excitation energies into the \emph{diabatic} representation. This yields $\{ E_{01}(t), E_{02}(t), \delta_{12}(t) \}$, the set of diabatic energies and couplings, along the MD trajectory, which can then be used to define the required spectral densities in Eqn.~\ref{eq:spectral-density-from-classical-md}. Additionally, the \emph{average} magnitudes of \emph{diabatic} transition dipole moments along the MD trajectory are used to define the constant transition dipole moments in the LVC Hamiltonian (see Eqn.~\ref{eqn:LVChamiltonian}), and the adiabatic energy gaps $\Delta_{0\alpha}$ are given by $\Delta_{0\alpha}=E_{0\alpha}^\textrm{av}-\lambda_{0\alpha}$. Here, $E_{0\alpha}^\textrm{av}$ is the average diabatic excitation energy along the MD trajectory and $\lambda_{0\alpha}$ is the reorganization energy contained in the spectral density $\mathcal{J}_{0\alpha}(\omega)$. 

 Similarly to the GNCT approach, some care has to be taken in the diabatization procedure, since the sign of $\boldsymbol{\mu}_{0\alpha} (t)$ is arbitrary, leading to arbitrary sign swaps for $\delta_{12}$ along the MD trajectory. We find that, when using the adiabatic states constructed following the same procedure as for the GNCT, i.e. performing an Eckart rotation and aligning dipoles with respect to reference states constructed at the ground state optimized geometries, we obtain well-defined correlation functions and coupling spectral densities $\mathcal{J}_{12}(\omega)$.

\subsubsection{Chain-mapping and finite temperature effects}

 Simulating finite-temperature quantum dynamics in the presence of a complex condensed phase environment is highly challenging, as an infinite number of modes described by the continuous spectral densities couple to the system. To overcome this challenge, we make use of the thermalized time-evolving density operator with orthogonal polynomials algorithm (T-TEDOPA). First, following Tamascelli and co-workers\cite{tamascelli2019efficient}, the exact quantum dynamics of an initially pure state coupled to a thermal bath is obtained through an effective, ``\emph{thermalized}'' system-bath coupling spectral density,
 \begin{equation}
     \mathcal{J}_\beta\left(\omega\right) = \frac{\mathrm{sign}\left(\omega\right)\,\mathcal{J}\left(\vert\omega\vert\right)}{2}\left[ 1 + \coth{\left(\frac{\beta\omega}{2}\right)} \right]\,.
     \label{eqn:thermalized_sd}
 \end{equation}
 constructed from the original spectral density $\mathcal{J}\left(\omega\right)$. This thermalized spectral density contains modes with negative frequency that account for finite-temperature effects.

Second, to make the simulations of system-bath interactions tractable, the continuous spectral densities are mapped onto a chain Hamiltonian of harmonic oscillators using unitary transformations defined through orthogonal polynomials\cite{Chin2010}. The three spectral densities for the tuning modes, $\mathcal{J}_{01}$ and $\mathcal{J}_{02}$, as well as the coupling modes, $\mathcal{J}_{12}$, are mapped to three independent chains. Bath correlations contained in $\mathcal{J}_\textrm{cross}$ are accounted for through long-range couplings in the chains describing the tuning modes (see Fig.~\ref{fig:fig_1_overview}, and Refs.~\onlinecite{Dunnett2021} and \onlinecite{Hunter2024}, as well as SI Sec.~IV for a detailed description of the chain mapping procedure). The chain mapping allows for the efficient representation of the many-body wavefunction in terms of a \emph{tree} matrix-product-state\cite{Haegeman2011,Haegeman2016,Schroder2019, kloss2018time} (tree-MPS), whose time-evolution is then calculated using the one-site Time-Dependent Variational Principle\cite{Haegeman2011,tdvp} (1TDVP). The accuracy of the MPS representation is controlled by an internal parameter known as the bond dimension $D$, and can thus be systematically improved. Additionally, two parameters determine the accuracy of the time-propagation: The chain length $N$ at which the semi-infinite chain resulting from the mapping procedure is truncated, and the local Fock space $d$ of each node in the chain. Typical values sufficient for the convergence of the spectra considered in this work were $N\sim 100$-150 and $d\sim 15$-30 (see also SI Sec.~VII).

All chain mappings in this work are performed with the \texttt{Python} code \texttt{spectral\_dens\_to\_chain} (see Data Availability), utilizing an algorithm based on the \texttt{ORTHOPOL} package\cite{gautschi_algorithm_1994}. The code takes as input the four spectral densities $\mathcal{J}_{0\alpha}$, $\mathcal{J}_{12}$, and $\mathcal{J}_\textrm{cross}$, constructs thermalized versions via Eqn.~\ref{eqn:thermalized_sd}, performs the chain mapping for the two tuning mode spectral densities and the single coupling spectral densities, and computes long-range couplings from the cross spectral density. Chain coefficients are stored in the HDF5 format.

Additionally, high-resolution spectral densities are needed to yield orthogonal polynomial coefficients free of numerical artifacts. To avoid numerical issues, we follow two strategies: First, diabatic energy gaps and couplings along the MD trajectory are interpolated between sampling time-steps using cubic splines. Second, we employ Fourier interpolation by padding the classical correlation functions of Eqn.~\ref{eq:spectral-density-from-classical-md} prior to the construction of spectral densities. The construction of the spectral densities required by the chain mapping is handled by the \texttt{Python} script \texttt{compute\_diabatic\_sds.py}, taking as input $\{E_{01}(t) , \delta_{12}(t), E_{01}(t), f_{01} (t), f_{02} (t) \}$, the list of diabatic energies, couplings and oscillator strengths sampled along the MD trajectory. 

\subsubsection{Computing optical spectra}
\label{subsec:optical_spectra}

To compute the relevant dipole-dipole response functions (Eqn.~\ref{eqn:abs_emi}) for absorption and emission lineshapes in the T-TEDOPA formalism, we employ the following procedure (see also Refs.~\onlinecite{Dunnett2021} and \onlinecite{Hunter2024}). For \emph{absorption}:
\begin{enumerate}
\item 
The initial electronic state is prepared as a superposition of the ground and the excited electronic states weighted by the transition dipole moments, such that
\begin{equation} \nonumber
|\Psi_\textrm{abs}(0)\rangle=N\left(|S_0\rangle+\mu_{01}|S_{1}\rangle+\mu_{02}|S_{2}\rangle\right),
\end{equation}
where $N$ is a normalization constant.
\item 
The many-body wavefunction is propagated from $t=0$ to $t=t_\textrm{max}^\textrm{abs}$ under the LVC Hamiltonian using the 1TDVP as implemented in \texttt{MPSDynamics.jl} \cite{lacroix2024mpsdynamicsjl}. 
\item 
The dipole-dipole response function is then given by the expectation value of the non-Hermitian operator $\hat{\mu}^-$ with respect to the time-evolved state $\Psi_{\textrm{abs}}(t)$: $\left \langle  \hat{\mu}^-(t) \hat{\mu}^+(0)\right\rangle_{\rho_0}=\langle\Psi_\textrm{abs}(t) | \hat{\mu}^- | \Psi_\textrm{abs} (t)\rangle$.
\end{enumerate}

For \emph{emission}: 
\begin{enumerate}
\item 
The same steps as for the absorption spectrum are followed to obtain a relaxed state $\left|\Psi_\textrm{abs}(t_\textrm{max}^\textrm{abs})\right\rangle$.
\item 
The state $|\phi_\textrm{R}\rangle =\hat{\mu}^{-}\left|\Psi_\textrm{abs}(t_\textrm{max}^\textrm{abs})\right\rangle$ is constructed. Since the operator $\hat{\mu}^-$ is independent of nuclear degrees of freedom, it does not act on the environmental (bath) states. 
\item 
The state $|\chi_\textrm{R}(0)\rangle$ is constructed as a superposition of $|\phi_\textrm{R}\rangle$ and the component of $\left|\Psi_\textrm{abs}(t_\textrm{max}^\textrm{abs})\right\rangle$ only involving the diabatic excited states (see SI Sec.~V for details on how the state is constructed in a MPS format).
\item 
The state $|\chi_\textrm{R}(0)\rangle$ is propagated from $t=0$ to $t=t_\textrm{max}^\textrm{emi}$. The dipole dipole response function for emission is given by $\left\langle \hat{\mu}^+(t) \hat{\mu}^-(0) \right\rangle_{\rho_\textrm{ex}}=\langle \chi_\textrm{R}(t)|\hat{\mu}^+|\chi_\textrm{R}(t)\rangle$.
\item 
Repeat steps 1.-4. for a number of different excited state relaxation times $t_\textrm{max}^\textrm{abs}$; average over the resulting emission spectra. 
\end{enumerate}

Thanks to the structure of the LVC Hamiltonian used, where the only couplings between the ground state and the two excited states considered are due to the transition dipole moment, it is straightforward to show that forming the expectation values of operators $\hat{\mu}^\pm$ yields the desired correlation functions\cite{Hunter2024}. 

In principle, the state $\left|\Psi_\textrm{abs}(t_\textrm{max}^\textrm{abs})\right\rangle$ used as the initial state for the computation of emission spectrum should represent a fully relaxed excited state, i.e. $t_\textrm{max}^\textrm{abs}\rightarrow \infty$. However, very long timescales are difficult to access within 1TDVP\cite{kloss2018time}, requiring both large bond dimensions $D$ and long chain lengths $N$ to prevent spurious system dynamics. To overcome this issue, we compute the emission lineshape as an average over different excited state relaxation times $t_\textrm{max}^\textrm{abs}$, chosen sufficiently long for excited state populations to have mostly reached a steady state. The convergence of optical spectra with respect to this averaging procedure is investigated in detail in Sec.~\ref{subsec:results_model}. 

\subsubsection{GPU-accelerated tensor-network propagation}
\label{subsec:GPUtensornetwork}

The 1TDVP algorithm used to propagate the system dynamics involves a variety of tensor operations, which benefit significantly from the utilization of GPUs rather than CPUs. In this work, all time-evolution is carried out using a modified version of the \texttt{MPSDynamics.jl} package by Dunnett and coworkers\cite{dunnett_angus_2021_5106435, lacroix2024mpsdynamicsjl}, making use of the in-built CUDA support in the \texttt{tensorOperations.jl} library. This modified version of \texttt{MPSDynamics.jl} is freely available on GitHub\cite{MPSDynamics_GPU} (see also Data Availability). 

All calculations presented in this work are performed on a single GPU, and we do not consider parallelizations across multiple GPUs. Since each time-step in the 1TDVP algorithm involves sequential optimization sweeps through the tensor network, it is not straightforwardly amenable to parallel execution. However, fully parallel implementations relying on splitting up the chain across multiple processes have recently been developed in the context of the density-matrix renormalization group (DMRG)\cite{Stoudenmire2013} and the two site time-dependent variational principle (2TDVP)\cite{Secular2020}. While such techniques could potentially be utilized to yield further speedup of the time-propagation in systems studied, they are beyond the scope of the current work. Detailed timing tests of the 1TDVP propagation on a single GPU in comparison with CPUs are presented in Sec.~\ref{sec:results_timing_gpu}. 

\section{Benchmark tests: Computational details}
\begin{figure}
\centering
\includegraphics[width=0.48\textwidth]{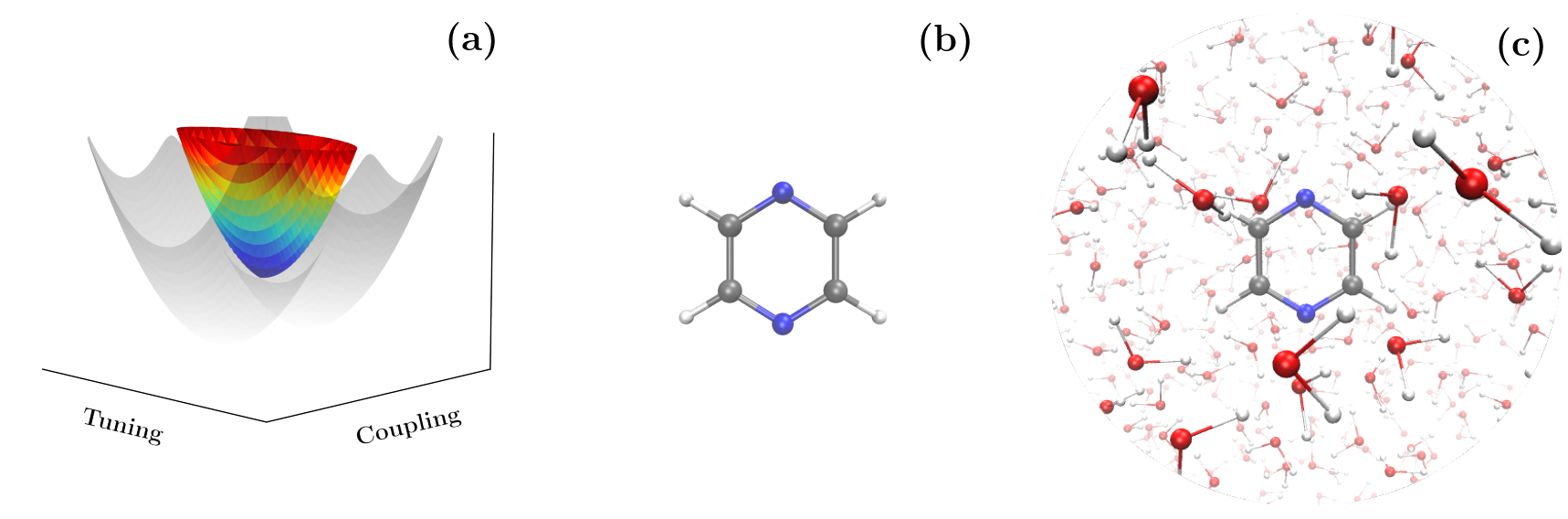}
\caption{Schematic showing the model systems considered in this work: (a) A 2D model system of a conical intersection, (b) pyrazine in vacuum, and (c) pyrazine in solution. }
\label{fig:fig_2_background}
\end{figure}

We demonstrate the strengths of the outlined computational workflow by studying the influence of a CI between two electronic excited states close to the Condon region on both absorption and fluorescence lineshapes. A specific emphasis is placed on the impact of complex condensed phase environments such as solvents on the quantum dynamics. We first focus on a two-mode model system, the simplest model for studying CIs. We then turn to the pyrazine molecule, whose well-characterized CI between S$_1$ and S$_2$ makes it a quintessential benchmark system for nonadiabatic quantum dynamics\cite{Raab1999} (see Fig.~\ref{fig:fig_2_background} for the systems studied in this work). In pyrazine, we contrast the GNCT approach with a full nonadiabatic treatment of the CI provided by the T-TEDOPA formalism. 

\subsection{Model systems}
\label{subsec:model_systems}

We consider model systems consisting of a ground state S$_0$, and two electronic excited states, S$_1$ and S$_2$, each having the same curvatures $\omega_t,\,\omega_c$ along a single tuning and coupling mode. The S$_1$ and S$_2$ states are coupled linearly through the coupling mode with coupling strength $\Lambda$. The system Hamiltonian is then given by the LVC model (see Eqn.~\ref{eqn:LVChamiltonian}), with
\begin{eqnarray}
H_0 &=& \frac{\hat{p}_t^2}{2} + \frac{\hat{p}_c^2}{2} + \frac{1}{2}\omega_t^2\hat{q}_t^2 + \frac{1}{2}\omega_c^2\hat{q}_c^2\,,
\\ \nonumber
H_1 &=& \frac{\hat{p}_t^2}{2} + \frac{\hat{p}_c^2}{2} + \frac{1}{2}\omega_t^2\left(\hat{q}_t - K_t^{(1)}\right)^2 + \frac{1}{2}\omega_c^2\hat{q}_c^2 +\Delta_{01},
\\ \nonumber
H_2 &=& \frac{\hat{p}_t^2}{2} + \frac{\hat{p}_c^2}{2} + \frac{1}{2}\omega_t^2\left(\hat{q}_t - K_t^{(2)}\right)^2 + \frac{1}{2}\omega_c^2\hat{q}_c^2 +\Delta_{02},
\end{eqnarray}
where $K_t^{(\alpha)}$ are the displacements between the ground and $\alpha^\text{th}$ excited state minima along the tuning mode, and $\Delta_{0\alpha}$ are the adiabatic energy gaps. We further assume that the transition dipole moments of diabatic states S$_1$ and S$_2$ are independent of tuning and coupling coordinates. Spectral densities of tuning and coupling modes are then given by:
\begin{eqnarray}\label{eq:delta_sds}\nonumber
\mathcal{J}_{0\alpha}\left(\omega\right) &=& \frac{\pi}{2}\omega^3\left(K_t^{(\alpha)}\right)^2\delta\left(\omega-\omega_t\right), \\
\mathcal{J}_{12}\left(\omega\right) &=& \frac{\pi}{2}\frac{\Lambda^2}{\omega}\delta\left(\omega-\omega_c\right).
\end{eqnarray}

In realistic systems, where the energy gap fluctuations are directly sampled from MD, spectral densities are continuous functions, and peaks due to molecular vibrations are broadened, rather than sharp $\delta$-functions. To incorporate this feature into our model systems, the peaks in the spectral densities are instead represented as Gaussians:
\begin{eqnarray} \nonumber
\mathcal{J}_{0\alpha}(\omega) &=& \mathcal{N}_\alpha\exp{\left[-\frac{(\omega-\omega_t)^2}{2\sigma_t^2}\right]}, \\
\mathcal{J}_c\left(\omega\right) &=& \mathcal{N}_c\exp{\left[-\frac{(\omega-\omega_c)^2}{2\sigma_c^2}\right]}.
\end{eqnarray}
Here, the constants $\mathcal{N}_\alpha$ and $\mathcal{N}_c$ are defined as
\begin{eqnarray} \nonumber
\mathcal{N}_{\alpha} = \lambda_{\alpha,t}\left(\frac{1}{\pi} \int_0^\infty \mathrm{d}\omega\, \omega^{-1}\exp{\left[-\frac{\left(\omega-\omega_t\right)^2}{2\sigma_t^2}\right]}\right)^{-1}, \\
\mathcal{N}_c = \lambda_c\left(\frac{1}{\pi} \int_0^\infty \mathrm{d}\omega\, \omega^{-1}\exp{\left[-\frac{\left(\omega-\omega_c\right)^2}{2\sigma_c^2}\right]}\right)^{-1},
\end{eqnarray}
such that, independent of broadening factors $\sigma_c$ and $\sigma_t$, the total reorganization energies $\lambda_c$ and $\lambda_{\alpha,t}$ contained in the spectral densities remain constant, where the reorganization energy is defined as  
\begin{equation}
\lambda^\textrm{reorg}\left[\mathcal{J}\right]=\frac{1}{\pi}\int_0^\infty \textrm{d}\omega\, \frac{\mathcal{J}(\omega)}{\omega}\,.
\end{equation}
The above spectral densities do not completely parameterize the model system Hamiltonians, however, as they do not distinguish the possible relative sign between $K_t^{(1)}$ and $K_t^{(2)}$. To account for this we define the cross spectral density,
\begin{equation}
\mathcal{J}_\mathrm{cross}\left(\omega\right) = c\left(\omega\right)\sqrt{\mathcal{J}_{01}\left(\omega\right)\mathcal{J}_{02}\left(\omega\right)}\,,
\end{equation}
where $c\left(\omega\right)=1$ corresponds to tuning modes which are fully positively correlated (FPC), and $c\left(\omega\right)=-1$ to ones which are fully negatively correlated (FNC).

To further incorporate environmental effects such as solvent interactions into the model system, we also include in the tuning and coupling spectral densities a continuous low frequency contribution of the Debye form,
\begin{equation}
    \mathcal{J}_\text{env}\left(\omega\right) = 2\lambda_\text{env}\frac{\omega_\text{env}\omega}{\omega_\text{env}^2 + \omega^2}\,,
\end{equation}
with $\omega_\text{env}\ll\omega_t,\,\omega_c$. This not only accounts phenomenologically for the presence of low frequency environmental modes, but serves to lower the necessary propagation time required to achieve convergence of the absorption and emission lineshapes in the simulated dynamics by providing a source of spectral broadening.

Even though the 2-mode model is conceptually simple, its parameter space is vast and can yield a wide variety of system dynamics and optical properties. For the purpose of this work, we focus on a bright S$_1$ ($\mu_{01}=2.54$~a.u.) and completely dark S$_2$ ($\mu_{02}=0$~a.u.) state, with vertical excitation energies of $E^{\textrm{vert}}_{01}=3.0$~eV and $E^{\textrm{vert}}_{01}=3.2$~eV in the Condon region. This model system allows us to probe intensity borrowing effects, one of the most prominent signatures of nonadiabatic effects in optical spectra due to CIs near the Condon region. A second model parameterization, where the dipole moments of S$_1$ and S$_2$ are swapped, such that a bright higher-lying state couples to a dark low-lying state, is considered in SI Sec.~XIV.

\subsection{Pyrazine}
Quantum mechanical (QM) and mixed quantum mechanical/molecular mechanical (QM/MM)\cite{Warshel1976} dynamics were carried out for pyrazine in vacuum and in solution respectively. For the QM dynamics, we utilized the GPU-accelerated \texttt{TeraChem}\cite{Ufimtsev2009} code, whereas for QM/MM simulations, the interface\cite{Isborn2012} between \texttt{TeraChem} and the \texttt{AMBER}\cite{amber} MD package was utilized. For all QM/MM dynamics, the QM region was limited to the pyrazine molecule during the propagation. For pyrazine in water, the TIP3P\cite{TIP3P} water model was used for the MM part, whereas for pyrazine in cyclohexane, a solvent model was constructed using \texttt{Antechamber}. A time-step of 0.5~fs was used throughout, the QM region was treated at the 6-31+G$^*$/CAM-B3LYP\cite{Dunning1990, Yanai2004} level of theory, and the temperature was kept at 300~K using a Langevin thermostat with a collision frequency of 1~ps$^{-1}$. Full simulation details can be found in SI Sec.~I. 

Upon equilibration (see SI Sec.~I), a 20~ps trajectory was generated for pyrazine in vacuum, cyclohexane and water. From this trajectory, snapshots were extracted every 2~fs, and vertical excitation energies were computed using time-dependent density-functional theory (TDDFT) within the Tamm-Dancoff approximation\cite{Hirata1999} as implemented in \texttt{TeraChem}\cite{Isborn2011}. For the solvated system, all solvent molecules were represented as classical point charges during the calculation of vertical excitations. The 6-31+G$^*$ basis set and the CAM-B3LYP functional were used throughout to avoid well-known artifacts in computed spectral densities that arise from a mismatch in Hamiltonians used for system propagation and the computation of energy gap fluctuations.\cite{Lee2016} 

The low energy absorption spectrum of pyrazine is strongly influenced by a CI between the S$_2$($\pi\pi^*$) and the S$_1$($n\pi^*$) states\cite{Raab1999,Burghardt2008,Yamazaki1983}. However, at the CAM-B3LYP level of theory, a state with predominantly ($\pi\pi^*$) character is not consistently predicted to be the second lowest excited state along the MD trajectory, often swapping places with higher lying states of different symmetry. In this work, we only explicitly consider two excited states in the quantum dynamics. The appropriate adiabatic states along the MD trajectory are selected by computing the four lowest excited states for each snapshot, and selecting the S$_2$($\pi\pi^*$) and S$_1$($n\pi^*$) states as the states that show the most overlap with reference transition dipole moments computed for those states in the ground state optimized geometry. We note that the transition dipole moment for the S$_2$($\pi\pi^*$) state lies within the plane of the molecule, perpendicular to the axis passing through the two N atoms, whereas for the S$_1$($n\pi^*$) state the transition dipole moment is perpendicular to the plane of the molecule. While state crossings do occur along the trajectory, little mixing with other adiabatic states is observed, such that the selection of the two excited states of interest from the low energy excited state manifold is straightforward. The adiabatic states and their transition dipole moments are then either used for constructing linear absorption spectra in the GNCT scheme (see Sec.~\ref{sec:Theory_GNCT}), or, following a diabatization procedure (see Sec.~\ref{sec:Theory_diabatization}), define the diabatic fluctuation and coupling spectral densities for the LVC Hamiltonian in the T-TEDOPA approach. Additional details on the diabatization procedure can be found in SI Sec.~II. 

The CAM-B3LYP functional overestimates the energy difference between the diabatic S$_1$ and S$_2$ states in the Condon region compared with experiment\cite{Halverson1951_pyrazine_solution,Yamazaki1983,Samir2020_pyrazine_vac}, predicting an average energy difference in the ground state equilibrium of 1.36~eV along the MD trajectory for pyrazine in vacuum. In order to obtain realistic lineshapes, in the tensor-network approach we adjust the adiabatic energy gap to 0.84~eV, in line with parameterizations carried out for MCTDH calculations\cite{Raab1999}. The same shift of $-0.52$~eV was also applied to the average S$_2$ energies of pyrazine in cyclohexane and water. For the GNCT, average S$_2$ energies are also adjusted to align the spectral lineshape with both simulated T-TEDOPA and experimental lineshapes. We stress that these are the only parameter adjustments made in the system Hamiltonian and all other quantities necessary for predicting optical spectra in the two approaches are directly derived from the MD sampling of the PES.

The total computational cost associated with parameterizing the LVC system Hamiltonian directly from MD (combined cost of ground state MD and the computation of vertical excitation energies) was 44 GPU hours for pyrazine in vacuum, 191 GPU hours for pyrazine in cyclohexane and 223 GPU hours for pyrazine in water on an NVIDIA~RTX~3080 GPU. 

\subsection{Quantum dynamics and spectra simulation}\label{sec:quantum_dynamics_spectra_simulation}
All tensor-network dynamics were performed using a modified version of the \texttt{MPSDynamics} Julia package\cite{dunnett_angus_2021_5106435, lacroix2024mpsdynamicsjl}, with modifications carried out to enable efficient time-propagation using the one-site Time-Dependent Variational Principle (1TDVP) on GPUs. All calculations on model systems and the pyrazine molecule were converged with respect to the bond dimension ($D$), the local Fock space ($d$) and the chain length ($N$). Convergence tests and converged parameters can be found in SI Sec.~VII. A time-step of 10~a.u. (0.242~fs) was used throughout for the time propagation, and typical propagation lengths were 150~fs (for absorption) or 250~fs (for fluorescence). For one of the model systems considered, namely MSB30 (Fig.~\ref{fig:fig_3_model_system_spectra}~b), a time-step of 10~a.u. led to a spurious negative feature in the absorption spectrum, and a time-step of 5~a.u. was used instead (see SI Sec.~XV). Typical computational efforts to construct absorption spectra were 6-8 hrs on a single NVIDA RTX 2080~Ti GPU. For fluorescence lineshapes, several individual spectra with different delay times have to be averaged (see Sec.~\ref{subsec:optical_spectra}), increasing typical computational efforts to approximately 100-150 GPU hours.

For GNCT calculations that include intensity-borrowing effects beyond the Condon approximation through dipole fluctuations but do not explicitly couple multiple excited states, all spectra were constructed using the \texttt{MolSpeckPy} \texttt{Python} package freely available on GitHub\cite{Spectroscopy_python_code}. The dipole-dipole response function was computed for adiabatic S$_1$ and S$_2$ states individually, and the total absorption spectrum was computed as a sum of both contributions. The S$_2$ spectrum was shifted down in energy by 0.48~eV (vacuum and cyclohexane) and 0.44~eV (water) to align with experiment.  All response functions were computed for 500~fs using a 0.25~fs time-step. The computational cost of evaluating the lineshape in the GNCT scheme is negligible, taking less than 1 minute on a single CPU.

For GCT, GNCT and T-TEDOPA, optical spectra were simulated at $T=300$~K, matching the temperature used in the MD sampling of the pyrazine system. 

\section{Results and discussion}
\subsection{Linear optical spectra in the presence of a CI: Model systems}
\label{subsec:results_model}

\begin{figure*}
\centering
\includegraphics[width=0.95\textwidth]{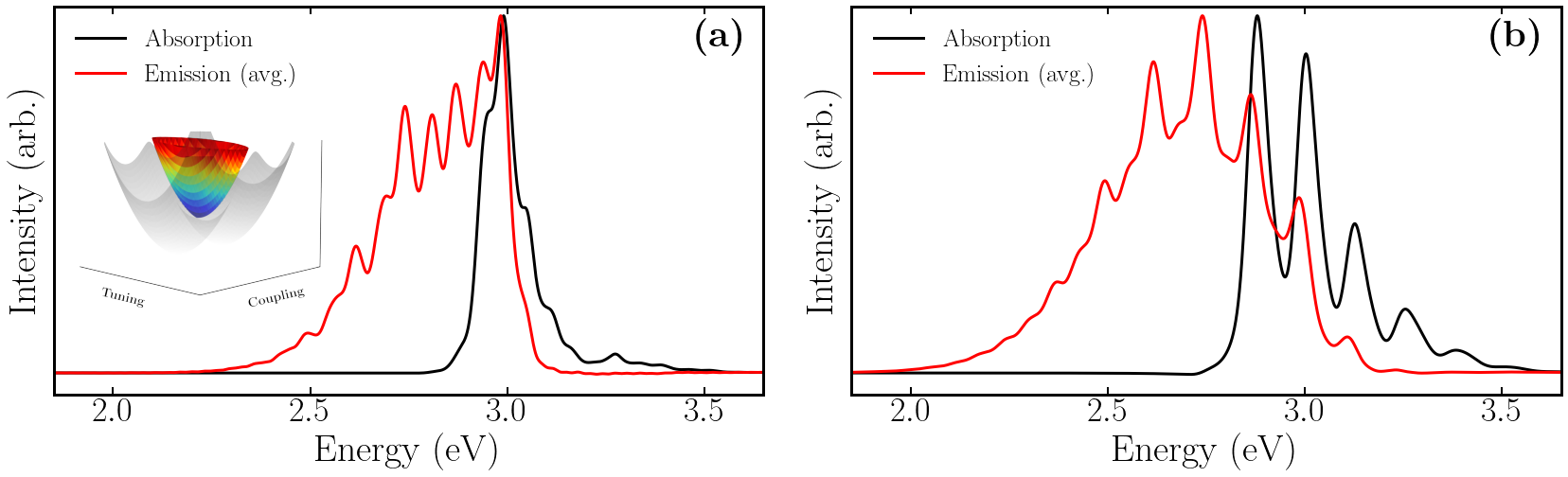}
\caption{Absorption and converged emission spectra for two model systems, MSA30 (a) and MSB30 (b). Parameters for each model system are provided in SI Sec.~VIII. The emission spectra are formed from the average of the spectra produced from 60 uniformly-spaced samplings of the trajectory between 116.1 and 203.2~fs. The inset in (a) is a schematic depiction of the adiabatic S$_1$ and S$_2$ potential energy surfaces formed from the diabatic surfaces of each model system.}
\label{fig:fig_3_model_system_spectra}
\end{figure*}

As described in Sec.~\ref{subsec:model_systems}, we restrict focus to a set of generic parameters which describe systems with an optically bright S$_1$ and dark S$_2$, and resulting intensity borrowing effects due to the presence of a CI near the Condon region. 

Specifically, we consider four model systems, MSA30, MSB30, MSA200 and MSA200, the explicit parameters of each are fully listed in SI Sec.~VIII, although here we also provide an overview. Common to all model systems are two tuning modes and one coupling mode of respective frequencies $\omega_t = 1,000$~cm$^{-1}$ and $\omega_c = 500$~cm$^{-1}$. These``molecular'' modes appear in the spectral densities parameterizing energy gap and coupling fluctuations as Gaussian profiles with widths of $\sigma_t=\sigma_c=8.493$~cm$^{-1}$. Additionally, included in each spectral density is an ``environmental'' Debye portion with cutoff frequencies $\omega_\text{env}$ characterizing the relaxation timescale of the low frequency environmental degrees of freedom coupled to the system. For the MSA30 (MSA200) and MSB30 (MSB200) model systems, we use $\omega_\textrm{env}=30$~cm$^{-1}$ (200~cm$^{-1}$). 

All four model systems employ identical coupling spectral densities, with a total reorganization energy of 124~meV distributed between molecular and environmental components, predominantly in the molecular contribution (111.6~meV). The key distinction between MSA and MSB model systems lies in the distribution of reorganization energy within the S$_1$ and S$_2$ tuning spectral densities. In the S$_1$ spectral density, MSB systems allocate 124~meV of reorganization energy to molecular and environmental modes, mirroring the coupling spectral density distribution. In contrast, MSA systems have a total energy of 24~meV in the S$_1$ spectral density, split evenly between molecular and environmental contributions. All S$_2$ spectral densities maintain a total reorganization energy of 124~meV, with MSA model systems more heavily weighted towards the molecular portion (117.8~meV) and MSB systems distributing energy evenly between molecular and environmental components. These energies are also displayed in Table I of the SI.

We focus on the case in which the energy gap fluctuations between S$_1$ and S$_2$ along the tuning mode $t$ are fully, positively correlated (FPC; $c\left(\omega\right)=1$). The impact of different correlations in the tuning modes is explored in SI Sec.~VI, where it is found that negatively correlated or uncorrelated fluctuations reduce observed nonadiabatic effects in the spectral lineshape for the chosen model system parameterizations.

\begin{figure*}
\centering
\includegraphics[width=0.95\textwidth]{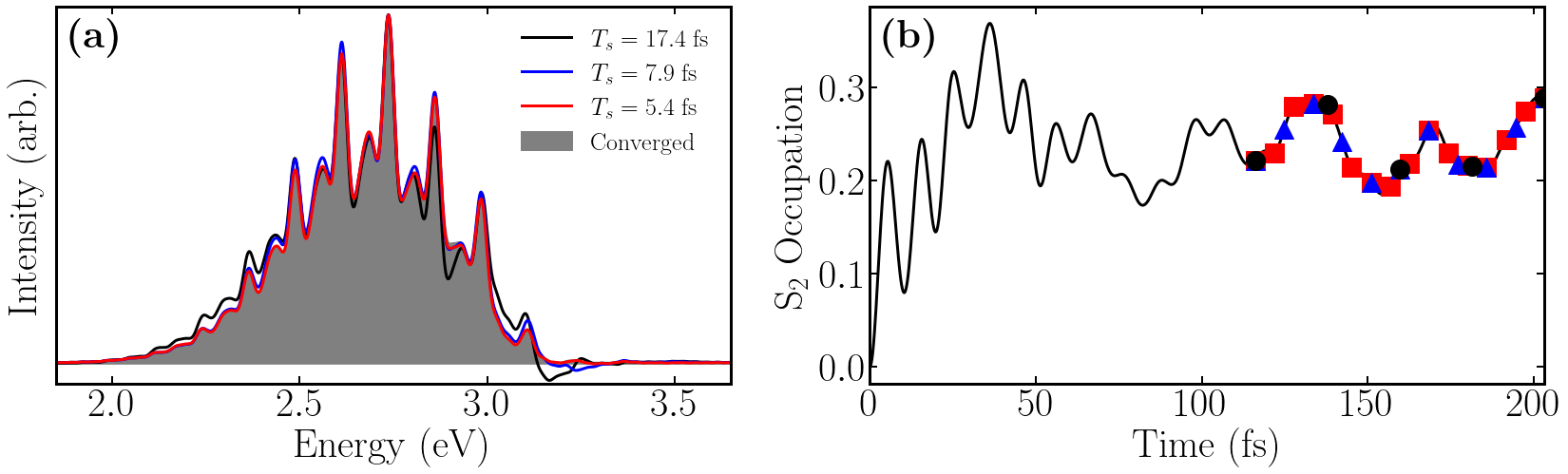}
\caption{Schematic of the procedure used for generating converged emission spectra with MSB200 as an example. (a) shows the convergence of the spectrum as the sampling interval $T_\mathrm{s}$ of the excited state trajectory between $116.1$ and $203.2$~fs is reduced. The markers in (b) indicate points in the excited state population dynamics from which samplings of the initial states $|\chi_\textrm{R}\rangle$ used in (a) were taken.}
\label{fig:fig_4_model_system_emission_convergence}
\end{figure*}

Fig.~\ref{fig:fig_3_model_system_spectra} shows computed absorption and emission lineshapes using the T-TEDOPA formalism (see Sec.~\ref{subsec:optical_spectra}) for the two model system parameterizations ``A'' and ``B'' coupled to the slow solvent bath of $\omega_\mathrm{env}=30$~cm$^{-1}$. In both system parameterizations, nonadiabatic effects on the optical spectra are clearly evident in the strongly broken mirror symmetry between absorption and emission lineshapes, with the emission spectrum becoming very broad and exhibiting vibronic progressions not present in the absorption spectrum. For the MSA30 model system, this symmetry breaking is particularly prominent, due to the low reorganization energy contained in the S$_1$ state coupling to the tuning mode. This causes the absorption lineshape to be very narrow, with almost no contribution from the dark S$_2$ state. Emission on the other hand occurs from a highly mixed state, leading to strong intensity borrowing of the S$_2$ state from S$_1$. For the MSB30 model, the increased reorganization energy in the high frequency tuning mode of S$_1$ leads to a pronounced vibronic progression in the absorption spectrum, again with little contribution from the dark S$_2$ state. The emission spectrum is, similar to the MSA parameterization, significantly broader than the absorption lineshape. A long, featureless tail in the spectrum indicates vibronic transitions that are highly mixed in character, and can no longer be assigned to either S$_1$ or S$_2$. The MSA and MSB parameterizations demonstrate that complex lineshapes can arise even for a small number of vibrational modes coupling to the system, as long as strong nonadiabatic effects mix the two electronic states. 

To compute emission lineshapes in the scheme outlined in Sec.~\ref{subsec:optical_spectra}, it is necessary to average over several emitting states obtained for different excited state relaxation times. Fig.~\ref{fig:fig_4_model_system_emission_convergence} outlines the convergence of the lineshape with respect to the averaging scheme for a specific model system parameterization. We note that upon excitation into S$_1$, the system undergoes ultrafast population transfer to S$_2$ via the CI. After nearly 100~fs, most of the wave packet dynamics is completed, but the excited state populations still undergo some fluctuations around a mean of $\approx25$\% of S$_2$ population, indicating some long-timescale influence of the CI on the system dynamics. The long-lived oscillations in population dynamics can likely be ascribed to the relatively low reorganization energy of 12~meV contained in the low frequency part of the S$_1$ and the coupling spectral density, providing limited relaxation pathways through ``environmental'' modes. Nevertheless, averaging over a range of different initial states, from 116.1 to 203.2~fs, leads to a rapid convergence of the computed emission lineshape including the vibronic fine-structure, with 16 samples ($T_s=5.4$~fs) found to be sufficient. We stress that in realistic systems with stronger low frequency couplings to condensed phase environmental modes, we often find that a significantly smaller number of sampling points is necessary to obtain converged emission lineshapes\cite{Hunter2024}. 

\begin{figure}
\centering
\includegraphics[width=0.95\columnwidth]{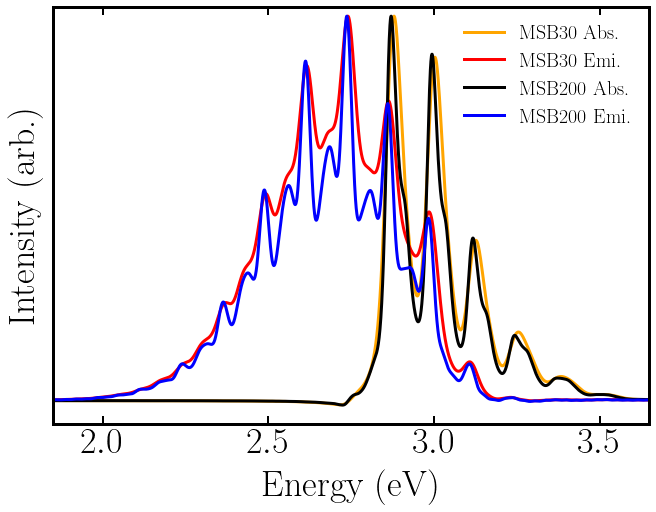}
\caption{Comparison of the absorption and emission spectra from model systems MSB30 and MSB200. MSB30 and MSB200 respectively have $\omega_\mathrm{env} = 30 \text{ cm}^{-1}$ and $\omega_\mathrm{env} = 200 \text{ cm}^{-1}$ but are otherwise identical.
The averaged emission spectra are formed from 60 uniformly-spaced samplings of the trajectory between 116.1 and 203.2~fs.}
\label{fig:fig_5_bath_speed}
\end{figure}

Simulating the system dynamics in the T-TEDOPA formalism allows us to model arbitrary spectral densities. In Fig.~\ref{fig:fig_5_bath_speed} we illustrate the effect of varying the characteristic cutoff frequency $\omega_\mathrm{env}$ controlling the speed of the low frequency solvent baths incorporated in the spectral densities on optical lineshapes for the MSB model. Specifically, we change $\omega_c$ from 30~cm$^{-1}$ (1.1~ps, corresponding to a slow bath), to 200~cm$^{-1}$ (167~fs). The absorption profile is almost independent with respect to the redistribution of reorganization energy into higher frequency environmental modes. In contrast, the emission profile is highly sensitive, with a \emph{slower} bath causing \emph{more} broadened features. The results again demonstrate that comparatively simple 2D model systems coupled to low frequency solvent baths can exhibit rich quantum dynamics and linear spectra due to the presence of a CI, and the computational workflow presented in this work is well suited to the study of these systems. 

\subsection{Realistic systems: The pyrazine molecule in solution}

\begin{figure*}
\centering
\includegraphics[width=0.95\textwidth]{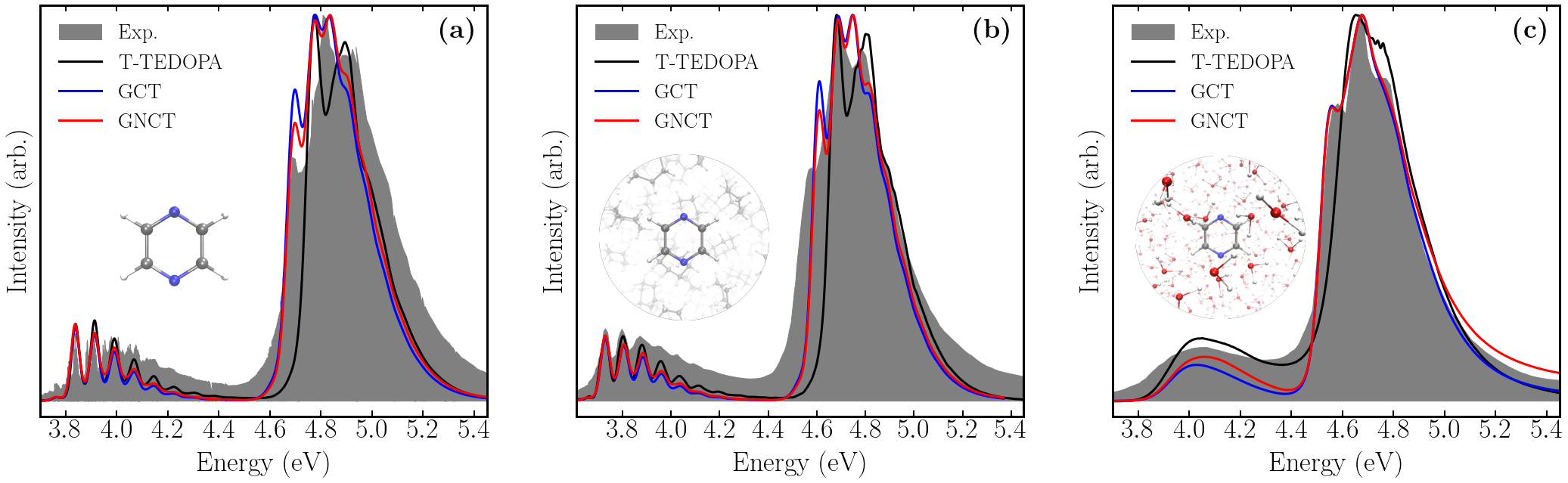}
\caption{Experimental linear absorption spectra for pyrazine in (a) vacuum\cite{Samir2020_pyrazine_vac}, (b) cyclohexane\cite{Halverson1951_pyrazine_solution} and (c) water\cite{Halverson1951_pyrazine_solution}, in comparison with spectra computed computed through the Gaussian Condon theory (GCT), the Gaussian Non-Condon Theory (GNCT) and the T-TEDOPA approach.}
\label{fig:fig_6_pyrazine_comparison}
\end{figure*}

We now turn to pyrazine, an ideal test system for the influence of strong couplings between excited states in the Condon region on optical spectra, since the optical lineshape and excited state relaxation dynamics of the molecule in the gas phase has been widely studied using a range of quantum dynamics approaches\cite{Raab1999,Burghardt2008,Neville2020}. Pyrazine also forms an interesting test case, as S$_1$ and S$_2$ are well-separated in the Condon region (by 0.84~eV, considerably larger than the 0.2~eV separation in all model systems studied in Sec.~\ref{subsec:results_model}), but are nevertheless strongly coupled, leading to ultrafast population transfer from the bright, higher-lying $\pi\pi^*$ to the $n\pi^*$ state upon excitation. Well-separated excited states are often taken to suggest that an approximate treatment in terms of independent excited states with non-Condon corrections to account for intensity-borrowing effects, such as provided by the GNCT scheme or the well-known Franck-Condon Herzberg-Teller approach, is justified. 

\begin{figure}
\centering
\includegraphics[width=0.95\columnwidth]{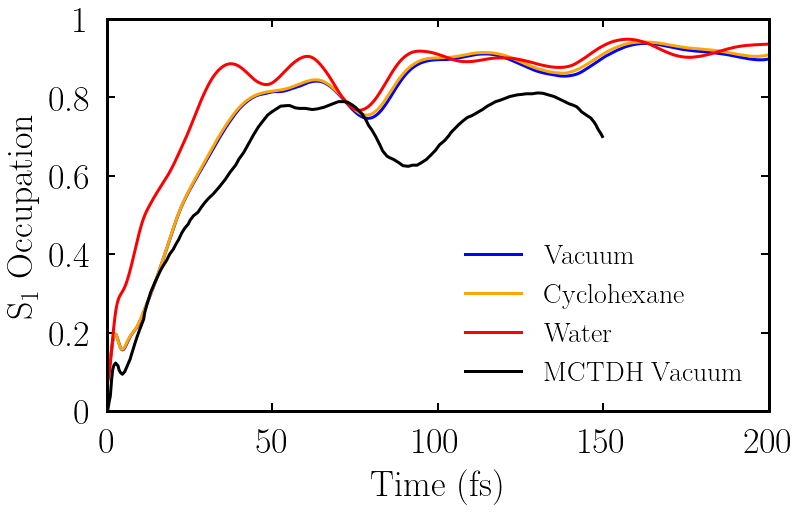}
\caption{A comparison of S$_1$ population dynamics for pyrazine in different solvents as predicted by T-TEDOPA, in comparison with benchmark MCTDH data from Ref.~\onlinecite{Raab1999}. For the MCTDH results, the initial S$_1$ population is zero, whereas for T-TEDOPA the initial wavefunction is a superposition of S$_1$ and S$_2$ weighted by their respective transition dipole moments.}
\label{fig:fig_7_quantum_dynamics}
\end{figure}

Simulated absorption lineshapes using GCT, GNCT and T-TEDOPA for pyrazine in vacuum, cyclohexane and water, in comparison with experimental spectra\cite{Samir2020_pyrazine_vac,Halverson1951_pyrazine_solution}, can be found in Fig.~\ref{fig:fig_6_pyrazine_comparison}. For pyrazine in vacuum and in cyclohexane, all approaches predict similar vibronic progressions in the low energy feature due to the $n\pi^*$ transition. Non-Condon effects incorporated in the GNCT lead to slightly more spectral weight on higher energy peaks in the vibronic progression compared with GCT, due to intensity borrowing driven by a 960~cm$^{-1}$ vibrational mode breaking the planar symmetry of the molecule (see SI Secs.~IX and X). The T-TEDOPA results due to fully coupled excited states show further enhanced intensities in the higher energy peaks in the vibronic progression. This is even more apparent for pyrazine in water, where the full nonadiabatic coupling between S$_1$ and S$_2$ accounted for in the tensor network formalism leads to significantly more intensity in the spectral range between the absorption maxima of the $n\pi^*$ and the $\pi\pi^*$ transition, in closer agreement with experiment. All approaches predict strongly broadened lineshapes for the $n\pi^*$ transition, which can be ascribed to low frequency solvent contribution in the S$_1$ spectral density (see SI Sec.~IX).  

The main differences between the GCT, GNCT and T-TEDOPA approaches in modeling the optical spectra for pyrazine in different solvents arise in the lineshape of the more intense high-energy $\pi\pi^*$ transition. The GCT consistently overestimates the intensity of the onset of the spectrum, and predicts wrong peak placements and spacing in the visible vibronic progression for the vacuum and cyclohexane results. Accounting for non-Condon effects in the transition dipole moment leads to some improvement compared with experiment, mainly in reducing the intensity of the first peak in the vibronic progression. Fully accounting for nonadiabatic effects in the coupling, however, strongly impacts the predicted lineshape. The first vibronic peak in the lineshape present in GCT, GNCT and the experimental lineshape disappears, leading to an overall narrower spectrum. Additionally, the intensity of the third vibronic peak in the GNCT lineshape is strongly suppressed, leading to a high-energy spectrum that is generally in better agreement with experiment. For pyrazine in water, all approaches yield more broadened vibronic features, and the GCT and GNCT lineshapes are in close agreement with each other. The spectrum due to fully coupled excited states mainly improves over the Gaussian theory by predicting a more pronounced vibronic shoulder in the high energies, but at the cost of reducing the intensity of the first vibronic peak. 

Additional calculations (see SI Secs.~IX, XI and XII) reveal that the spectral lineshape is strongly influenced by a 960~cm$^{-1}$ coupling mode breaking the planar symmetry of the molecule and an intense in-plane tuning mode at 610~cm$^{-1}$ that drives \emph{anti-correlated} energy gap fluctuations in the $n\pi^*$ and the $\pi\pi^*$ transition. The anti-correlated fluctuations in the tuning mode are particularly important, as enforcing correlated fluctuations results in a lineshape that is highly similar to that predicted by GNCT. Additionally, this tuning mode couples strongly to environmental interactions, undergoing a significant frequency shift in solution, likely due to hydrogen bonding interactions with the nitrogen atoms in the pyrazine ring. 

Fig.~\ref{fig:fig_7_quantum_dynamics} shows the population dynamics of the $n\pi^*$ state upon excitation as computed for the T-TEDOPA formalism for pyrazine in vacuum, cylcohexane and water, in comparison with MCTDH results obtained by Raab \emph{et al.}\cite{Raab1999} for a full 24-mode parameterization of pyrazine in vacuum. We note that the MCTDH results initialize the dynamics at zero population for the S$_1$ states, whereas the T-TEDOPA results generated in this work start the dynamics in a superposition of both electronic excited states, weighted by their respective transition dipole moment. However, T-TEDOPA results for cyclohexane and vacuum show a similar ultrafast ($\approx$~50~fs) population transfer from the $\pi\pi^*$ to the $n\pi^*$ transition as the MCTDH benchmark results. The main difference can be seen in a slightly faster overall transfer and faster oscillations in the population in our calculations when compared to the benchmark data. Pyrazine in water, on the other hand, shows a considerably more rapid population transfer, which we ascribe to the reduced S$_1$-S$_2$ gap in the Condon region due to environmental polarization effects predicted by the MD sampling of the energy gap fluctuations (see SI Sec.~IX). 

We find that both the GNCT relying on approximating the spectrum in terms of two decoupled adiabatic states and the tensor network formalism explicitly accounting for couplings between the two states do not yield spectra in exact agreement with experimental data, in contrast with benchmark MCTDH results for a 24-mode model in vacuum of Raab and co-workers\cite{Raab1999}. However, we note that in our calculations, the only adjustment of the Hamiltonian parameters obtained from MD sampling is the average S$_1$-S$_2$ gap in the Condon region, whereas the 24-mode model did require additional reparameterization and the use of experimental rather than calculated vibrational frequencies to achieve agreement with experiment\cite{Raab1999}. In SI Sec.~XIII we demonstrate that adjusting the spectral densities parameterized from TDDFT/CAM-B3LYP sampling to align with couplings used in a simplified four-mode model of pyrazine\cite{Krempl1994} derived from multi-reference calculations indeed restores the low-energy shoulder absent in the S$_2$ band in the T-TEDOPA formalism and yields a spectrum in close agreement with experiment. These results suggest that the main source of error in the calculations presented here is the quality of the PESs as sampled with our chosen TDDFT method. 

The results presented here demonstrate the importance of accounting for environmental interactions in determining spectral lineshapes, correctly predicting fully broadened lineshapes devoid of vibronic fine structure for the $n\pi^*$ transition in water. We stress that no phenomenological broadening was added to any of the lineshapes presented in Fig.~\ref{fig:fig_6_pyrazine_comparison}, and all spectral broadening effects arise directly from the MD sampling of anharmonicities and solvent interactions, as well as the inclusion of finite temperature effects. The considerable differences between the GNCT and T-TEDOPA also demonstrate that strong non-Condon effects in which transitions cannot be treated as decoupled, exist even for ``well-separated'' excited states. Both the increased intensity in the $n\pi^*$ transition and the reorganization of vibronic peaks in the $\pi\pi^*$ transition can be directly ascribed to nonadiabatic effects and the ultrafast population transfer to the $n\pi^*$ state upon excitation.

\subsection{Timing tests on GPUs}
\label{sec:results_timing_gpu}

\begin{figure}
\centering
\includegraphics[width=0.95\columnwidth]{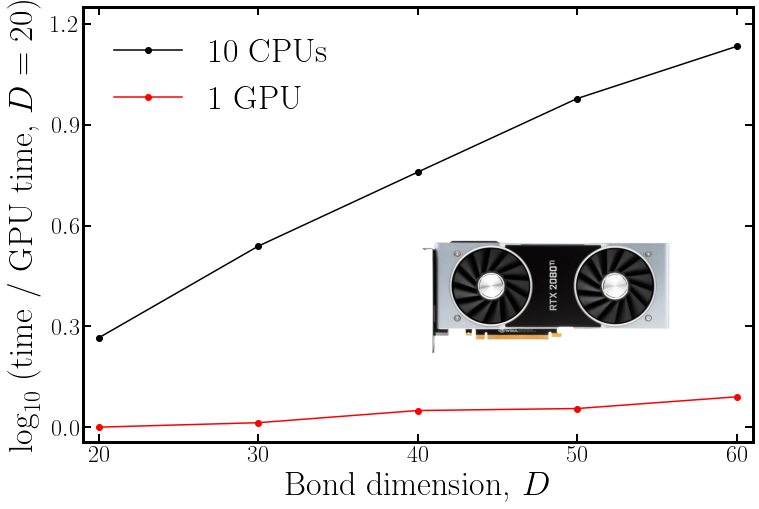}
\caption{Relative time taken for a single time-step of the 1TDVP propagation algorithm between 10~CPUs on an Intel Xeon Gold 6230R processor and 1 NVIDIA RTX~2080~Ti GPU for different choices of the bond dimension $D$, relative to the time taken for $D=20$ on the single GPU. All timing tests were performed for pyrazine in vacuum, with a chain length $N=150$ and a local Fock space $d=30.$}
\label{fig:fig_8_gpus}
\end{figure}

The chain mapping procedure in the T-TEDOPA scheme enables numerically exact finite-temperature quantum dynamics and the calculation of spectroscopic observables in complex condensed phase systems. However, these calculations are still computationally expensive, especially when computing fluorescence lineshapes, as these generally require longer propagation times, larger bond dimensions and an averaging over several initial states obtained through varying excited state relaxation times. In this work, all tensor network calculations were run on GPUs using the modified 1TDVP algorithm described in Sec.~\ref{subsec:GPUtensornetwork}, significantly accelerating calculation times. 

Relative timings between the CPU-only and GPU-accelerated versions of the \texttt{MPSDynamics.jl} package can be found in Fig.~\ref{fig:fig_8_gpus}. Timing tests were performed for 20 time steps of 1TDVP propagation for pyrazine in vacuum, using a chain length $N=150$ and a local Fock space of $d=30$, varying the bond dimension $D$ from 20 to 60. The number of CPUs used in the benchmark reference results was obtained by determining the maximum number of CPUs for which a parallel speedup could still be achieved, as tested for $D=50$. Fig.~\ref{fig:fig_8_gpus} demonstrates over an order of magnitude speedup in the 1TDVP propagation carried out on a single NVIDIA RTX~2080~Ti GPU in comparison with ten CPUs on a single Intel Xeon Gold 6230R processor. The resulting speed-up translates to a few hours of wall-time for computing a converged linear absorption spectrum on a single GPU, with the same calculation requiring 4-5 days of wall-time on a CPU node. GPU calculations are mainly limited not by wall-time but rather by memory requirements of the calculation. However, the relatively modest 11~GB of GPU RAM on the RTX~2080~Ti GPU was found to be sufficient for obtaining fully converged spectra and population dynamics for all systems studied in this work. 

\subsection{Computational effort in realistic systems}
As demonstrated in Sec.~\ref{sec:results_timing_gpu}, computing optical properties of systems with coupled excited states through MD sampling and the T-TEDOPA formalism is highly efficient when utilizing GPUs. For the pyrazine molecule, the test system chosen for the purpose of this work, the total computational cost for constructing an absorption spectrum due to two coupled excited states is approximately 50~GPU hrs (in vacuum) or 200~GPU hrs (in solution). A large fraction of this computational cost is identical for the GCT and GNCT approaches, only saving 6-8~GPU hours per calculation by avoiding the cost associated with performing explicit wavefunction dynamics of coupled excited states.

Even for a small, rigid molecule like pyrazine, the dominant fraction of the computational cost lies not in the quantum dynamics to compute the appropriate dipole-dipole response function for coupled excited states, but rather the MD sampling of the ground-state dynamics and the computation of vertical excitation energies to obtain well-converged spectral densities of system-bath coupling. We stress that, thanks to the chain mapping procedure, the computational cost associated with the quantum dynamics is independent of the number of \emph{physical} vibrational degrees of freedom in the system. This allows the outlined approach to account for arbitrary complex condensed phase environment interactions and arbitrarily large chromophores, without increasing the cost of the quantum dynamics, as long as the system of interest can be reduced to two coupled excited states. 

For pyrazine in vacuum, parameterizing an LVC Hamiltonian directly from MD is necessarily computationally inefficient compared to a direct parameterization based on normal mode calculations of the molecule, as was done in parameterizations for MCTDH calculations\cite{Raab1999,Burghardt2008}. For significantly larger molecules, especially molecules undergoing low-frequency large-amplitude motion, we expect the MD-based approach to be much more comparable in cost, with an additional advantage that the explicit sampling of the PES approximately accounts for anharmonic effects. Additionally, computing spectral densities directly from MD allows for the straightforward inclusion of solute-solvent interactions and environmental polarization effects, which can significantly impact computed lineshapes\cite{Zuehlsdorff2020,Hunter2024}. Computational cost associated with MD sampling and vertical excitation energies generally increases with the size of the QM region (with most DFT-based approaches showing an $\mathcal{O}(N^3)$ scaling with system size). However, we note that the cost associated with computing vertical excitation energies can be reduced by more than an order of magnitude by employing machine-learning (ML) techniques\cite{Chen2020,Chen2023}, making the outlined approach computationally feasible even for large systems embedded in complex environments.

\section{Conclusion}

In this work, we have compared two approaches for computing linear optical spectra of molecules in complex environments in the presence of non-Condon and nonadiabatic effects. Both approaches rely on sampling excitation energies and transition dipole moments directly from molecular dynamics simulations, such that molecular vibrations and couplings to the condensed phase environment are treated on an equal footing, and energy-gap and dipole fluctuations are encoded in spectral densities of system-bath coupling. In the regime of weakly coupled excited states, non-Condon effects such as intensity borrowing are accounted for through the transition dipole fluctuations of \emph{adiabatic} electronic states through a recently introduced Gaussian Non-Condon Theory (GNCT). For strongly coupled excited states, transition dipoles are instead used to construct \emph{diabatic} states and couplings and the finite-temperature quantum dynamics of the resulting Hamiltonian is then computed using GPU-accelerated tensor-network approaches (T-TEDOPA). 

The performance of the approaches was studied on a 2-mode model system of a CI coupled to a low-frequency solvent bath (for T-TEDOPA) and the pyrazine molecule in different environments (for GNCT and T-TEDOPA). For pyrazine, both approaches were found to predict similar lineshapes for the low energy $n\pi^*$ state, and significant solvent broadening effects in water, generally reproducing the main features of the experimental lineshape. For the high-energy $\pi\pi^*$ transition, significant differences in the predicted lineshapes emerge, demonstrating the presence of nonadiabatic effects which go beyond a coupling of the transition dipole moment to nuclear degrees of freedom. Since both approaches originate from the same sampling of nuclear degrees of freedom in an MD trajectory, comparing the resulting lineshapes can yield direct insight into the types of non-Condon effects that can be described through independent \emph{adiabatic} states, and spectral features that can only arise from considering the explicit quantum dynamics of strongly coupled \emph{diabatic} states. All calculations in this work were performed on GPUs, using GPU-accelerated electronic structure methods for the MD and computation of energy-gap fluctuations, as well as a GPU-accelerated implementation of the 1TDVP for the quantum dynamics. 

All calculations of optical spectra presented in this work were performed using open-source software packages, and we have here provided a range of \texttt{Python} tools to facilitate all calculation steps starting from raw data generated from MD. We expect the methodology outlined here to be applicable to a wide range of systems embedded in complex condensed phase environments, from solvated dyes to pigment-protein complexes.

\section{Supplementary material}
See supplementary material for additional computational details of the MD dynamics, the diabatization procedure, chain mapping schemes, convergence tests, and a comparison of the computed spectral densities for pyrazine in vacuum and in solution. 

\section{Data availability}
The underlying data of this publication is made available under the following persistent DOI: 10.5281/zenodo.11786828. 
The \texttt{Python} scripts used for performing Eckart rotations, generating the spectral densities from molecular dynamics data, performing the chain mapping of the LVC Hamiltonian, as well as post-processing of the dipole-dipole correlation functions can be found the same DOI. 
All tensor network simulations were performed using a modified version of the \texttt{MPSDynamics.jl} Julia package\cite{dunnett_angus_2021_5106435} capable of running on GPUs that is available on GitHub\cite{MPSDynamics_GPU} and under the following DOI: 10.5281/zenodo.10712009. Gaussian Non-Condon Theory (GNCT) calculations were carried out using the \texttt{MolSpeckPy} package available on GitHub\cite{Spectroscopy_python_code}.

\acknowledgements{T. J. Z. acknowledges startup funding provided by Oregon State University. }

\bibliography{bibliography}

\begin{thebibliography}{96}%
\makeatletter
\providecommand \@ifxundefined [1]{%
 \@ifx{#1\undefined}
}%
\providecommand \@ifnum [1]{%
 \ifnum #1\expandafter \@firstoftwo
 \else \expandafter \@secondoftwo
 \fi
}%
\providecommand \@ifx [1]{%
 \ifx #1\expandafter \@firstoftwo
 \else \expandafter \@secondoftwo
 \fi
}%
\providecommand \natexlab [1]{#1}%
\providecommand \enquote  [1]{``#1''}%
\providecommand \bibnamefont  [1]{#1}%
\providecommand \bibfnamefont [1]{#1}%
\providecommand \citenamefont [1]{#1}%
\providecommand \href@noop [0]{\@secondoftwo}%
\providecommand \href [0]{\begingroup \@sanitize@url \@href}%
\providecommand \@href[1]{\@@startlink{#1}\@@href}%
\providecommand \@@href[1]{\endgroup#1\@@endlink}%
\providecommand \@sanitize@url [0]{\catcode `\\12\catcode `\$12\catcode
  `\&12\catcode `\#12\catcode `\^12\catcode `\_12\catcode `\%12\relax}%
\providecommand \@@startlink[1]{}%
\providecommand \@@endlink[0]{}%
\providecommand \url  [0]{\begingroup\@sanitize@url \@url }%
\providecommand \@url [1]{\endgroup\@href {#1}{\urlprefix }}%
\providecommand \urlprefix  [0]{URL }%
\providecommand \Eprint [0]{\href }%
\providecommand \doibase [0]{https://doi.org/}%
\providecommand \selectlanguage [0]{\@gobble}%
\providecommand \bibinfo  [0]{\@secondoftwo}%
\providecommand \bibfield  [0]{\@secondoftwo}%
\providecommand \translation [1]{[#1]}%
\providecommand \BibitemOpen [0]{}%
\providecommand \bibitemStop [0]{}%
\providecommand \bibitemNoStop [0]{.\EOS\space}%
\providecommand \EOS [0]{\spacefactor3000\relax}%
\providecommand \BibitemShut  [1]{\csname bibitem#1\endcsname}%
\let\auto@bib@innerbib\@empty
\bibitem [{\citenamefont {Loco}\ and\ \citenamefont
  {Cupellini}(2018)}]{Loco2018b}%
  \BibitemOpen
  \bibfield  {author} {\bibinfo {author} {\bibfnamefont {D.}~\bibnamefont
  {Loco}}\ and\ \bibinfo {author} {\bibfnamefont {L.}~\bibnamefont
  {Cupellini}},\ }\bibfield  {title} {\enquote {\bibinfo {title} {{Modeling the
  absorption lineshape of embedded systems from molecular dynamics: A tutorial
  review}},}\ }\href {https://doi.org/10.1002/qua.25726} {\bibfield  {journal}
  {\bibinfo  {journal} {Int. J. Quantum Chem.}\ }\textbf {\bibinfo {volume}
  {119}},\ \bibinfo {pages} {e25726} (\bibinfo {year} {2018})}\BibitemShut
  {NoStop}%
\bibitem [{\citenamefont {Loco}\ \emph {et~al.}(2018)\citenamefont {Loco},
  \citenamefont {Jurinovich}, \citenamefont {Cupellini}, \citenamefont
  {Menger},\ and\ \citenamefont {Mennucci}}]{Loco_2018}%
  \BibitemOpen
  \bibfield  {author} {\bibinfo {author} {\bibfnamefont {D.}~\bibnamefont
  {Loco}}, \bibinfo {author} {\bibfnamefont {S.}~\bibnamefont {Jurinovich}},
  \bibinfo {author} {\bibfnamefont {L.}~\bibnamefont {Cupellini}}, \bibinfo
  {author} {\bibfnamefont {M.~F. S.~J.}\ \bibnamefont {Menger}},\ and\ \bibinfo
  {author} {\bibfnamefont {B.}~\bibnamefont {Mennucci}},\ }\bibfield  {title}
  {\enquote {\bibinfo {title} {{The modeling of the absorption lineshape for
  embedded molecules through a polarizable QM/MM approach}},}\ }\href
  {https://doi.org/10.1039/c8pp00033f} {\bibfield  {journal} {\bibinfo
  {journal} {Photochem. Photobiol. Sci.}\ }\textbf {\bibinfo {volume} {17}},\
  \bibinfo {pages} {552--560} (\bibinfo {year} {2018})}\BibitemShut {NoStop}%
\bibitem [{\citenamefont {Zuehlsdorff}\ and\ \citenamefont
  {Isborn}(2019)}]{Zuehlsdorff_IJQC_2019}%
  \BibitemOpen
  \bibfield  {author} {\bibinfo {author} {\bibfnamefont {T.~J.}\ \bibnamefont
  {Zuehlsdorff}}\ and\ \bibinfo {author} {\bibfnamefont {C.~M.}\ \bibnamefont
  {Isborn}},\ }\bibfield  {title} {\enquote {\bibinfo {title} {Modeling
  absorption spectra of molecules in solution},}\ }\href
  {https://doi.org/10.1002/qua.25719} {\bibfield  {journal} {\bibinfo
  {journal} {Int. J. Quantum Chem.}\ }\textbf {\bibinfo {volume} {119}},\
  \bibinfo {pages} {e25719} (\bibinfo {year} {2019})}\BibitemShut {NoStop}%
\bibitem [{\citenamefont {Zuehlsdorff}\ \emph {et~al.}(2021)\citenamefont
  {Zuehlsdorff}, \citenamefont {Shedge}, \citenamefont {Lu}, \citenamefont
  {Hong}, \citenamefont {Aguirre}, \citenamefont {Shi},\ and\ \citenamefont
  {Isborn}}]{Zuehlsdorff2021}%
  \BibitemOpen
  \bibfield  {author} {\bibinfo {author} {\bibfnamefont {T.~J.}\ \bibnamefont
  {Zuehlsdorff}}, \bibinfo {author} {\bibfnamefont {S.~V.}\ \bibnamefont
  {Shedge}}, \bibinfo {author} {\bibfnamefont {S.-Y.}\ \bibnamefont {Lu}},
  \bibinfo {author} {\bibfnamefont {H.}~\bibnamefont {Hong}}, \bibinfo {author}
  {\bibfnamefont {V.~P.}\ \bibnamefont {Aguirre}}, \bibinfo {author}
  {\bibfnamefont {L.}~\bibnamefont {Shi}},\ and\ \bibinfo {author}
  {\bibfnamefont {C.~M.}\ \bibnamefont {Isborn}},\ }\bibfield  {title}
  {\enquote {\bibinfo {title} {Vibronic and environmental effects in
  simulations of optical spectroscopy},}\ }\href
  {https://doi.org/10.1146/annurev-physchem-090419-051350} {\bibfield
  {journal} {\bibinfo  {journal} {Annu. Rev. Phys. Chem.}\ }\textbf {\bibinfo
  {volume} {72}},\ \bibinfo {pages} {165--188} (\bibinfo {year}
  {2021})}\BibitemShut {NoStop}%
\bibitem [{\citenamefont {Orlandi}\ and\ \citenamefont
  {Siebrand}(1973)}]{Orlandi1973}%
  \BibitemOpen
  \bibfield  {author} {\bibinfo {author} {\bibfnamefont {G.}~\bibnamefont
  {Orlandi}}\ and\ \bibinfo {author} {\bibfnamefont {W.}~\bibnamefont
  {Siebrand}},\ }\bibfield  {title} {\enquote {\bibinfo {title} {{Theory of
  vibronic intensity borrowing. Comparison of Herzberg-Teller and
  Born-Oppenheimer coupling}},}\ }\href {https://doi.org/10.1063/1.1679014}
  {\bibfield  {journal} {\bibinfo  {journal} {J. Chem. Phys.}\ }\textbf
  {\bibinfo {volume} {4513}},\ \bibinfo {pages} {4513--4523} (\bibinfo {year}
  {1973})}\BibitemShut {NoStop}%
\bibitem [{\citenamefont {Curchod}\ and\ \citenamefont
  {Mart\'{i}nez}(2018)}]{Curchod2018}%
  \BibitemOpen
  \bibfield  {author} {\bibinfo {author} {\bibfnamefont {B.~F.~E.}\
  \bibnamefont {Curchod}}\ and\ \bibinfo {author} {\bibfnamefont {T.~J.}\
  \bibnamefont {Mart\'{i}nez}},\ }\bibfield  {title} {\enquote {\bibinfo
  {title} {Ab initio nonadiabatic quantum molecular dynamics},}\ }\href
  {https://doi.org/10.1021/acs.chemrev.7b00423} {\bibfield  {journal} {\bibinfo
   {journal} {Chem. Rev.}\ }\textbf {\bibinfo {volume} {118}},\ \bibinfo
  {pages} {3305--3336} (\bibinfo {year} {2018})}\BibitemShut {NoStop}%
\bibitem [{\citenamefont {Santoro}\ \emph {et~al.}(2008)\citenamefont
  {Santoro}, \citenamefont {Lami}, \citenamefont {Improta}, \citenamefont
  {Bloino},\ and\ \citenamefont {Barone}}]{Santoro_2008}%
  \BibitemOpen
  \bibfield  {author} {\bibinfo {author} {\bibfnamefont {F.}~\bibnamefont
  {Santoro}}, \bibinfo {author} {\bibfnamefont {A.}~\bibnamefont {Lami}},
  \bibinfo {author} {\bibfnamefont {R.}~\bibnamefont {Improta}}, \bibinfo
  {author} {\bibfnamefont {J.}~\bibnamefont {Bloino}},\ and\ \bibinfo {author}
  {\bibfnamefont {V.}~\bibnamefont {Barone}},\ }\bibfield  {title} {\enquote
  {\bibinfo {title} {Effective method for the computation of optical spectra of
  large molecules at finite temperature including the {D}uschinsky and
  {H}erzberg-{T}eller effect: The {Q}x band of porphyrin as a case study},}\
  }\href {https://doi.org/10.1063/1.2929846} {\bibfield  {journal} {\bibinfo
  {journal} {J. Chem. Phys.}\ }\textbf {\bibinfo {volume} {128}},\ \bibinfo
  {pages} {224311} (\bibinfo {year} {2008})}\BibitemShut {NoStop}%
\bibitem [{\citenamefont {Baiardi}, \citenamefont {Bloino},\ and\ \citenamefont
  {Barone}(2013)}]{Baiardi2013}%
  \BibitemOpen
  \bibfield  {author} {\bibinfo {author} {\bibfnamefont {A.}~\bibnamefont
  {Baiardi}}, \bibinfo {author} {\bibfnamefont {J.}~\bibnamefont {Bloino}},\
  and\ \bibinfo {author} {\bibfnamefont {V.}~\bibnamefont {Barone}},\
  }\bibfield  {title} {\enquote {\bibinfo {title} {General time dependent
  approach to vibronic spectroscopy including {F}ranck--{C}ondon,
  {H}erzberg--{T}eller, and {D}uschinky effects},}\ }\href
  {https://doi.org/10.1021/ct400450k} {\bibfield  {journal} {\bibinfo
  {journal} {J. Chem. Theory Comput.}\ }\textbf {\bibinfo {volume} {9}},\
  \bibinfo {pages} {4097--4115} (\bibinfo {year} {2013})}\BibitemShut {NoStop}%
\bibitem [{\citenamefont {de~Souza}, \citenamefont {Neese},\ and\ \citenamefont
  {Izs\'{a}k}(2018)}]{deSouza2018}%
  \BibitemOpen
  \bibfield  {author} {\bibinfo {author} {\bibfnamefont {B.}~\bibnamefont
  {de~Souza}}, \bibinfo {author} {\bibfnamefont {F.}~\bibnamefont {Neese}},\
  and\ \bibinfo {author} {\bibfnamefont {R.}~\bibnamefont {Izs\'{a}k}},\
  }\bibfield  {title} {\enquote {\bibinfo {title} {On the theoretical
  prediction of fluorescence rates from first principles using the path
  integral approach},}\ }\href {https://doi.org/10.1063/1.5010895} {\bibfield
  {journal} {\bibinfo  {journal} {J. Chem. Phys.}\ }\textbf {\bibinfo {volume}
  {148}},\ \bibinfo {pages} {034104} (\bibinfo {year} {2018})}\BibitemShut
  {NoStop}%
\bibitem [{\citenamefont {Born}\ and\ \citenamefont
  {Oppenheimer}(1927)}]{Born1927}%
  \BibitemOpen
  \bibfield  {author} {\bibinfo {author} {\bibfnamefont {M.}~\bibnamefont
  {Born}}\ and\ \bibinfo {author} {\bibfnamefont {R.}~\bibnamefont
  {Oppenheimer}},\ }\bibfield  {title} {\enquote {\bibinfo {title} {Zur
  quantentheorie der molekeln},}\ }\href@noop {} {\bibfield  {journal}
  {\bibinfo  {journal} {Ann. Phys.}\ }\textbf {\bibinfo {volume} {389}},\
  \bibinfo {pages} {457--484} (\bibinfo {year} {1927})}\BibitemShut {NoStop}%
\bibitem [{\citenamefont {Ferrer}\ \emph {et~al.}(2014)\citenamefont {Ferrer},
  \citenamefont {Davari}, \citenamefont {Morozov}, \citenamefont {Groenhof},\
  and\ \citenamefont {Santoro}}]{Ferrer2014}%
  \BibitemOpen
  \bibfield  {author} {\bibinfo {author} {\bibfnamefont {F.~J.~A.}\
  \bibnamefont {Ferrer}}, \bibinfo {author} {\bibfnamefont {M.~D.}\
  \bibnamefont {Davari}}, \bibinfo {author} {\bibfnamefont {D.}~\bibnamefont
  {Morozov}}, \bibinfo {author} {\bibfnamefont {G.}~\bibnamefont {Groenhof}},\
  and\ \bibinfo {author} {\bibfnamefont {F.}~\bibnamefont {Santoro}},\
  }\bibfield  {title} {\enquote {\bibinfo {title} {The lineshape of the
  electronic spectrum of the green fluorescent protein chromophore, part ii:
  Solution phase},}\ }\href {https://doi.org/10.1002/cphc.201402485} {\bibfield
   {journal} {\bibinfo  {journal} {ChemPhysChem}\ }\textbf {\bibinfo {volume}
  {15}},\ \bibinfo {pages} {3246--3257} (\bibinfo {year} {2014})}\BibitemShut
  {NoStop}%
\bibitem [{\citenamefont {Cerezo}\ \emph {et~al.}()\citenamefont {Cerezo},
  \citenamefont {Gierschner}, \citenamefont {Santoro},\ and\ \citenamefont
  {Prampolini}}]{Cerezo2024}%
  \BibitemOpen
  \bibfield  {author} {\bibinfo {author} {\bibfnamefont {J.}~\bibnamefont
  {Cerezo}}, \bibinfo {author} {\bibfnamefont {J.}~\bibnamefont {Gierschner}},
  \bibinfo {author} {\bibfnamefont {F.}~\bibnamefont {Santoro}},\ and\ \bibinfo
  {author} {\bibfnamefont {G.}~\bibnamefont {Prampolini}},\ }\bibfield  {title}
  {\enquote {\bibinfo {title} {Explicit modelling of spectral bandshapes by a
  mixed quantum-classical approach: Solvent order and temperature effects in
  the optical spectra of distryrylbenzene},}\ }\href
  {https://doi.org/https://doi.org/10.1002/cphc.202400307} {\bibfield
  {journal} {\bibinfo  {journal} {ChemPhysChem}\ }\textbf {\bibinfo {volume}
  {n/a}},\ \bibinfo {pages} {e202400307}}\BibitemShut {NoStop}%
\bibitem [{\citenamefont {Cammi}\ \emph {et~al.}(2005)\citenamefont {Cammi},
  \citenamefont {Corni}, \citenamefont {Mennucci},\ and\ \citenamefont
  {Tomasi}}]{Cammi_2005}%
  \BibitemOpen
  \bibfield  {author} {\bibinfo {author} {\bibfnamefont {R.}~\bibnamefont
  {Cammi}}, \bibinfo {author} {\bibfnamefont {S.}~\bibnamefont {Corni}},
  \bibinfo {author} {\bibfnamefont {B.}~\bibnamefont {Mennucci}},\ and\
  \bibinfo {author} {\bibfnamefont {J.}~\bibnamefont {Tomasi}},\ }\bibfield
  {title} {\enquote {\bibinfo {title} {Electronic excitation energies of
  molecules in solution: State specific and linear response methods for
  nonequilibrium continuum solvation models},}\ }\href
  {https://doi.org/10.1063/1.1867373} {\bibfield  {journal} {\bibinfo
  {journal} {J. Chem. Phys.}\ }\textbf {\bibinfo {volume} {122}},\ \bibinfo
  {pages} {104513} (\bibinfo {year} {2005})}\BibitemShut {NoStop}%
\bibitem [{\citenamefont {Cerezo}\ \emph {et~al.}(2015)\citenamefont {Cerezo},
  \citenamefont {Avila~Ferrer}, \citenamefont {Prampolini},\ and\ \citenamefont
  {Santoro}}]{Cerezo2015}%
  \BibitemOpen
  \bibfield  {author} {\bibinfo {author} {\bibfnamefont {J.}~\bibnamefont
  {Cerezo}}, \bibinfo {author} {\bibfnamefont {F.~J.}\ \bibnamefont
  {Avila~Ferrer}}, \bibinfo {author} {\bibfnamefont {G.}~\bibnamefont
  {Prampolini}},\ and\ \bibinfo {author} {\bibfnamefont {F.}~\bibnamefont
  {Santoro}},\ }\bibfield  {title} {\enquote {\bibinfo {title} {Modeling
  solvent broadening on the vibronic spectra of a series of coumarin dyes. from
  implicit to explicit solvent models},}\ }\href
  {https://doi.org/10.1021/acs.jctc.5b00870} {\bibfield  {journal} {\bibinfo
  {journal} {J. Chem. Theory Comput.}\ }\textbf {\bibinfo {volume} {11}},\
  \bibinfo {pages} {5810--5825} (\bibinfo {year} {2015})}\BibitemShut {NoStop}%
\bibitem [{\citenamefont {Cerezo}\ \emph {et~al.}(2019)\citenamefont {Cerezo},
  \citenamefont {Aranda}, \citenamefont {Avila~Ferrer}, \citenamefont
  {Prampolini},\ and\ \citenamefont {Santoro}}]{Cerezo2019}%
  \BibitemOpen
  \bibfield  {author} {\bibinfo {author} {\bibfnamefont {J.}~\bibnamefont
  {Cerezo}}, \bibinfo {author} {\bibfnamefont {D.}~\bibnamefont {Aranda}},
  \bibinfo {author} {\bibfnamefont {F.~J.}\ \bibnamefont {Avila~Ferrer}},
  \bibinfo {author} {\bibfnamefont {G.}~\bibnamefont {Prampolini}},\ and\
  \bibinfo {author} {\bibfnamefont {F.}~\bibnamefont {Santoro}},\ }\bibfield
  {title} {\enquote {\bibinfo {title} {Adiabatic-molecular dynamics generalized
  vertical hessian approach: a mixed quantum classical method to compute
  electronic spectra of flexible molecules in the condensed phase},}\
  }\href@noop {} {\bibfield  {journal} {\bibinfo  {journal} {J. Chem. Theory
  Comput.}\ }\textbf {\bibinfo {volume} {16}},\ \bibinfo {pages} {1215--1231}
  (\bibinfo {year} {2019})}\BibitemShut {NoStop}%
\bibitem [{\citenamefont {Meyer}, \citenamefont {Manthe},\ and\ \citenamefont
  {Cederbaum}(1990)}]{Meyer1990}%
  \BibitemOpen
  \bibfield  {author} {\bibinfo {author} {\bibfnamefont {H.~D.}\ \bibnamefont
  {Meyer}}, \bibinfo {author} {\bibfnamefont {U.}~\bibnamefont {Manthe}},\ and\
  \bibinfo {author} {\bibfnamefont {L.~S.}\ \bibnamefont {Cederbaum}},\
  }\bibfield  {title} {\enquote {\bibinfo {title} {{The multi-configurational
  time-dependent Hartree approach}},}\ }\href
  {https://doi.org/10.1016/0009-2614(90)87014-I} {\bibfield  {journal}
  {\bibinfo  {journal} {Chemical Physics Letters}\ }\textbf {\bibinfo {volume}
  {165}},\ \bibinfo {pages} {73--78} (\bibinfo {year} {1990})}\BibitemShut
  {NoStop}%
\bibitem [{\citenamefont {Beck}\ \emph {et~al.}(2000)\citenamefont {Beck},
  \citenamefont {J\"{a}ckle}, \citenamefont {Worth},\ and\ \citenamefont
  {Meyer}}]{Beck2000}%
  \BibitemOpen
  \bibfield  {author} {\bibinfo {author} {\bibfnamefont {M.~H.}\ \bibnamefont
  {Beck}}, \bibinfo {author} {\bibfnamefont {A.}~\bibnamefont {J\"{a}ckle}},
  \bibinfo {author} {\bibfnamefont {G.~A.}\ \bibnamefont {Worth}},\ and\
  \bibinfo {author} {\bibfnamefont {H.~D.}\ \bibnamefont {Meyer}},\ }\bibfield
  {title} {\enquote {\bibinfo {title} {The multiconfiguration time-dependent
  {H}artree ({MCTDH}) method: A highly efficient algorithm for propagating
  wavepackets},}\ }\href {https://doi.org/10.1016/S0370-1573(99)00047-2}
  {\bibfield  {journal} {\bibinfo  {journal} {Phys. Rep.}\ }\textbf {\bibinfo
  {volume} {324}},\ \bibinfo {pages} {1--105} (\bibinfo {year}
  {2000})}\BibitemShut {NoStop}%
\bibitem [{\citenamefont {Gatti}\ and\ \citenamefont
  {Worth}(2009)}]{Meyer2009}%
  \BibitemOpen
  \bibfield  {author} {\bibinfo {author} {\bibfnamefont {F.}~\bibnamefont
  {Gatti}}\ and\ \bibinfo {author} {\bibfnamefont {G.~A.}\ \bibnamefont
  {Worth}},\ }\href {https://doi.org/10.1002/9783527627400} {\emph {\bibinfo
  {title} {Multidimensional quantum dynamics: MCTDH theory and
  applications}}},\ edited by\ \bibinfo {editor} {\bibfnamefont {H.~D.}\
  \bibnamefont {Meyer}}\ (\bibinfo  {publisher} {Wiley VCH},\ \bibinfo {year}
  {2009})\BibitemShut {NoStop}%
\bibitem [{\citenamefont {Wang}(2015)}]{Wang2015}%
  \BibitemOpen
  \bibfield  {author} {\bibinfo {author} {\bibfnamefont {H.}~\bibnamefont
  {Wang}},\ }\bibfield  {title} {\enquote {\bibinfo {title} {Multilayer
  multiconfiguration time-dependent {H}artree theory},}\ }\href@noop {}
  {\bibfield  {journal} {\bibinfo  {journal} {J. Phys. Chem. A}\ }\textbf
  {\bibinfo {volume} {119}},\ \bibinfo {pages} {7951--7965} (\bibinfo {year}
  {2015})}\BibitemShut {NoStop}%
\bibitem [{\citenamefont {Worth}(2020)}]{Worth2020}%
  \BibitemOpen
  \bibfield  {author} {\bibinfo {author} {\bibfnamefont {G.}~\bibnamefont
  {Worth}},\ }\bibfield  {title} {\enquote {\bibinfo {title} {Quantics: A
  general purpose package for quantum molecular dynamics simulations},}\
  }\href@noop {} {\bibfield  {journal} {\bibinfo  {journal} {Comput. Phys.
  Commun.}\ }\textbf {\bibinfo {volume} {248}},\ \bibinfo {pages} {107040}
  (\bibinfo {year} {2020})}\BibitemShut {NoStop}%
\bibitem [{\citenamefont {Green}\ \emph {et~al.}(2021)\citenamefont {Green},
  \citenamefont {Yaghoubi~Jouybari}, \citenamefont {Aranda}, \citenamefont
  {Improta},\ and\ \citenamefont {Santoro}}]{Green2021}%
  \BibitemOpen
  \bibfield  {author} {\bibinfo {author} {\bibfnamefont {J.~A.}\ \bibnamefont
  {Green}}, \bibinfo {author} {\bibfnamefont {M.}~\bibnamefont
  {Yaghoubi~Jouybari}}, \bibinfo {author} {\bibfnamefont {D.}~\bibnamefont
  {Aranda}}, \bibinfo {author} {\bibfnamefont {R.}~\bibnamefont {Improta}},\
  and\ \bibinfo {author} {\bibfnamefont {F.}~\bibnamefont {Santoro}},\
  }\bibfield  {title} {\enquote {\bibinfo {title} {Nonadiabatic absorption
  spectra and ultrafast dynamics of {DNA} and {RNA} photoexcited
  nucleobases},}\ }\href@noop {} {\bibfield  {journal} {\bibinfo  {journal}
  {Molecules}\ }\textbf {\bibinfo {volume} {26}},\ \bibinfo {pages} {1743}
  (\bibinfo {year} {2021})}\BibitemShut {NoStop}%
\bibitem [{\citenamefont {Segalina}\ \emph {et~al.}(2022)\citenamefont
  {Segalina}, \citenamefont {Aranda}, \citenamefont {Green}, \citenamefont
  {Cristino}, \citenamefont {Caramori}, \citenamefont {Prampolini},
  \citenamefont {Pastore},\ and\ \citenamefont {Santoro}}]{Segalina2022}%
  \BibitemOpen
  \bibfield  {author} {\bibinfo {author} {\bibfnamefont {A.}~\bibnamefont
  {Segalina}}, \bibinfo {author} {\bibfnamefont {D.}~\bibnamefont {Aranda}},
  \bibinfo {author} {\bibfnamefont {J.~A.}\ \bibnamefont {Green}}, \bibinfo
  {author} {\bibfnamefont {V.}~\bibnamefont {Cristino}}, \bibinfo {author}
  {\bibfnamefont {S.}~\bibnamefont {Caramori}}, \bibinfo {author}
  {\bibfnamefont {G.}~\bibnamefont {Prampolini}}, \bibinfo {author}
  {\bibfnamefont {M.}~\bibnamefont {Pastore}},\ and\ \bibinfo {author}
  {\bibfnamefont {F.}~\bibnamefont {Santoro}},\ }\bibfield  {title} {\enquote
  {\bibinfo {title} {How the interplay among conformational disorder,
  solvation, local, and charge-transfer excitations affects the absorption
  spectrum and photoinduced dynamics of perylene diimide dimers: A molecular
  dynamics/quantum vibronic approach},}\ }\href@noop {} {\bibfield  {journal}
  {\bibinfo  {journal} {J. Chem. Theory Comput}\ }\textbf {\bibinfo {volume}
  {18}},\ \bibinfo {pages} {3718--3736} (\bibinfo {year} {2022})}\BibitemShut
  {NoStop}%
\bibitem [{\citenamefont {Cerezo}\ \emph {et~al.}(2023)\citenamefont {Cerezo},
  \citenamefont {García-Iriepa}, \citenamefont {Santoro}, \citenamefont
  {Navizet},\ and\ \citenamefont {Prampolini}}]{Cerezo2023}%
  \BibitemOpen
  \bibfield  {author} {\bibinfo {author} {\bibfnamefont {J.}~\bibnamefont
  {Cerezo}}, \bibinfo {author} {\bibfnamefont {C.}~\bibnamefont
  {García-Iriepa}}, \bibinfo {author} {\bibfnamefont {F.}~\bibnamefont
  {Santoro}}, \bibinfo {author} {\bibfnamefont {I.}~\bibnamefont {Navizet}},\
  and\ \bibinfo {author} {\bibfnamefont {G.}~\bibnamefont {Prampolini}},\
  }\bibfield  {title} {\enquote {\bibinfo {title} {Unraveling the contributions
  to the spectral shape of flexible dyes in solution: insights on the
  absorption spectrum of an oxyluciferin analogue},}\ }\href@noop {} {\bibfield
   {journal} {\bibinfo  {journal} {Phys. Chem. Chem. Phys.}\ }\textbf {\bibinfo
  {volume} {25}},\ \bibinfo {pages} {5007--5020} (\bibinfo {year}
  {2023})}\BibitemShut {NoStop}%
\bibitem [{\citenamefont {Dunnett}\ \emph {et~al.}(2021)\citenamefont
  {Dunnett}, \citenamefont {Gowland}, \citenamefont {Isborn}, \citenamefont
  {Chin},\ and\ \citenamefont {Zuehlsdorff}}]{Dunnett2021}%
  \BibitemOpen
  \bibfield  {author} {\bibinfo {author} {\bibfnamefont {A.~J.}\ \bibnamefont
  {Dunnett}}, \bibinfo {author} {\bibfnamefont {D.}~\bibnamefont {Gowland}},
  \bibinfo {author} {\bibfnamefont {C.~M.}\ \bibnamefont {Isborn}}, \bibinfo
  {author} {\bibfnamefont {A.~W.}\ \bibnamefont {Chin}},\ and\ \bibinfo
  {author} {\bibfnamefont {T.~J.}\ \bibnamefont {Zuehlsdorff}},\ }\bibfield
  {title} {\enquote {\bibinfo {title} {Influence of non-adiabatic effects on
  linear absorption spectra in the condensed phase: Methylene blue},}\ }\href
  {https://doi.org/10.1063/5.0062950} {\bibfield  {journal} {\bibinfo
  {journal} {J. Chem. Phys.}\ }\textbf {\bibinfo {volume} {155}},\ \bibinfo
  {pages} {144112} (\bibinfo {year} {2021})}\BibitemShut {NoStop}%
\bibitem [{\citenamefont {Wiethorn}\ \emph {et~al.}(2023)\citenamefont
  {Wiethorn}, \citenamefont {Hunter}, \citenamefont {Zuehlsdorff},\ and\
  \citenamefont {Montoya-Castillo}}]{Wiethorn2023}%
  \BibitemOpen
  \bibfield  {author} {\bibinfo {author} {\bibfnamefont {Z.~R.}\ \bibnamefont
  {Wiethorn}}, \bibinfo {author} {\bibfnamefont {K.~E.}\ \bibnamefont
  {Hunter}}, \bibinfo {author} {\bibfnamefont {T.~J.}\ \bibnamefont
  {Zuehlsdorff}},\ and\ \bibinfo {author} {\bibfnamefont {A.}~\bibnamefont
  {Montoya-Castillo}},\ }\bibfield  {title} {\enquote {\bibinfo {title}
  {{Beyond the Condon limit: Condensed phase optical spectra from atomistic
  simulations}},}\ }\href@noop {} {\bibfield  {journal} {\bibinfo  {journal}
  {J. Chem. Phys.}\ }\textbf {\bibinfo {volume} {159}},\ \bibinfo {pages}
  {244114} (\bibinfo {year} {2023})}\BibitemShut {NoStop}%
\bibitem [{\citenamefont {Hunter}\ \emph {et~al.}(2024)\citenamefont {Hunter},
  \citenamefont {Mao}, \citenamefont {Chin},\ and\ \citenamefont
  {Zuehlsdorff}}]{Hunter2024}%
  \BibitemOpen
  \bibfield  {author} {\bibinfo {author} {\bibfnamefont {K.~E.}\ \bibnamefont
  {Hunter}}, \bibinfo {author} {\bibfnamefont {Y.}~\bibnamefont {Mao}},
  \bibinfo {author} {\bibfnamefont {A.~W.}\ \bibnamefont {Chin}},\ and\
  \bibinfo {author} {\bibfnamefont {T.~J.}\ \bibnamefont {Zuehlsdorff}},\
  }\bibfield  {title} {\enquote {\bibinfo {title} {Environmentally driven
  symmetry breaking quenches dual fluorescence in proflavine},}\ }\href@noop {}
  {\bibfield  {journal} {\bibinfo  {journal} {J. Phys. Chem. Lett.}\ }\textbf
  {\bibinfo {volume} {15}},\ \bibinfo {pages} {4623--4632} (\bibinfo {year}
  {2024})}\BibitemShut {NoStop}%
\bibitem [{\citenamefont {Prior}\ \emph {et~al.}(2010)\citenamefont {Prior},
  \citenamefont {Chin}, \citenamefont {Huelga},\ and\ \citenamefont
  {Plenio}}]{Prior2010}%
  \BibitemOpen
  \bibfield  {author} {\bibinfo {author} {\bibfnamefont {J.}~\bibnamefont
  {Prior}}, \bibinfo {author} {\bibfnamefont {A.~W.}\ \bibnamefont {Chin}},
  \bibinfo {author} {\bibfnamefont {S.~F.}\ \bibnamefont {Huelga}},\ and\
  \bibinfo {author} {\bibfnamefont {M.~B.}\ \bibnamefont {Plenio}},\ }\bibfield
   {title} {\enquote {\bibinfo {title} {{Efficient simulation of strong
  system-environment interactions}},}\ }\href
  {https://doi.org/10.1103/PhysRevLett.105.050404} {\bibfield  {journal}
  {\bibinfo  {journal} {Physical Review Letters}\ }\textbf {\bibinfo {volume}
  {105}},\ \bibinfo {pages} {1--4} (\bibinfo {year} {2010})}\BibitemShut
  {NoStop}%
\bibitem [{\citenamefont {Schr\"oder}\ and\ \citenamefont
  {Chin}(2016)}]{Schroder2016}%
  \BibitemOpen
  \bibfield  {author} {\bibinfo {author} {\bibfnamefont {F.~A. Y.~N.}\
  \bibnamefont {Schr\"oder}}\ and\ \bibinfo {author} {\bibfnamefont {A.~W.}\
  \bibnamefont {Chin}},\ }\bibfield  {title} {\enquote {\bibinfo {title}
  {Simulating open quantum dynamics with time-dependent variational matrix
  product states: Towards microscopic correlation of environment dynamics and
  reduced system evolution},}\ }\href@noop {} {\bibfield  {journal} {\bibinfo
  {journal} {Phys. Rev. B}\ }\textbf {\bibinfo {volume} {93}},\ \bibinfo
  {pages} {075105} (\bibinfo {year} {2016})}\BibitemShut {NoStop}%
\bibitem [{\citenamefont {Schr{\"{o}}der}\ \emph {et~al.}(2019)\citenamefont
  {Schr{\"{o}}der}, \citenamefont {Turban}, \citenamefont {Musser},
  \citenamefont {Hine},\ and\ \citenamefont {Chin}}]{Schroder2019}%
  \BibitemOpen
  \bibfield  {author} {\bibinfo {author} {\bibfnamefont {F.~A.}\ \bibnamefont
  {Schr{\"{o}}der}}, \bibinfo {author} {\bibfnamefont {D.~H.}\ \bibnamefont
  {Turban}}, \bibinfo {author} {\bibfnamefont {A.~J.}\ \bibnamefont {Musser}},
  \bibinfo {author} {\bibfnamefont {N.~D.}\ \bibnamefont {Hine}},\ and\
  \bibinfo {author} {\bibfnamefont {A.~W.}\ \bibnamefont {Chin}},\ }\bibfield
  {title} {\enquote {\bibinfo {title} {{Tensor network simulation of
  multi-environmental open quantum dynamics via machine learning and
  entanglement renormalisation}},}\ }\href
  {https://doi.org/10.1038/s41467-019-09039-7} {\bibfield  {journal} {\bibinfo
  {journal} {Nature Commun.}\ }\textbf {\bibinfo {volume} {10}},\ \bibinfo
  {pages} {1--10} (\bibinfo {year} {2019})}\BibitemShut {NoStop}%
\bibitem [{\citenamefont {Mukamel}(1995{\natexlab{a}})}]{mukamel_book}%
  \BibitemOpen
  \bibfield  {author} {\bibinfo {author} {\bibfnamefont {S.}~\bibnamefont
  {Mukamel}},\ }\href@noop {} {\emph {\bibinfo {title} {Principals of Nonlinear
  Optical Spectroscopy}}}\ (\bibinfo  {publisher} {Oxford University Press},\
  \bibinfo {year} {1995})\BibitemShut {NoStop}%
\bibitem [{\citenamefont {Condon}(1926)}]{Condon1926}%
  \BibitemOpen
  \bibfield  {author} {\bibinfo {author} {\bibfnamefont {E.}~\bibnamefont
  {Condon}},\ }\bibfield  {title} {\enquote {\bibinfo {title} {A theory of
  intensity distribution in band systems},}\ }\href@noop {} {\bibfield
  {journal} {\bibinfo  {journal} {Phys. Rev.}\ }\textbf {\bibinfo {volume}
  {28}},\ \bibinfo {pages} {1182--1201} (\bibinfo {year} {1926})}\BibitemShut
  {NoStop}%
\bibitem [{\citenamefont {Condon}(1928)}]{Condon1928}%
  \BibitemOpen
  \bibfield  {author} {\bibinfo {author} {\bibfnamefont {E.}~\bibnamefont
  {Condon}},\ }\bibfield  {title} {\enquote {\bibinfo {title} {{Nuclear motion
  associated with electron transitions in diatomic molecules}},}\ }\href@noop
  {} {\bibfield  {journal} {\bibinfo  {journal} {Phys. Rev.}\ }\textbf
  {\bibinfo {volume} {32}},\ \bibinfo {pages} {858--872} (\bibinfo {year}
  {1928})}\BibitemShut {NoStop}%
\bibitem [{\citenamefont {Raab}\ \emph {et~al.}(1999)\citenamefont {Raab},
  \citenamefont {Worth}, \citenamefont {Meyer},\ and\ \citenamefont
  {Cederbaum}}]{Raab1999}%
  \BibitemOpen
  \bibfield  {author} {\bibinfo {author} {\bibfnamefont {A.}~\bibnamefont
  {Raab}}, \bibinfo {author} {\bibfnamefont {G.~A.}\ \bibnamefont {Worth}},
  \bibinfo {author} {\bibfnamefont {H.-D.}\ \bibnamefont {Meyer}},\ and\
  \bibinfo {author} {\bibfnamefont {L.~S.}\ \bibnamefont {Cederbaum}},\
  }\bibfield  {title} {\enquote {\bibinfo {title} {{Molecular dynamics of
  pyrazine after excitation to the S2 electronic state using a realistic
  24-mode model Hamiltonian}},}\ }\href@noop {} {\bibfield  {journal} {\bibinfo
   {journal} {J. Chem. Phys.}\ }\textbf {\bibinfo {volume} {110}},\ \bibinfo
  {pages} {936--946} (\bibinfo {year} {1999})}\BibitemShut {NoStop}%
\bibitem [{\citenamefont {Burghardt}, \citenamefont {Giri},\ and\ \citenamefont
  {Worth}(2008)}]{Burghardt2008}%
  \BibitemOpen
  \bibfield  {author} {\bibinfo {author} {\bibfnamefont {I.}~\bibnamefont
  {Burghardt}}, \bibinfo {author} {\bibfnamefont {K.}~\bibnamefont {Giri}},\
  and\ \bibinfo {author} {\bibfnamefont {G.~A.}\ \bibnamefont {Worth}},\
  }\bibfield  {title} {\enquote {\bibinfo {title} {{Multimode quantum dynamics
  using Gaussian wavepackets: The Gaussian-based multiconfiguration
  time-dependent Hartree (G-MCTDH) method applied to the absorption spectrum of
  pyrazine}},}\ }\href@noop {} {\bibfield  {journal} {\bibinfo  {journal} {J.
  Chem. Phys.}\ }\textbf {\bibinfo {volume} {129}},\ \bibinfo {pages} {174104}
  (\bibinfo {year} {2008})}\BibitemShut {NoStop}%
\bibitem [{\citenamefont {Halverson}\ and\ \citenamefont
  {Hirt}(1951)}]{Halverson1951_pyrazine_solution}%
  \BibitemOpen
  \bibfield  {author} {\bibinfo {author} {\bibfnamefont {F.}~\bibnamefont
  {Halverson}}\ and\ \bibinfo {author} {\bibfnamefont {R.~C.}\ \bibnamefont
  {Hirt}},\ }\bibfield  {title} {\enquote {\bibinfo {title} {Near ultraviolet
  solution spectra of the diazines},}\ }\href@noop {} {\bibfield  {journal}
  {\bibinfo  {journal} {J. Chem. Phys.}\ }\textbf {\bibinfo {volume} {19}},\
  \bibinfo {pages} {711--718} (\bibinfo {year} {1951})}\BibitemShut {NoStop}%
\bibitem [{\citenamefont {Samir}\ \emph {et~al.}(2020)\citenamefont {Samir},
  \citenamefont {Kalalian}, \citenamefont {Roth}, \citenamefont {Salghi},\ and\
  \citenamefont {Chakir}}]{Samir2020_pyrazine_vac}%
  \BibitemOpen
  \bibfield  {author} {\bibinfo {author} {\bibfnamefont {B.}~\bibnamefont
  {Samir}}, \bibinfo {author} {\bibfnamefont {C.}~\bibnamefont {Kalalian}},
  \bibinfo {author} {\bibfnamefont {E.}~\bibnamefont {Roth}}, \bibinfo {author}
  {\bibfnamefont {R.}~\bibnamefont {Salghi}},\ and\ \bibinfo {author}
  {\bibfnamefont {A.}~\bibnamefont {Chakir}},\ }\bibfield  {title} {\enquote
  {\bibinfo {title} {Gas-phase {UV} absorption spectra of pyrazine, pyrimidine
  and pyridazine},}\ }\href@noop {} {\bibfield  {journal} {\bibinfo  {journal}
  {Chem. Phys. Lett.}\ }\textbf {\bibinfo {volume} {751}},\ \bibinfo {pages}
  {137469} (\bibinfo {year} {2020})}\BibitemShut {NoStop}%
\bibitem [{\citenamefont {Kasha}(1950)}]{Kasha1950}%
  \BibitemOpen
  \bibfield  {author} {\bibinfo {author} {\bibfnamefont {M.}~\bibnamefont
  {Kasha}},\ }\bibfield  {title} {\enquote {\bibinfo {title} {Characterization
  of electronic transitions in complex molecules},}\ }\href
  {https://doi.org/10.1039/DF9500900014} {\bibfield  {journal} {\bibinfo
  {journal} {Discuss. Faraday Soc.}\ }\textbf {\bibinfo {volume} {9}},\
  \bibinfo {pages} {14--19} (\bibinfo {year} {1950})}\BibitemShut {NoStop}%
\bibitem [{\citenamefont {Mukamel}(1995{\natexlab{b}})}]{Mukamel-book}%
  \BibitemOpen
  \bibfield  {author} {\bibinfo {author} {\bibfnamefont {S.}~\bibnamefont
  {Mukamel}},\ }\href@noop {} {\emph {\bibinfo {title} {{Principles of
  nonlinear optical spectroscopy}}}}\ (\bibinfo  {publisher} {Oxford University
  Press},\ \bibinfo {address} {New York},\ \bibinfo {year} {1995})\BibitemShut
  {NoStop}%
\bibitem [{\citenamefont {Mukamel}\ and\ \citenamefont
  {Abramavicius}(2004)}]{Mukamel04}%
  \BibitemOpen
  \bibfield  {author} {\bibinfo {author} {\bibfnamefont {S.}~\bibnamefont
  {Mukamel}}\ and\ \bibinfo {author} {\bibfnamefont {D.}~\bibnamefont
  {Abramavicius}},\ }\bibfield  {title} {\enquote {\bibinfo {title} {Many-body
  approaches for simulating coherent nonlinear spectroscopies for electronic
  and vibrational excitons},}\ }\href {https://doi.org/10.1021/cr020681b}
  {\bibfield  {journal} {\bibinfo  {journal} {Chem. Rev.}\ }\textbf {\bibinfo
  {volume} {104}},\ \bibinfo {pages} {2073--2098} (\bibinfo {year}
  {2004})}\BibitemShut {NoStop}%
\bibitem [{\citenamefont {Zuehlsdorff}\ \emph {et~al.}(2019)\citenamefont
  {Zuehlsdorff}, \citenamefont {Montoya-Castillo}, \citenamefont {Napoli},
  \citenamefont {Markland},\ and\ \citenamefont {Isborn}}]{Zuehlsdorff2019b}%
  \BibitemOpen
  \bibfield  {author} {\bibinfo {author} {\bibfnamefont {T.~J.}\ \bibnamefont
  {Zuehlsdorff}}, \bibinfo {author} {\bibfnamefont {A.}~\bibnamefont
  {Montoya-Castillo}}, \bibinfo {author} {\bibfnamefont {J.~A.}\ \bibnamefont
  {Napoli}}, \bibinfo {author} {\bibfnamefont {T.~E.}\ \bibnamefont
  {Markland}},\ and\ \bibinfo {author} {\bibfnamefont {C.~M.}\ \bibnamefont
  {Isborn}},\ }\bibfield  {title} {\enquote {\bibinfo {title} {Optical spectra
  in the condensed phase: Capturing anharmonic and vibronic features using
  dynamic and static approaches},}\ }\href@noop {} {\bibfield  {journal}
  {\bibinfo  {journal} {J. Chem. Phys.}\ }\textbf {\bibinfo {volume} {151}},\
  \bibinfo {pages} {074111} (\bibinfo {year} {2019})}\BibitemShut {NoStop}%
\bibitem [{\citenamefont {Fidler}\ and\ \citenamefont
  {Engel}(2013)}]{Fidler2013}%
  \BibitemOpen
  \bibfield  {author} {\bibinfo {author} {\bibfnamefont {A.~F.}\ \bibnamefont
  {Fidler}}\ and\ \bibinfo {author} {\bibfnamefont {G.~S.}\ \bibnamefont
  {Engel}},\ }\bibfield  {title} {\enquote {\bibinfo {title} {Nonlinear
  spectroscopy theory of displaced harmonic oscillators with differing
  curvatures: A correlation function approach},}\ }\href
  {https://doi.org/10.1021/jp311713x} {\bibfield  {journal} {\bibinfo
  {journal} {J. Phys. Chem. A}\ }\textbf {\bibinfo {volume} {117}},\ \bibinfo
  {pages} {9444--9453} (\bibinfo {year} {2013})}\BibitemShut {NoStop}%
\bibitem [{\citenamefont {Allan}\ and\ \citenamefont
  {Zuehlsdorff}(2024)}]{Allan2024}%
  \BibitemOpen
  \bibfield  {author} {\bibinfo {author} {\bibfnamefont {L.}~\bibnamefont
  {Allan}}\ and\ \bibinfo {author} {\bibfnamefont {T.~J.}\ \bibnamefont
  {Zuehlsdorff}},\ }\bibfield  {title} {\enquote {\bibinfo {title} {{Taming the
  third order cumulant approximation to linear optical spectroscopy}},}\
  }\href@noop {} {\bibfield  {journal} {\bibinfo  {journal} {J. Chem. Phys.}\
  }\textbf {\bibinfo {volume} {160}},\ \bibinfo {pages} {074108} (\bibinfo
  {year} {2024})}\BibitemShut {NoStop}%
\bibitem [{\citenamefont {Egorov}, \citenamefont {Everitt},\ and\ \citenamefont
  {Skinner}(1999)}]{Egorov1999}%
  \BibitemOpen
  \bibfield  {author} {\bibinfo {author} {\bibfnamefont {S.~A.}\ \bibnamefont
  {Egorov}}, \bibinfo {author} {\bibfnamefont {K.~F.}\ \bibnamefont
  {Everitt}},\ and\ \bibinfo {author} {\bibfnamefont {J.~L.}\ \bibnamefont
  {Skinner}},\ }\bibfield  {title} {\enquote {\bibinfo {title} {{Quantum
  dynamics and vibrational relaxation}},}\ }\href@noop {} {\bibfield  {journal}
  {\bibinfo  {journal} {J. Phys. Chem. A}\ }\textbf {\bibinfo {volume} {103}},\
  \bibinfo {pages} {9494–9499} (\bibinfo {year} {1999})}\BibitemShut
  {NoStop}%
\bibitem [{\citenamefont {Craig}\ and\ \citenamefont
  {Manolopoulos}(2004)}]{Craig2004}%
  \BibitemOpen
  \bibfield  {author} {\bibinfo {author} {\bibfnamefont {I.~R.}\ \bibnamefont
  {Craig}}\ and\ \bibinfo {author} {\bibfnamefont {D.~E.}\ \bibnamefont
  {Manolopoulos}},\ }\bibfield  {title} {\enquote {\bibinfo {title} {{Quantum
  statistics and classical mechanics: Real time correlation functions from ring
  polymer molecular dynamics}},}\ }\href {https://doi.org/10.1063/1.1777575}
  {\bibfield  {journal} {\bibinfo  {journal} {J. Chem. Phys.}\ }\textbf
  {\bibinfo {volume} {121}},\ \bibinfo {pages} {3368} (\bibinfo {year}
  {2004})}\BibitemShut {NoStop}%
\bibitem [{\citenamefont {Ramirez}\ \emph {et~al.}(2004)\citenamefont
  {Ramirez}, \citenamefont {Lopez-Ciudad}, \citenamefont {P},\ and\
  \citenamefont {Marx}}]{Ramirez2004}%
  \BibitemOpen
  \bibfield  {author} {\bibinfo {author} {\bibfnamefont {R.}~\bibnamefont
  {Ramirez}}, \bibinfo {author} {\bibfnamefont {T.}~\bibnamefont
  {Lopez-Ciudad}}, \bibinfo {author} {\bibfnamefont {P.~K.}\ \bibnamefont
  {P}},\ and\ \bibinfo {author} {\bibfnamefont {D.}~\bibnamefont {Marx}},\
  }\bibfield  {title} {\enquote {\bibinfo {title} {{Quantum corrections to
  classical time-correlation functions: Hydrogen bonding and anharmonic floppy
  modes}},}\ }\href {https://doi.org/10.1063/1.1774986} {\bibfield  {journal}
  {\bibinfo  {journal} {J. Chem. Phys.}\ }\textbf {\bibinfo {volume} {121}},\
  \bibinfo {pages} {3973} (\bibinfo {year} {2004})}\BibitemShut {NoStop}%
\bibitem [{\citenamefont {Valleau}, \citenamefont {Eisfeld},\ and\
  \citenamefont {Aspuru-Guzik}(2012)}]{Valleau2012}%
  \BibitemOpen
  \bibfield  {author} {\bibinfo {author} {\bibfnamefont {S.}~\bibnamefont
  {Valleau}}, \bibinfo {author} {\bibfnamefont {A.}~\bibnamefont {Eisfeld}},\
  and\ \bibinfo {author} {\bibfnamefont {A.}~\bibnamefont {Aspuru-Guzik}},\
  }\bibfield  {title} {\enquote {\bibinfo {title} {{On the alternatives for
  bath correlators and spectral densities from mixed quantum-classical
  simulations}},}\ }\href {https://doi.org/10.1063/1.4769079} {\bibfield
  {journal} {\bibinfo  {journal} {J. Chem. Phys.}\ }\textbf {\bibinfo {volume}
  {137}},\ \bibinfo {pages} {224103} (\bibinfo {year} {2012})}\BibitemShut
  {NoStop}%
\bibitem [{\citenamefont {Shim}\ \emph {et~al.}(2012)\citenamefont {Shim},
  \citenamefont {Rebentrost}, \citenamefont {Valleau},\ and\ \citenamefont
  {Aspuru-Guzik}}]{Shim2012}%
  \BibitemOpen
  \bibfield  {author} {\bibinfo {author} {\bibfnamefont {S.}~\bibnamefont
  {Shim}}, \bibinfo {author} {\bibfnamefont {P.}~\bibnamefont {Rebentrost}},
  \bibinfo {author} {\bibfnamefont {S.}~\bibnamefont {Valleau}},\ and\ \bibinfo
  {author} {\bibfnamefont {A.}~\bibnamefont {Aspuru-Guzik}},\ }\bibfield
  {title} {\enquote {\bibinfo {title} {{Atomistic study of the long-lived
  quantum coherences in the Fenna-Matthews-Olson complex}},}\ }\href
  {https://doi.org/10.1016/j.bpj.2011.12.021} {\bibfield  {journal} {\bibinfo
  {journal} {Biophys. J.}\ }\textbf {\bibinfo {volume} {102}},\ \bibinfo
  {pages} {649--660} (\bibinfo {year} {2012})}\BibitemShut {NoStop}%
\bibitem [{\citenamefont {Lee}\ and\ \citenamefont {Coker}(2016)}]{Lee2016}%
  \BibitemOpen
  \bibfield  {author} {\bibinfo {author} {\bibfnamefont {M.~K.}\ \bibnamefont
  {Lee}}\ and\ \bibinfo {author} {\bibfnamefont {D.~F.}\ \bibnamefont
  {Coker}},\ }\bibfield  {title} {\enquote {\bibinfo {title} {{Modeling
  electronic-nuclear interactions for excitation energy transfer processes in
  light-harvesting complexes}},}\ }\href
  {https://doi.org/10.1021/acs.jpclett.6b01440} {\bibfield  {journal} {\bibinfo
   {journal} {J. Phys. Chem. Lett.}\ }\textbf {\bibinfo {volume} {7}},\
  \bibinfo {pages} {3171--3178} (\bibinfo {year} {2016})}\BibitemShut {NoStop}%
\bibitem [{\citenamefont {Lee}, \citenamefont {Huo},\ and\ \citenamefont
  {Coker}(2016)}]{Lee2016b}%
  \BibitemOpen
  \bibfield  {author} {\bibinfo {author} {\bibfnamefont {M.~K.}\ \bibnamefont
  {Lee}}, \bibinfo {author} {\bibfnamefont {P.}~\bibnamefont {Huo}},\ and\
  \bibinfo {author} {\bibfnamefont {D.~F.}\ \bibnamefont {Coker}},\ }\bibfield
  {title} {\enquote {\bibinfo {title} {{Semiclassical path integral dynamics:
  Photosynthetic energy transfer with realistic environment interactions}},}\
  }\href {https://doi.org/10.1146/annurev-physchem-040215-112252} {\bibfield
  {journal} {\bibinfo  {journal} {Annu. Rev. Phys. Chem.}\ }\textbf {\bibinfo
  {volume} {67}},\ \bibinfo {pages} {639--668} (\bibinfo {year}
  {2016})}\BibitemShut {NoStop}%
\bibitem [{\citenamefont {Cignoni}\ \emph {et~al.}(2022)\citenamefont
  {Cignoni}, \citenamefont {Slama}, \citenamefont {Cupellini},\ and\
  \citenamefont {Mennucci}}]{Cignoni2022}%
  \BibitemOpen
  \bibfield  {author} {\bibinfo {author} {\bibfnamefont {E.}~\bibnamefont
  {Cignoni}}, \bibinfo {author} {\bibfnamefont {V.}~\bibnamefont {Slama}},
  \bibinfo {author} {\bibfnamefont {L.}~\bibnamefont {Cupellini}},\ and\
  \bibinfo {author} {\bibfnamefont {B.}~\bibnamefont {Mennucci}},\ }\bibfield
  {title} {\enquote {\bibinfo {title} {{The atomistic modeling of
  light-harvesting complexes from the physical models to the computational
  protocol}},}\ }\href@noop {} {\bibfield  {journal} {\bibinfo  {journal} {J.
  Chem. Phys.}\ }\textbf {\bibinfo {volume} {156}},\ \bibinfo {pages} {120901}
  (\bibinfo {year} {2022})}\BibitemShut {NoStop}%
\bibitem [{\citenamefont {Runge}\ and\ \citenamefont
  {Gross}(1984)}]{Runge1984}%
  \BibitemOpen
  \bibfield  {author} {\bibinfo {author} {\bibfnamefont {E.}~\bibnamefont
  {Runge}}\ and\ \bibinfo {author} {\bibfnamefont {E.~K.~U.}\ \bibnamefont
  {Gross}},\ }\bibfield  {title} {\enquote {\bibinfo {title}
  {Density-functional theory for time-dependent systems},}\ }\href
  {https://doi.org/10.1103/PhysRevLett.52.997} {\bibfield  {journal} {\bibinfo
  {journal} {Phys. Rev. Lett.}\ }\textbf {\bibinfo {volume} {52}},\ \bibinfo
  {pages} {997--1000} (\bibinfo {year} {1984})}\BibitemShut {NoStop}%
\bibitem [{\citenamefont {Casida}(1995)}]{Casida1995}%
  \BibitemOpen
  \bibfield  {author} {\bibinfo {author} {\bibfnamefont {M.~E.}\ \bibnamefont
  {Casida}},\ }\enquote {\bibinfo {title} {Time-dependent density functional
  response theory for molecules},}\ in\ \href@noop {} {\emph {\bibinfo
  {booktitle} {Recent advances in density functional methods}}}\ (\bibinfo
  {year} {1995})\ pp.\ \bibinfo {pages} {155--192}\BibitemShut {NoStop}%
\bibitem [{\citenamefont {Eckart}(1935)}]{Eckart1935}%
  \BibitemOpen
  \bibfield  {author} {\bibinfo {author} {\bibfnamefont {C.}~\bibnamefont
  {Eckart}},\ }\bibfield  {title} {\enquote {\bibinfo {title} {Some studies
  concerning rotating axes and polyatomic molecules},}\ }\href
  {https://doi.org/10.1103/PhysRev.47.552} {\bibfield  {journal} {\bibinfo
  {journal} {Phys. Rev.}\ }\textbf {\bibinfo {volume} {47}},\ \bibinfo {pages}
  {552--558} (\bibinfo {year} {1935})}\BibitemShut {NoStop}%
\bibitem [{\citenamefont {Krasnoshchekov}, \citenamefont {Isayeva},\ and\
  \citenamefont {Stepanov}(2014)}]{Krasnoshchekov2014}%
  \BibitemOpen
  \bibfield  {author} {\bibinfo {author} {\bibfnamefont {S.~V.}\ \bibnamefont
  {Krasnoshchekov}}, \bibinfo {author} {\bibfnamefont {E.~V.}\ \bibnamefont
  {Isayeva}},\ and\ \bibinfo {author} {\bibfnamefont {N.~F.}\ \bibnamefont
  {Stepanov}},\ }\bibfield  {title} {\enquote {\bibinfo {title} {{Determination
  of the Eckart molecule-fixed frame by use of the apparatus of quaternion
  algebra}},}\ }\href {https://doi.org/10.1063/1.4870936} {\bibfield  {journal}
  {\bibinfo  {journal} {J. Chem. Phys.}\ }\textbf {\bibinfo {volume} {140}},\
  \bibinfo {pages} {154104} (\bibinfo {year} {2014})}\BibitemShut {NoStop}%
\bibitem [{\citenamefont {Zuehlsdorff}\ \emph
  {et~al.}(2020{\natexlab{a}})\citenamefont {Zuehlsdorff}, \citenamefont
  {Hong}, \citenamefont {Shi},\ and\ \citenamefont {Isborn}}]{Zuehlsdorff2020}%
  \BibitemOpen
  \bibfield  {author} {\bibinfo {author} {\bibfnamefont {T.~J.}\ \bibnamefont
  {Zuehlsdorff}}, \bibinfo {author} {\bibfnamefont {H.}~\bibnamefont {Hong}},
  \bibinfo {author} {\bibfnamefont {L.}~\bibnamefont {Shi}},\ and\ \bibinfo
  {author} {\bibfnamefont {C.~M.}\ \bibnamefont {Isborn}},\ }\bibfield  {title}
  {\enquote {\bibinfo {title} {{Influence of electronic polarization on the
  spectral density}},}\ }\href {https://doi.org/10.1021/acs.jpcb.9b10250}
  {\bibfield  {journal} {\bibinfo  {journal} {J. Phys. Chem. B}\ }\textbf
  {\bibinfo {volume} {124}},\ \bibinfo {pages} {531--543} (\bibinfo {year}
  {2020}{\natexlab{a}})}\BibitemShut {NoStop}%
\bibitem [{\citenamefont {Zuehlsdorff}(2024)}]{Spectroscopy_python_code}%
  \BibitemOpen
  \bibfield  {author} {\bibinfo {author} {\bibfnamefont {T.~J.}\ \bibnamefont
  {Zuehlsdorff}},\ }\href@noop {} {\enquote {\bibinfo {title} {Molspeckpy},}\
  }\bibinfo {howpublished}
  {\url{https://github.com/tjz21/Spectroscopy_python_code}} (\bibinfo {year}
  {2024}),\ \bibinfo {note} {accessed: 2024-06-13}\BibitemShut {NoStop}%
\bibitem [{\citenamefont {Domcke}\ and\ \citenamefont
  {Yarkony}(2012)}]{Domcke2012}%
  \BibitemOpen
  \bibfield  {author} {\bibinfo {author} {\bibfnamefont {W.}~\bibnamefont
  {Domcke}}\ and\ \bibinfo {author} {\bibfnamefont {D.~R.}\ \bibnamefont
  {Yarkony}},\ }\bibfield  {title} {\enquote {\bibinfo {title} {{Role of
  conical intersections in molecular spectroscopy and photoinduced chemical
  dynamics}},}\ }\href {https://doi.org/10.1146/annurev-physchem-032210-103522}
  {\bibfield  {journal} {\bibinfo  {journal} {Annu. Rev. Phys. Chem.}\ }\textbf
  {\bibinfo {volume} {63}},\ \bibinfo {pages} {325--352} (\bibinfo {year}
  {2012})}\BibitemShut {NoStop}%
\bibitem [{\citenamefont {Tamascelli}\ \emph {et~al.}(2019)\citenamefont
  {Tamascelli}, \citenamefont {Smirne}, \citenamefont {Lim}, \citenamefont
  {Huelga},\ and\ \citenamefont {Plenio}}]{tamascelli2019efficient}%
  \BibitemOpen
  \bibfield  {author} {\bibinfo {author} {\bibfnamefont {D.}~\bibnamefont
  {Tamascelli}}, \bibinfo {author} {\bibfnamefont {A.}~\bibnamefont {Smirne}},
  \bibinfo {author} {\bibfnamefont {J.}~\bibnamefont {Lim}}, \bibinfo {author}
  {\bibfnamefont {S.~F.}\ \bibnamefont {Huelga}},\ and\ \bibinfo {author}
  {\bibfnamefont {M.~B.}\ \bibnamefont {Plenio}},\ }\bibfield  {title}
  {\enquote {\bibinfo {title} {Efficient simulation of finite-temperature open
  quantum systems},}\ }\href@noop {} {\bibfield  {journal} {\bibinfo  {journal}
  {Phys. Rev. Lett.}\ }\textbf {\bibinfo {volume} {123}},\ \bibinfo {pages}
  {090402} (\bibinfo {year} {2019})}\BibitemShut {NoStop}%
\bibitem [{\citenamefont {K{\"{o}}ppel}, \citenamefont {Domcke},\ and\
  \citenamefont {Cederbaum}(1984)}]{Koppel1984}%
  \BibitemOpen
  \bibfield  {author} {\bibinfo {author} {\bibfnamefont {H.}~\bibnamefont
  {K{\"{o}}ppel}}, \bibinfo {author} {\bibfnamefont {W.}~\bibnamefont
  {Domcke}},\ and\ \bibinfo {author} {\bibfnamefont {L.~S.}\ \bibnamefont
  {Cederbaum}},\ }\href {https://doi.org/10.1002/9780470142813.ch2} {\emph
  {\bibinfo {title} {{Multimode molecular dynamics beyond the Born-Oppenheimer
  approximation}}}},\ edited by\ \bibinfo {editor} {\bibfnamefont
  {I.}~\bibnamefont {Prigogine}}\ and\ \bibinfo {editor} {\bibfnamefont
  {S.~A.}\ \bibnamefont {Rice}},\ Vol.\ \bibinfo {volume} {Advances in Chemical
  Physics, LVII}\ (\bibinfo  {publisher} {Wiley},\ \bibinfo {year} {1984})\
  pp.\ \bibinfo {pages} {59--246}\BibitemShut {NoStop}%
\bibitem [{\citenamefont {Zuehlsdorff}\ \emph
  {et~al.}(2020{\natexlab{b}})\citenamefont {Zuehlsdorff}, \citenamefont
  {Hong}, \citenamefont {Shi},\ and\ \citenamefont
  {Isborn}}]{Zuehlsdorff2020b}%
  \BibitemOpen
  \bibfield  {author} {\bibinfo {author} {\bibfnamefont {T.~J.}\ \bibnamefont
  {Zuehlsdorff}}, \bibinfo {author} {\bibfnamefont {H.}~\bibnamefont {Hong}},
  \bibinfo {author} {\bibfnamefont {L.}~\bibnamefont {Shi}},\ and\ \bibinfo
  {author} {\bibfnamefont {C.~M.}\ \bibnamefont {Isborn}},\ }\bibfield  {title}
  {\enquote {\bibinfo {title} {Nonlinear spectroscopy in the condensed phase:
  The role of duschinsky rotations and third order cumulant contributions},}\
  }\href@noop {} {\bibfield  {journal} {\bibinfo  {journal} {The Journal of
  Chemical Physics}\ }\textbf {\bibinfo {volume} {153}},\ \bibinfo {pages}
  {044127} (\bibinfo {year} {2020}{\natexlab{b}})}\BibitemShut {NoStop}%
\bibitem [{\citenamefont {Domcke}, \citenamefont {K{\"{o}}ppel},\ and\
  \citenamefont {Cederbaum}(1981)}]{Domcke1981}%
  \BibitemOpen
  \bibfield  {author} {\bibinfo {author} {\bibfnamefont {W.}~\bibnamefont
  {Domcke}}, \bibinfo {author} {\bibfnamefont {H.}~\bibnamefont
  {K{\"{o}}ppel}},\ and\ \bibinfo {author} {\bibfnamefont {L.~S.}\ \bibnamefont
  {Cederbaum}},\ }\bibfield  {title} {\enquote {\bibinfo {title}
  {{Spectroscopic effects of conical intersections of molecular potential
  energy surfaces}},}\ }\href {https://doi.org/10.1080/00268978100101721}
  {\bibfield  {journal} {\bibinfo  {journal} {Mol. Phys.}\ }\textbf {\bibinfo
  {volume} {43}},\ \bibinfo {pages} {851--875} (\bibinfo {year}
  {1981})}\BibitemShut {NoStop}%
\bibitem [{\citenamefont {Worth}, \citenamefont {Meyer},\ and\ \citenamefont
  {Cederbaum}(1996)}]{Worth1996}%
  \BibitemOpen
  \bibfield  {author} {\bibinfo {author} {\bibfnamefont {G.~A.}\ \bibnamefont
  {Worth}}, \bibinfo {author} {\bibfnamefont {H.~D.}\ \bibnamefont {Meyer}},\
  and\ \bibinfo {author} {\bibfnamefont {L.~S.}\ \bibnamefont {Cederbaum}},\
  }\bibfield  {title} {\enquote {\bibinfo {title} {{The effect of a model
  environment on the S2 absorption spectrum of pyrazine: A wave packet study
  treating all 24 vibrational modes}},}\ }\href
  {https://doi.org/10.1063/1.472327} {\bibfield  {journal} {\bibinfo  {journal}
  {J. Chem. Phys.}\ }\textbf {\bibinfo {volume} {105}},\ \bibinfo {pages}
  {4412--4426} (\bibinfo {year} {1996})}\BibitemShut {NoStop}%
\bibitem [{\citenamefont {Capano}\ \emph {et~al.}(2014)\citenamefont {Capano},
  \citenamefont {Chergui}, \citenamefont {Rothlisberger}, \citenamefont
  {Tavernelli},\ and\ \citenamefont {Penfold}}]{Capano2014}%
  \BibitemOpen
  \bibfield  {author} {\bibinfo {author} {\bibfnamefont {G.}~\bibnamefont
  {Capano}}, \bibinfo {author} {\bibfnamefont {M.}~\bibnamefont {Chergui}},
  \bibinfo {author} {\bibfnamefont {U.}~\bibnamefont {Rothlisberger}}, \bibinfo
  {author} {\bibfnamefont {I.}~\bibnamefont {Tavernelli}},\ and\ \bibinfo
  {author} {\bibfnamefont {T.~J.}\ \bibnamefont {Penfold}},\ }\bibfield
  {title} {\enquote {\bibinfo {title} {A quantum dynamics study of the
  ultrafast relaxation in a prototypical cu(i)–phenanthroline},}\ }\href
  {https://doi.org/10.1021/jp509728m} {\bibfield  {journal} {\bibinfo
  {journal} {J. Phys. Chem. A}\ }\textbf {\bibinfo {volume} {118}},\ \bibinfo
  {pages} {9861--9869} (\bibinfo {year} {2014})}\BibitemShut {NoStop}%
\bibitem [{\citenamefont {P\'{a}pai}\ \emph {et~al.}(2016)\citenamefont
  {P\'{a}pai}, \citenamefont {Vank\'{o}}, \citenamefont {Rozgonyi},\ and\
  \citenamefont {Penfold}}]{Papai2016}%
  \BibitemOpen
  \bibfield  {author} {\bibinfo {author} {\bibfnamefont {M.}~\bibnamefont
  {P\'{a}pai}}, \bibinfo {author} {\bibfnamefont {G.}~\bibnamefont
  {Vank\'{o}}}, \bibinfo {author} {\bibfnamefont {T.}~\bibnamefont
  {Rozgonyi}},\ and\ \bibinfo {author} {\bibfnamefont {T.~J.}\ \bibnamefont
  {Penfold}},\ }\bibfield  {title} {\enquote {\bibinfo {title} {High-efficiency
  iron photosensitizer explained with quantum wavepacket dynamics},}\ }\href
  {https://doi.org/10.1021/acs.jpclett.6b00711} {\bibfield  {journal} {\bibinfo
   {journal} {J. Phys. Chem. Lett.}\ }\textbf {\bibinfo {volume} {7}},\
  \bibinfo {pages} {2009--2014} (\bibinfo {year} {2016})}\BibitemShut {NoStop}%
\bibitem [{\citenamefont {Neville}, \citenamefont {Stolow},\ and\ \citenamefont
  {Schuurman}(2018)}]{Neville2018}%
  \BibitemOpen
  \bibfield  {author} {\bibinfo {author} {\bibfnamefont {S.~P.}\ \bibnamefont
  {Neville}}, \bibinfo {author} {\bibfnamefont {A.}~\bibnamefont {Stolow}},\
  and\ \bibinfo {author} {\bibfnamefont {M.~S.}\ \bibnamefont {Schuurman}},\
  }\bibfield  {title} {\enquote {\bibinfo {title} {{Vacuum ultraviolet excited
  state dynamics of the smallest ring, cyclopropane. I. A reinterpretation of
  the electronic spectrum and the effect of intensity borrowing}},}\ }\href
  {http://dx.doi.org/10.1063/1.5044392} {\bibfield  {journal} {\bibinfo
  {journal} {J. Chem. Phys.}\ }\textbf {\bibinfo {volume} {149}} (\bibinfo
  {year} {2018})}\BibitemShut {NoStop}%
\bibitem [{\citenamefont {Aranda}\ and\ \citenamefont
  {Santoro}(2021)}]{Aranda2021}%
  \BibitemOpen
  \bibfield  {author} {\bibinfo {author} {\bibfnamefont {D.}~\bibnamefont
  {Aranda}}\ and\ \bibinfo {author} {\bibfnamefont {F.}~\bibnamefont
  {Santoro}},\ }\bibfield  {title} {\enquote {\bibinfo {title} {Vibronic
  spectra of pi-conjugated systems with a multitude of coupled states: A
  protocol based on linear vibronic coupling models and quantum dynamics tested
  on hexahelicene},}\ }\href {https://doi.org/10.1021/acs.jctc.1c00022}
  {\bibfield  {journal} {\bibinfo  {journal} {J. Chem. Theory Comput.}\
  }\textbf {\bibinfo {volume} {17}},\ \bibinfo {pages} {1691--1700} (\bibinfo
  {year} {2021})}\BibitemShut {NoStop}%
\bibitem [{\citenamefont {Zobel}\ \emph {et~al.}(2021)\citenamefont {Zobel},
  \citenamefont {Heindl}, \citenamefont {Plasser}, \citenamefont {Mai},\ and\
  \citenamefont {González}}]{Zobel2021}%
  \BibitemOpen
  \bibfield  {author} {\bibinfo {author} {\bibfnamefont {J.~P.}\ \bibnamefont
  {Zobel}}, \bibinfo {author} {\bibfnamefont {M.}~\bibnamefont {Heindl}},
  \bibinfo {author} {\bibfnamefont {F.}~\bibnamefont {Plasser}}, \bibinfo
  {author} {\bibfnamefont {S.}~\bibnamefont {Mai}},\ and\ \bibinfo {author}
  {\bibfnamefont {L.}~\bibnamefont {González}},\ }\bibfield  {title} {\enquote
  {\bibinfo {title} {Surface hopping dynamics on vibronic coupling models},}\
  }\href {https://doi.org/10.1021/acs.accounts.1c00485} {\bibfield  {journal}
  {\bibinfo  {journal} {Acc. Chem. Res.}\ }\textbf {\bibinfo {volume} {54}},\
  \bibinfo {pages} {3760--3771} (\bibinfo {year} {2021})}\BibitemShut {NoStop}%
\bibitem [{\citenamefont {Segatta}\ \emph {et~al.}(2023)\citenamefont
  {Segatta}, \citenamefont {Ruiz}, \citenamefont {Aleotti}, \citenamefont
  {Yaghoubi}, \citenamefont {Mukamel}, \citenamefont {Garavelli}, \citenamefont
  {Santoro},\ and\ \citenamefont {Nenov}}]{Segatta2023}%
  \BibitemOpen
  \bibfield  {author} {\bibinfo {author} {\bibfnamefont {F.}~\bibnamefont
  {Segatta}}, \bibinfo {author} {\bibfnamefont {D.~A.}\ \bibnamefont {Ruiz}},
  \bibinfo {author} {\bibfnamefont {F.}~\bibnamefont {Aleotti}}, \bibinfo
  {author} {\bibfnamefont {M.}~\bibnamefont {Yaghoubi}}, \bibinfo {author}
  {\bibfnamefont {S.}~\bibnamefont {Mukamel}}, \bibinfo {author} {\bibfnamefont
  {M.}~\bibnamefont {Garavelli}}, \bibinfo {author} {\bibfnamefont
  {F.}~\bibnamefont {Santoro}},\ and\ \bibinfo {author} {\bibfnamefont
  {A.}~\bibnamefont {Nenov}},\ }\bibfield  {title} {\enquote {\bibinfo {title}
  {Nonlinear molecular electronic spectroscopy via {MCTDH} quantum dynamics:
  From exact to approximate expressions},}\ }\href@noop {} {\bibfield
  {journal} {\bibinfo  {journal} {J. Chem. Theory Comput.}\ }\textbf {\bibinfo
  {volume} {19}},\ \bibinfo {pages} {2075--2091} (\bibinfo {year}
  {2023})}\BibitemShut {NoStop}%
\bibitem [{\citenamefont {Penfold}\ and\ \citenamefont
  {Eng}(2023)}]{Penfold2022}%
  \BibitemOpen
  \bibfield  {author} {\bibinfo {author} {\bibfnamefont {T.~J.}\ \bibnamefont
  {Penfold}}\ and\ \bibinfo {author} {\bibfnamefont {J.}~\bibnamefont {Eng}},\
  }\bibfield  {title} {\enquote {\bibinfo {title} {Mind the gap: quantifying
  the breakdown of the linear vibronic coupling hamiltonian},}\ }\href@noop {}
  {\bibfield  {journal} {\bibinfo  {journal} {Phys. Chem. Chem. Phys.}\
  }\textbf {\bibinfo {volume} {25}},\ \bibinfo {pages} {7195--7204} (\bibinfo
  {year} {2023})}\BibitemShut {NoStop}%
\bibitem [{\citenamefont {Subotnik}\ \emph {et~al.}(2015)\citenamefont
  {Subotnik}, \citenamefont {Alguire}, \citenamefont {Ou}, \citenamefont
  {Landry},\ and\ \citenamefont {Fatehi}}]{Subotnik2015}%
  \BibitemOpen
  \bibfield  {author} {\bibinfo {author} {\bibfnamefont {J.~E.}\ \bibnamefont
  {Subotnik}}, \bibinfo {author} {\bibfnamefont {E.~C.}\ \bibnamefont
  {Alguire}}, \bibinfo {author} {\bibfnamefont {Q.}~\bibnamefont {Ou}},
  \bibinfo {author} {\bibfnamefont {B.~R.}\ \bibnamefont {Landry}},\ and\
  \bibinfo {author} {\bibfnamefont {S.}~\bibnamefont {Fatehi}},\ }\bibfield
  {title} {\enquote {\bibinfo {title} {The requisite electronic structure
  theory to describe photoexcited nonadiabatic dynamics: Nonadiabatic
  derivative couplings and diabatic electronic couplings},}\ }\href@noop {}
  {\bibfield  {journal} {\bibinfo  {journal} {Acc. Chem. Res.}\ }\textbf
  {\bibinfo {volume} {48}},\ \bibinfo {pages} {1340--1350} (\bibinfo {year}
  {2015})}\BibitemShut {NoStop}%
\bibitem [{\citenamefont {Medders}\ \emph {et~al.}(2017)\citenamefont
  {Medders}, \citenamefont {Alguire}, \citenamefont {Jain},\ and\ \citenamefont
  {Subotnik}}]{Medders2017}%
  \BibitemOpen
  \bibfield  {author} {\bibinfo {author} {\bibfnamefont {G.~R.}\ \bibnamefont
  {Medders}}, \bibinfo {author} {\bibfnamefont {E.~C.}\ \bibnamefont
  {Alguire}}, \bibinfo {author} {\bibfnamefont {A.}~\bibnamefont {Jain}},\ and\
  \bibinfo {author} {\bibfnamefont {J.~E.}\ \bibnamefont {Subotnik}},\
  }\bibfield  {title} {\enquote {\bibinfo {title} {Ultrafast electronic
  relaxation through a conical intersection: Nonadiabatic dynamics disentangled
  through an oscillator strength-based diabatization framework},}\ }\href@noop
  {} {\bibfield  {journal} {\bibinfo  {journal} {J. Phys. Chem. A}\ }\textbf
  {\bibinfo {volume} {121}},\ \bibinfo {pages} {1425--1434} (\bibinfo {year}
  {2017})}\BibitemShut {NoStop}%
\bibitem [{\citenamefont {Chin}\ \emph {et~al.}(2010)\citenamefont {Chin},
  \citenamefont {Rivas}, \citenamefont {Huelga},\ and\ \citenamefont
  {Plenio}}]{Chin2010}%
  \BibitemOpen
  \bibfield  {author} {\bibinfo {author} {\bibfnamefont {A.~W.}\ \bibnamefont
  {Chin}}, \bibinfo {author} {\bibfnamefont {{\'{A}}.}~\bibnamefont {Rivas}},
  \bibinfo {author} {\bibfnamefont {S.~F.}\ \bibnamefont {Huelga}},\ and\
  \bibinfo {author} {\bibfnamefont {M.~B.}\ \bibnamefont {Plenio}},\ }\bibfield
   {title} {\enquote {\bibinfo {title} {{Exact mapping between system-reservoir
  quantum models and semi-infinite discrete chains using orthogonal
  polynomials}},}\ }\href@noop {} {\bibfield  {journal} {\bibinfo  {journal}
  {J. Math. Phys.}\ }\textbf {\bibinfo {volume} {51}} (\bibinfo {year}
  {2010})},\ \Eprint {https://arxiv.org/abs/1006.4507} {arXiv:1006.4507}
  \BibitemShut {NoStop}%
\bibitem [{\citenamefont {Haegeman}\ \emph {et~al.}(2011)\citenamefont
  {Haegeman}, \citenamefont {Cirac}, \citenamefont {Osborne}, \citenamefont
  {Pi\ifmmode~\check{z}\else \v{z}\fi{}orn}, \citenamefont {Verschelde},\ and\
  \citenamefont {Verstraete}}]{Haegeman2011}%
  \BibitemOpen
  \bibfield  {author} {\bibinfo {author} {\bibfnamefont {J.}~\bibnamefont
  {Haegeman}}, \bibinfo {author} {\bibfnamefont {J.~I.}\ \bibnamefont {Cirac}},
  \bibinfo {author} {\bibfnamefont {T.~J.}\ \bibnamefont {Osborne}}, \bibinfo
  {author} {\bibfnamefont {I.}~\bibnamefont {Pi\ifmmode~\check{z}\else
  \v{z}\fi{}orn}}, \bibinfo {author} {\bibfnamefont {H.}~\bibnamefont
  {Verschelde}},\ and\ \bibinfo {author} {\bibfnamefont {F.}~\bibnamefont
  {Verstraete}},\ }\bibfield  {title} {\enquote {\bibinfo {title}
  {Time-dependent variational principle for quantum lattices},}\ }\href@noop {}
  {\bibfield  {journal} {\bibinfo  {journal} {Phys. Rev. Lett.}\ }\textbf
  {\bibinfo {volume} {107}},\ \bibinfo {pages} {070601} (\bibinfo {year}
  {2011})}\BibitemShut {NoStop}%
\bibitem [{\citenamefont {Haegeman}\ \emph
  {et~al.}(2016{\natexlab{a}})\citenamefont {Haegeman}, \citenamefont {Lubich},
  \citenamefont {Oseledets}, \citenamefont {Vandereycken},\ and\ \citenamefont
  {Verstraete}}]{Haegeman2016}%
  \BibitemOpen
  \bibfield  {author} {\bibinfo {author} {\bibfnamefont {J.}~\bibnamefont
  {Haegeman}}, \bibinfo {author} {\bibfnamefont {C.}~\bibnamefont {Lubich}},
  \bibinfo {author} {\bibfnamefont {I.}~\bibnamefont {Oseledets}}, \bibinfo
  {author} {\bibfnamefont {B.}~\bibnamefont {Vandereycken}},\ and\ \bibinfo
  {author} {\bibfnamefont {F.}~\bibnamefont {Verstraete}},\ }\bibfield  {title}
  {\enquote {\bibinfo {title} {Unifying time evolution and optimization with
  matrix product states},}\ }\href@noop {} {\bibfield  {journal} {\bibinfo
  {journal} {Phys. Rev. B}\ }\textbf {\bibinfo {volume} {94}},\ \bibinfo
  {pages} {165116} (\bibinfo {year} {2016}{\natexlab{a}})}\BibitemShut
  {NoStop}%
\bibitem [{\citenamefont {Kloss}, \citenamefont {Lev},\ and\ \citenamefont
  {Reichman}(2018)}]{kloss2018time}%
  \BibitemOpen
  \bibfield  {author} {\bibinfo {author} {\bibfnamefont {B.}~\bibnamefont
  {Kloss}}, \bibinfo {author} {\bibfnamefont {Y.~B.}\ \bibnamefont {Lev}},\
  and\ \bibinfo {author} {\bibfnamefont {D.}~\bibnamefont {Reichman}},\
  }\bibfield  {title} {\enquote {\bibinfo {title} {Time-dependent variational
  principle in matrix-product state manifolds: Pitfalls and potential},}\
  }\href@noop {} {\bibfield  {journal} {\bibinfo  {journal} {Phys. Rev. B}\
  }\textbf {\bibinfo {volume} {97}},\ \bibinfo {pages} {024307} (\bibinfo
  {year} {2018})}\BibitemShut {NoStop}%
\bibitem [{\citenamefont {Haegeman}\ \emph
  {et~al.}(2016{\natexlab{b}})\citenamefont {Haegeman}, \citenamefont {Lubich},
  \citenamefont {Oseledets}, \citenamefont {Vandereycken},\ and\ \citenamefont
  {Verstraete}}]{tdvp}%
  \BibitemOpen
  \bibfield  {author} {\bibinfo {author} {\bibfnamefont {J.}~\bibnamefont
  {Haegeman}}, \bibinfo {author} {\bibfnamefont {C.}~\bibnamefont {Lubich}},
  \bibinfo {author} {\bibfnamefont {I.}~\bibnamefont {Oseledets}}, \bibinfo
  {author} {\bibfnamefont {B.}~\bibnamefont {Vandereycken}},\ and\ \bibinfo
  {author} {\bibfnamefont {F.}~\bibnamefont {Verstraete}},\ }\bibfield  {title}
  {\enquote {\bibinfo {title} {Unifying time evolution and optimization with
  matrix product states},}\ }\href {https://doi.org/10.1103/PhysRevB.94.165116}
  {\bibfield  {journal} {\bibinfo  {journal} {Phys. Rev. B}\ }\textbf {\bibinfo
  {volume} {94}},\ \bibinfo {pages} {165116} (\bibinfo {year}
  {2016}{\natexlab{b}})}\BibitemShut {NoStop}%
\bibitem [{\citenamefont {Gautschi}(1994)}]{gautschi_algorithm_1994}%
  \BibitemOpen
  \bibfield  {author} {\bibinfo {author} {\bibfnamefont {W.}~\bibnamefont
  {Gautschi}},\ }\bibfield  {title} {\enquote {\bibinfo {title} {Algorithm 726:
  {ORTHPOL}–a package of routines for generating orthogonal polynomials and
  gauss-type quadrature rules},}\ }\href
  {https://doi.org/10.1145/174603.174605} {\bibfield  {journal} {\bibinfo
  {journal} {ACM Trans. Math. Softw.}\ }\textbf {\bibinfo {volume} {20}},\
  \bibinfo {pages} {21--62} (\bibinfo {year} {1994})}\BibitemShut {NoStop}%
\bibitem [{\citenamefont {Lacroix}\ \emph {et~al.}(2024)\citenamefont
  {Lacroix}, \citenamefont {Dé}, \citenamefont {Riva}, \citenamefont
  {Dunnett},\ and\ \citenamefont {Chin}}]{lacroix2024mpsdynamicsjl}%
  \BibitemOpen
  \bibfield  {author} {\bibinfo {author} {\bibfnamefont {T.}~\bibnamefont
  {Lacroix}}, \bibinfo {author} {\bibfnamefont {B.~L.}\ \bibnamefont {Dé}},
  \bibinfo {author} {\bibfnamefont {A.}~\bibnamefont {Riva}}, \bibinfo {author}
  {\bibfnamefont {A.~J.}\ \bibnamefont {Dunnett}},\ and\ \bibinfo {author}
  {\bibfnamefont {A.~W.}\ \bibnamefont {Chin}},\ }\href@noop {} {\enquote
  {\bibinfo {title} {{MPSD}ynamics.jl: Tensor network simulations for
  finite-temperature (non-{M}arkovian) open quantum system dynamics},}\ }
  (\bibinfo {year} {2024}),\ \Eprint {https://arxiv.org/abs/2406.07052}
  {arXiv:2406.07052 [quant-ph]} \BibitemShut {NoStop}%
\bibitem [{\citenamefont {Dunnett}(2021)}]{dunnett_angus_2021_5106435}%
  \BibitemOpen
  \bibfield  {author} {\bibinfo {author} {\bibfnamefont {A.}~\bibnamefont
  {Dunnett}},\ }\href@noop {} {\enquote {\bibinfo {title} {{MPSDynamics}},}\
  }\bibinfo {howpublished} {\url{https://github.com/shareloqs/MPSDynamics}}
  (\bibinfo {year} {2021})\BibitemShut {NoStop}%
\bibitem [{\citenamefont {Hunter}\ and\ \citenamefont
  {Zuehlsdorff}(2024)}]{MPSDynamics_GPU}%
  \BibitemOpen
  \bibfield  {author} {\bibinfo {author} {\bibfnamefont {K.~E.}\ \bibnamefont
  {Hunter}}\ and\ \bibinfo {author} {\bibfnamefont {T.~J.}\ \bibnamefont
  {Zuehlsdorff}},\ }\href@noop {} {\enquote {\bibinfo {title} {Mpsdynamics.jl
  (gpu version)},}\ }\bibinfo {howpublished}
  {\url{https://github.com/tjz21/MPSDynamics_GPU}} (\bibinfo {year} {2024}),\
  \bibinfo {note} {accessed: 2024-06-13}\BibitemShut {NoStop}%
\bibitem [{\citenamefont {Stoudenmire}\ and\ \citenamefont
  {White}(2013)}]{Stoudenmire2013}%
  \BibitemOpen
  \bibfield  {author} {\bibinfo {author} {\bibfnamefont {E.~M.}\ \bibnamefont
  {Stoudenmire}}\ and\ \bibinfo {author} {\bibfnamefont {S.~R.}\ \bibnamefont
  {White}},\ }\bibfield  {title} {\enquote {\bibinfo {title} {Real-space
  parallel density matrix renormalization group},}\ }\href@noop {} {\bibfield
  {journal} {\bibinfo  {journal} {Phys. Rev. B}\ }\textbf {\bibinfo {volume}
  {87}},\ \bibinfo {pages} {155137} (\bibinfo {year} {2013})}\BibitemShut
  {NoStop}%
\bibitem [{\citenamefont {Secular}\ \emph {et~al.}(2020)\citenamefont
  {Secular}, \citenamefont {Gourianov}, \citenamefont {Lubasch}, \citenamefont
  {Dolgov}, \citenamefont {Clark},\ and\ \citenamefont {Jaksch}}]{Secular2020}%
  \BibitemOpen
  \bibfield  {author} {\bibinfo {author} {\bibfnamefont {P.}~\bibnamefont
  {Secular}}, \bibinfo {author} {\bibfnamefont {N.}~\bibnamefont {Gourianov}},
  \bibinfo {author} {\bibfnamefont {M.}~\bibnamefont {Lubasch}}, \bibinfo
  {author} {\bibfnamefont {S.}~\bibnamefont {Dolgov}}, \bibinfo {author}
  {\bibfnamefont {S.~R.}\ \bibnamefont {Clark}},\ and\ \bibinfo {author}
  {\bibfnamefont {D.}~\bibnamefont {Jaksch}},\ }\bibfield  {title} {\enquote
  {\bibinfo {title} {Parallel time-dependent variational principle algorithm
  for matrix product states},}\ }\href@noop {} {\bibfield  {journal} {\bibinfo
  {journal} {Phys. Rev. B}\ }\textbf {\bibinfo {volume} {101}},\ \bibinfo
  {pages} {235123} (\bibinfo {year} {2020})}\BibitemShut {NoStop}%
\bibitem [{\citenamefont {Warshel}\ and\ \citenamefont
  {Levitt}(1976)}]{Warshel1976}%
  \BibitemOpen
  \bibfield  {author} {\bibinfo {author} {\bibfnamefont {A.}~\bibnamefont
  {Warshel}}\ and\ \bibinfo {author} {\bibfnamefont {M.}~\bibnamefont
  {Levitt}},\ }\bibfield  {title} {\enquote {\bibinfo {title} {Theoretical
  studies of enzymic reactions: Dielectric, electrostatic and steric
  stabilization of the carbonium ion in the reaction of lysozyme},}\
  }\href@noop {} {\bibfield  {journal} {\bibinfo  {journal} {J. Mol. Biol.}\
  }\textbf {\bibinfo {volume} {103}},\ \bibinfo {pages} {227--249} (\bibinfo
  {year} {1976})}\BibitemShut {NoStop}%
\bibitem [{\citenamefont {Ufimtsev}\ and\ \citenamefont
  {Martinez}(2009)}]{Ufimtsev2009}%
  \BibitemOpen
  \bibfield  {author} {\bibinfo {author} {\bibfnamefont {I.~S.}\ \bibnamefont
  {Ufimtsev}}\ and\ \bibinfo {author} {\bibfnamefont {T.~J.}\ \bibnamefont
  {Martinez}},\ }\bibfield  {title} {\enquote {\bibinfo {title} {{Quantum
  chemistry on graphical processing units. 3. Analytical energy gradients and
  first principles molecular dynamics}},}\ }\href
  {https://doi.org/http://pubs.acs.org/doi/pdf/10.1021/ct9003004} {\bibfield
  {journal} {\bibinfo  {journal} {J. Chem. Theory Comput.}\ }\textbf {\bibinfo
  {volume} {5}},\ \bibinfo {pages} {2619--2628} (\bibinfo {year}
  {2009})}\BibitemShut {NoStop}%
\bibitem [{\citenamefont {Isborn}\ \emph {et~al.}(2012)\citenamefont {Isborn},
  \citenamefont {G\"{o}tz}, \citenamefont {Clark}, \citenamefont {Walker},\
  and\ \citenamefont {Mart\'{i}nez}}]{Isborn2012}%
  \BibitemOpen
  \bibfield  {author} {\bibinfo {author} {\bibfnamefont {C.~M.}\ \bibnamefont
  {Isborn}}, \bibinfo {author} {\bibfnamefont {A.~W.}\ \bibnamefont
  {G\"{o}tz}}, \bibinfo {author} {\bibfnamefont {M.~A.}\ \bibnamefont {Clark}},
  \bibinfo {author} {\bibfnamefont {R.~C.}\ \bibnamefont {Walker}},\ and\
  \bibinfo {author} {\bibfnamefont {T.~J.}\ \bibnamefont {Mart\'{i}nez}},\
  }\bibfield  {title} {\enquote {\bibinfo {title} {Electronic absorption
  spectra from {MM} and ab initio {QM/MM} molecular dynamics: Environmental
  effects on the absorption spectrum of photoactive yellow protein},}\ }\href
  {https://doi.org/10.1021/ct3006826} {\bibfield  {journal} {\bibinfo
  {journal} {J. Chem. Theory Comput.}\ }\textbf {\bibinfo {volume} {8}},\
  \bibinfo {pages} {5092--5106} (\bibinfo {year} {2012})}\BibitemShut {NoStop}%
\bibitem [{\citenamefont {Case}\ \emph {et~al.}(2012)\citenamefont {Case},
  \citenamefont {Darden}, \citenamefont {Cheatham}, \citenamefont {Simmerling},
  \citenamefont {Wang}, \citenamefont {Duke}, \citenamefont {Luo},
  \citenamefont {Walker}, \citenamefont {Zhang}, \citenamefont {Merz},
  \citenamefont {Roberts}, \citenamefont {Hayik}, \citenamefont {Roitberg},
  \citenamefont {Seabra}, \citenamefont {Swails}, \citenamefont {G\"{o}tz},
  \citenamefont {Kolossvari}, \citenamefont {Wong}, \citenamefont {Paesani},
  \citenamefont {Vanicek}, \citenamefont {Wolf}, \citenamefont {Liu},
  \citenamefont {Wu}, \citenamefont {Brozell}, \citenamefont {Steinbrecher},
  \citenamefont {Gohlke}, \citenamefont {Cai}, \citenamefont {Ye},
  \citenamefont {Wang}, \citenamefont {Hsieh}, \citenamefont {Cui},
  \citenamefont {Roe}, \citenamefont {Mathews}, \citenamefont {Seetin},
  \citenamefont {Salomon-Ferrer}, \citenamefont {Sagui}, \citenamefont {Babin},
  \citenamefont {Luchko}, \citenamefont {Gusarov}, \citenamefont {Kovalenko},\
  and\ \citenamefont {Kollman}}]{amber}%
  \BibitemOpen
  \bibfield  {author} {\bibinfo {author} {\bibfnamefont {D.~A.}\ \bibnamefont
  {Case}}, \bibinfo {author} {\bibfnamefont {T.~A.}\ \bibnamefont {Darden}},
  \bibinfo {author} {\bibfnamefont {I.}~\bibnamefont {Cheatham}}, \bibinfo
  {author} {\bibfnamefont {C.~L.}\ \bibnamefont {Simmerling}}, \bibinfo
  {author} {\bibfnamefont {J.}~\bibnamefont {Wang}}, \bibinfo {author}
  {\bibfnamefont {R.~E.}\ \bibnamefont {Duke}}, \bibinfo {author}
  {\bibfnamefont {R.}~\bibnamefont {Luo}}, \bibinfo {author} {\bibfnamefont
  {R.~C.}\ \bibnamefont {Walker}}, \bibinfo {author} {\bibfnamefont
  {W.}~\bibnamefont {Zhang}}, \bibinfo {author} {\bibfnamefont {K.~M.}\
  \bibnamefont {Merz}}, \bibinfo {author} {\bibfnamefont {B.}~\bibnamefont
  {Roberts}}, \bibinfo {author} {\bibfnamefont {S.}~\bibnamefont {Hayik}},
  \bibinfo {author} {\bibfnamefont {A.}~\bibnamefont {Roitberg}}, \bibinfo
  {author} {\bibfnamefont {G.}~\bibnamefont {Seabra}}, \bibinfo {author}
  {\bibfnamefont {J.}~\bibnamefont {Swails}}, \bibinfo {author} {\bibfnamefont
  {A.~W.}\ \bibnamefont {G\"{o}tz}}, \bibinfo {author} {\bibfnamefont
  {I.}~\bibnamefont {Kolossvari}}, \bibinfo {author} {\bibfnamefont {K.~F.}\
  \bibnamefont {Wong}}, \bibinfo {author} {\bibfnamefont {F.}~\bibnamefont
  {Paesani}}, \bibinfo {author} {\bibfnamefont {J.}~\bibnamefont {Vanicek}},
  \bibinfo {author} {\bibfnamefont {R.~M.}\ \bibnamefont {Wolf}}, \bibinfo
  {author} {\bibfnamefont {J.}~\bibnamefont {Liu}}, \bibinfo {author}
  {\bibfnamefont {X.}~\bibnamefont {Wu}}, \bibinfo {author} {\bibfnamefont
  {S.~R.}\ \bibnamefont {Brozell}}, \bibinfo {author} {\bibfnamefont
  {T.}~\bibnamefont {Steinbrecher}}, \bibinfo {author} {\bibfnamefont
  {H.}~\bibnamefont {Gohlke}}, \bibinfo {author} {\bibfnamefont
  {Q.}~\bibnamefont {Cai}}, \bibinfo {author} {\bibfnamefont {X.}~\bibnamefont
  {Ye}}, \bibinfo {author} {\bibfnamefont {J.}~\bibnamefont {Wang}}, \bibinfo
  {author} {\bibfnamefont {M.-J.}\ \bibnamefont {Hsieh}}, \bibinfo {author}
  {\bibfnamefont {G.}~\bibnamefont {Cui}}, \bibinfo {author} {\bibfnamefont
  {D.~R.}\ \bibnamefont {Roe}}, \bibinfo {author} {\bibfnamefont {D.~H.}\
  \bibnamefont {Mathews}}, \bibinfo {author} {\bibfnamefont {M.~G.}\
  \bibnamefont {Seetin}}, \bibinfo {author} {\bibfnamefont {R.}~\bibnamefont
  {Salomon-Ferrer}}, \bibinfo {author} {\bibfnamefont {C.}~\bibnamefont
  {Sagui}}, \bibinfo {author} {\bibfnamefont {V.}~\bibnamefont {Babin}},
  \bibinfo {author} {\bibfnamefont {T.}~\bibnamefont {Luchko}}, \bibinfo
  {author} {\bibfnamefont {S.}~\bibnamefont {Gusarov}}, \bibinfo {author}
  {\bibfnamefont {A.}~\bibnamefont {Kovalenko}},\ and\ \bibinfo {author}
  {\bibfnamefont {P.~A.}\ \bibnamefont {Kollman}},\ }\href@noop {} {\enquote
  {\bibinfo {title} {{AMBER} 12},}\ } (\bibinfo {year} {2012}),\ \bibinfo
  {note} {{University of California: San Francisco, CA}}\BibitemShut {NoStop}%
\bibitem [{\citenamefont {Jorgensen}, \citenamefont {Chandrasekhar},\ and\
  \citenamefont {Madura}(1983)}]{TIP3P}%
  \BibitemOpen
  \bibfield  {author} {\bibinfo {author} {\bibfnamefont {W.~L.}\ \bibnamefont
  {Jorgensen}}, \bibinfo {author} {\bibfnamefont {J.}~\bibnamefont
  {Chandrasekhar}},\ and\ \bibinfo {author} {\bibfnamefont {J.~D.}\
  \bibnamefont {Madura}},\ }\bibfield  {title} {\enquote {\bibinfo {title}
  {{Comparison of simple potential functions for simulating liquid water}},}\
  }\href {https://doi.org/10.1063/1.445869} {\bibfield  {journal} {\bibinfo
  {journal} {J. Chem. Phys.}\ }\textbf {\bibinfo {volume} {79}},\ \bibinfo
  {pages} {926} (\bibinfo {year} {1983})}\BibitemShut {NoStop}%
\bibitem [{\citenamefont {Dunning~Jr.}(1989)}]{Dunning1990}%
  \BibitemOpen
  \bibfield  {author} {\bibinfo {author} {\bibfnamefont {T.~H.}\ \bibnamefont
  {Dunning~Jr.}},\ }\bibfield  {title} {\enquote {\bibinfo {title} {{Gaussian
  basis sets for use in correlated molecular calculations. I. The atoms Boron
  through Neon and Hydrogen}},}\ }\href {https://doi.org/10.1063/1.456153}
  {\bibfield  {journal} {\bibinfo  {journal} {J. Chem. Phys.}\ }\textbf
  {\bibinfo {volume} {90}},\ \bibinfo {pages} {1007} (\bibinfo {year}
  {1989})}\BibitemShut {NoStop}%
\bibitem [{\citenamefont {Yanai}, \citenamefont {Tew},\ and\ \citenamefont
  {Handy}(2004)}]{Yanai2004}%
  \BibitemOpen
  \bibfield  {author} {\bibinfo {author} {\bibfnamefont {T.}~\bibnamefont
  {Yanai}}, \bibinfo {author} {\bibfnamefont {D.~P.}\ \bibnamefont {Tew}},\
  and\ \bibinfo {author} {\bibfnamefont {N.~C.}\ \bibnamefont {Handy}},\
  }\bibfield  {title} {\enquote {\bibinfo {title} {{A new hybrid
  exchange-correlation functional using the Coulomb-attenuating method
  (CAM-B3LYP)}},}\ }\href {https://doi.org/10.1016/j.cplett.2004.06.011}
  {\bibfield  {journal} {\bibinfo  {journal} {Chem. Phys. Lett.}\ }\textbf
  {\bibinfo {volume} {393}},\ \bibinfo {pages} {51--57} (\bibinfo {year}
  {2004})}\BibitemShut {NoStop}%
\bibitem [{\citenamefont {Hirata}\ and\ \citenamefont
  {Head-Gordon}(1999)}]{Hirata1999}%
  \BibitemOpen
  \bibfield  {author} {\bibinfo {author} {\bibfnamefont {S.}~\bibnamefont
  {Hirata}}\ and\ \bibinfo {author} {\bibfnamefont {M.}~\bibnamefont
  {Head-Gordon}},\ }\bibfield  {title} {\enquote {\bibinfo {title}
  {{Time-dependent density functional theory within the Tamm–Dancoff
  approximation}},}\ }\href {https://doi.org/10.1016/S0009-2614(99)01149-5}
  {\bibfield  {journal} {\bibinfo  {journal} {Chem. Phys. Lett.}\ }\textbf
  {\bibinfo {volume} {314}},\ \bibinfo {pages} {291--299} (\bibinfo {year}
  {1999})}\BibitemShut {NoStop}%
\bibitem [{\citenamefont {Isborn}\ \emph {et~al.}(2011)\citenamefont {Isborn},
  \citenamefont {Luehr}, \citenamefont {Ufimtsev},\ and\ \citenamefont
  {Mart\'{i}nez}}]{Isborn2011}%
  \BibitemOpen
  \bibfield  {author} {\bibinfo {author} {\bibfnamefont {C.~M.}\ \bibnamefont
  {Isborn}}, \bibinfo {author} {\bibfnamefont {N.}~\bibnamefont {Luehr}},
  \bibinfo {author} {\bibfnamefont {I.~S.}\ \bibnamefont {Ufimtsev}},\ and\
  \bibinfo {author} {\bibfnamefont {T.~J.}\ \bibnamefont {Mart\'{i}nez}},\
  }\bibfield  {title} {\enquote {\bibinfo {title} {{Excited-state electronic
  structure with configuration interaction singles and Tamm--Dancoff
  time-dependent density functional Theory on graphical processing units}},}\
  }\href {https://doi.org/10.1021/ct200030k} {\bibfield  {journal} {\bibinfo
  {journal} {J. Chem. Theory Comput.}\ }\textbf {\bibinfo {volume} {7}},\
  \bibinfo {pages} {1814--1823} (\bibinfo {year} {2011})}\BibitemShut {NoStop}%
\bibitem [{\citenamefont {Yamazaki}\ \emph {et~al.}(1983)\citenamefont
  {Yamazaki}, \citenamefont {Murao}, \citenamefont {Yamanaka},\ and\
  \citenamefont {Yoshihara}}]{Yamazaki1983}%
  \BibitemOpen
  \bibfield  {author} {\bibinfo {author} {\bibfnamefont {I.}~\bibnamefont
  {Yamazaki}}, \bibinfo {author} {\bibfnamefont {T.}~\bibnamefont {Murao}},
  \bibinfo {author} {\bibfnamefont {T.}~\bibnamefont {Yamanaka}},\ and\
  \bibinfo {author} {\bibfnamefont {K.}~\bibnamefont {Yoshihara}},\ }\bibfield
  {title} {\enquote {\bibinfo {title} {Intramolecular electronic relaxation and
  photoisomerization processes in the isolated azabenzene molecules pyridine{,}
  pyrazine and pyrimidine},}\ }\href@noop {} {\bibfield  {journal} {\bibinfo
  {journal} {Faraday Discuss. Chem. Soc.}\ }\textbf {\bibinfo {volume} {75}},\
  \bibinfo {pages} {395--405} (\bibinfo {year} {1983})}\BibitemShut {NoStop}%
\bibitem [{\citenamefont {Neville}, \citenamefont {Seidu},\ and\ \citenamefont
  {Schuurman}(2020)}]{Neville2020}%
  \BibitemOpen
  \bibfield  {author} {\bibinfo {author} {\bibfnamefont {S.~P.}\ \bibnamefont
  {Neville}}, \bibinfo {author} {\bibfnamefont {I.}~\bibnamefont {Seidu}},\
  and\ \bibinfo {author} {\bibfnamefont {M.~S.}\ \bibnamefont {Schuurman}},\
  }\bibfield  {title} {\enquote {\bibinfo {title} {{Propagative block
  diagonalization diabatization of DFT/MRCI electronic states}},}\ }\href@noop
  {} {\bibfield  {journal} {\bibinfo  {journal} {J. Chem. Phys.}\ }\textbf
  {\bibinfo {volume} {152}},\ \bibinfo {pages} {114110} (\bibinfo {year}
  {2020})}\BibitemShut {NoStop}%
\bibitem [{\citenamefont {Krempl}\ \emph {et~al.}(1994)\citenamefont {Krempl},
  \citenamefont {Winterstetter}, \citenamefont {Plöhn},\ and\ \citenamefont
  {Domcke}}]{Krempl1994}%
  \BibitemOpen
  \bibfield  {author} {\bibinfo {author} {\bibfnamefont {S.}~\bibnamefont
  {Krempl}}, \bibinfo {author} {\bibfnamefont {M.}~\bibnamefont
  {Winterstetter}}, \bibinfo {author} {\bibfnamefont {H.}~\bibnamefont
  {Plöhn}},\ and\ \bibinfo {author} {\bibfnamefont {W.}~\bibnamefont
  {Domcke}},\ }\bibfield  {title} {\enquote {\bibinfo {title} {{Path‐integral
  treatment of multi‐mode vibronic coupling}},}\ }\href
  {https://doi.org/10.1063/1.467253} {\bibfield  {journal} {\bibinfo  {journal}
  {J. Chem. Phys.}\ }\textbf {\bibinfo {volume} {100}},\ \bibinfo {pages}
  {926--937} (\bibinfo {year} {1994})}\BibitemShut {NoStop}%
\bibitem [{\citenamefont {Chen}\ \emph {et~al.}(2020)\citenamefont {Chen},
  \citenamefont {Zuehlsdorff}, \citenamefont {Morawietz}, \citenamefont
  {Isborn},\ and\ \citenamefont {Markland}}]{Chen2020}%
  \BibitemOpen
  \bibfield  {author} {\bibinfo {author} {\bibfnamefont {M.~S.}\ \bibnamefont
  {Chen}}, \bibinfo {author} {\bibfnamefont {T.~J.}\ \bibnamefont
  {Zuehlsdorff}}, \bibinfo {author} {\bibfnamefont {T.}~\bibnamefont
  {Morawietz}}, \bibinfo {author} {\bibfnamefont {C.~M.}\ \bibnamefont
  {Isborn}},\ and\ \bibinfo {author} {\bibfnamefont {T.~E.}\ \bibnamefont
  {Markland}},\ }\bibfield  {title} {\enquote {\bibinfo {title} {Exploiting
  machine learning to efficiently predict multidimensional optical spectra in
  complex environments},}\ }\href {https://doi.org/10.1021/acs.jpclett.0c02168}
  {\bibfield  {journal} {\bibinfo  {journal} {J. Phys. Chem. Lett.}\ }\textbf
  {\bibinfo {volume} {11}},\ \bibinfo {pages} {7559--7568} (\bibinfo {year}
  {2020})}\BibitemShut {NoStop}%
\bibitem [{\citenamefont {Chen}\ \emph {et~al.}(2023)\citenamefont {Chen},
  \citenamefont {Mao}, \citenamefont {Snider}, \citenamefont {Gupta},
  \citenamefont {Montoya-Castillo}, \citenamefont {Zuehlsdorff}, \citenamefont
  {Isborn},\ and\ \citenamefont {Markland}}]{Chen2023}%
  \BibitemOpen
  \bibfield  {author} {\bibinfo {author} {\bibfnamefont {M.~S.}\ \bibnamefont
  {Chen}}, \bibinfo {author} {\bibfnamefont {Y.}~\bibnamefont {Mao}}, \bibinfo
  {author} {\bibfnamefont {A.}~\bibnamefont {Snider}}, \bibinfo {author}
  {\bibfnamefont {P.}~\bibnamefont {Gupta}}, \bibinfo {author} {\bibfnamefont
  {A.}~\bibnamefont {Montoya-Castillo}}, \bibinfo {author} {\bibfnamefont
  {T.~J.}\ \bibnamefont {Zuehlsdorff}}, \bibinfo {author} {\bibfnamefont
  {C.~M.}\ \bibnamefont {Isborn}},\ and\ \bibinfo {author} {\bibfnamefont
  {T.~E.}\ \bibnamefont {Markland}},\ }\bibfield  {title} {\enquote {\bibinfo
  {title} {Elucidating the role of hydrogen bonding in the optical spectroscopy
  of the solvated green fluorescent protein chromophore: Using machine learning
  to establish the importance of high-level electronic structure},}\
  }\href@noop {} {\bibfield  {journal} {\bibinfo  {journal} {J. Phys. Chem.
  Lett.}\ }\textbf {\bibinfo {volume} {14}},\ \bibinfo {pages} {6610--6619}
  (\bibinfo {year} {2023})}\BibitemShut {NoStop}%
\end{thebibliography}%


\begin{thebibliography}{25}%
\makeatletter
\providecommand \@ifxundefined [1]{%
 \@ifx{#1\undefined}
}%
\providecommand \@ifnum [1]{%
 \ifnum #1\expandafter \@firstoftwo
 \else \expandafter \@secondoftwo
 \fi
}%
\providecommand \@ifx [1]{%
 \ifx #1\expandafter \@firstoftwo
 \else \expandafter \@secondoftwo
 \fi
}%
\providecommand \natexlab [1]{#1}%
\providecommand \enquote  [1]{``#1''}%
\providecommand \bibnamefont  [1]{#1}%
\providecommand \bibfnamefont [1]{#1}%
\providecommand \citenamefont [1]{#1}%
\providecommand \href@noop [0]{\@secondoftwo}%
\providecommand \href [0]{\begingroup \@sanitize@url \@href}%
\providecommand \@href[1]{\@@startlink{#1}\@@href}%
\providecommand \@@href[1]{\endgroup#1\@@endlink}%
\providecommand \@sanitize@url [0]{\catcode `\\12\catcode `\$12\catcode
  `\&12\catcode `\#12\catcode `\^12\catcode `\_12\catcode `\%12\relax}%
\providecommand \@@startlink[1]{}%
\providecommand \@@endlink[0]{}%
\providecommand \url  [0]{\begingroup\@sanitize@url \@url }%
\providecommand \@url [1]{\endgroup\@href {#1}{\urlprefix }}%
\providecommand \urlprefix  [0]{URL }%
\providecommand \Eprint [0]{\href }%
\providecommand \doibase [0]{https://doi.org/}%
\providecommand \selectlanguage [0]{\@gobble}%
\providecommand \bibinfo  [0]{\@secondoftwo}%
\providecommand \bibfield  [0]{\@secondoftwo}%
\providecommand \translation [1]{[#1]}%
\providecommand \BibitemOpen [0]{}%
\providecommand \bibitemStop [0]{}%
\providecommand \bibitemNoStop [0]{.\EOS\space}%
\providecommand \EOS [0]{\spacefactor3000\relax}%
\providecommand \BibitemShut  [1]{\csname bibitem#1\endcsname}%
\let\auto@bib@innerbib\@empty
\bibitem [{\citenamefont {Ufimtsev}\ and\ \citenamefont
  {Martinez}(2009)}]{Ufimtsev2009}%
  \BibitemOpen
  \bibfield  {author} {\bibinfo {author} {\bibfnamefont {I.~S.}\ \bibnamefont
  {Ufimtsev}}\ and\ \bibinfo {author} {\bibfnamefont {T.~J.}\ \bibnamefont
  {Martinez}},\ }\bibfield  {title} {\enquote {\bibinfo {title} {{Quantum
  chemistry on graphical processing units. 3. Analytical energy gradients and
  first principles molecular dynamics}},}\ }\href
  {https://doi.org/http://pubs.acs.org/doi/pdf/10.1021/ct9003004} {\bibfield
  {journal} {\bibinfo  {journal} {J. Chem. Theory Comput.}\ }\textbf {\bibinfo
  {volume} {5}},\ \bibinfo {pages} {2619--2628} (\bibinfo {year}
  {2009})}\BibitemShut {NoStop}%
\bibitem [{\citenamefont {Dunning~Jr.}(1989)}]{Dunning1990}%
  \BibitemOpen
  \bibfield  {author} {\bibinfo {author} {\bibfnamefont {T.~H.}\ \bibnamefont
  {Dunning~Jr.}},\ }\bibfield  {title} {\enquote {\bibinfo {title} {{Gaussian
  basis sets for use in correlated molecular calculations. I. The atoms Boron
  through Neon and Hydrogen}},}\ }\href {https://doi.org/10.1063/1.456153}
  {\bibfield  {journal} {\bibinfo  {journal} {J. Chem. Phys.}\ }\textbf
  {\bibinfo {volume} {90}},\ \bibinfo {pages} {1007} (\bibinfo {year}
  {1989})}\BibitemShut {NoStop}%
\bibitem [{\citenamefont {Yanai}, \citenamefont {Tew},\ and\ \citenamefont
  {Handy}(2004)}]{Yanai2004}%
  \BibitemOpen
  \bibfield  {author} {\bibinfo {author} {\bibfnamefont {T.}~\bibnamefont
  {Yanai}}, \bibinfo {author} {\bibfnamefont {D.~P.}\ \bibnamefont {Tew}},\
  and\ \bibinfo {author} {\bibfnamefont {N.~C.}\ \bibnamefont {Handy}},\
  }\bibfield  {title} {\enquote {\bibinfo {title} {{A new hybrid
  exchange-correlation functional using the Coulomb-attenuating method
  (CAM-B3LYP)}},}\ }\href {https://doi.org/10.1016/j.cplett.2004.06.011}
  {\bibfield  {journal} {\bibinfo  {journal} {Chem. Phys. Lett.}\ }\textbf
  {\bibinfo {volume} {393}},\ \bibinfo {pages} {51--57} (\bibinfo {year}
  {2004})}\BibitemShut {NoStop}%
\bibitem [{\citenamefont {Case}\ \emph {et~al.}(2012)\citenamefont {Case},
  \citenamefont {Darden}, \citenamefont {Cheatham}, \citenamefont {Simmerling},
  \citenamefont {Wang}, \citenamefont {Duke}, \citenamefont {Luo},
  \citenamefont {Walker}, \citenamefont {Zhang}, \citenamefont {Merz},
  \citenamefont {Roberts}, \citenamefont {Hayik}, \citenamefont {Roitberg},
  \citenamefont {Seabra}, \citenamefont {Swails}, \citenamefont {G\"{o}tz},
  \citenamefont {Kolossvari}, \citenamefont {Wong}, \citenamefont {Paesani},
  \citenamefont {Vanicek}, \citenamefont {Wolf}, \citenamefont {Liu},
  \citenamefont {Wu}, \citenamefont {Brozell}, \citenamefont {Steinbrecher},
  \citenamefont {Gohlke}, \citenamefont {Cai}, \citenamefont {Ye},
  \citenamefont {Wang}, \citenamefont {Hsieh}, \citenamefont {Cui},
  \citenamefont {Roe}, \citenamefont {Mathews}, \citenamefont {Seetin},
  \citenamefont {Salomon-Ferrer}, \citenamefont {Sagui}, \citenamefont {Babin},
  \citenamefont {Luchko}, \citenamefont {Gusarov}, \citenamefont {Kovalenko},\
  and\ \citenamefont {Kollman}}]{amber}%
  \BibitemOpen
  \bibfield  {author} {\bibinfo {author} {\bibfnamefont {D.~A.}\ \bibnamefont
  {Case}}, \bibinfo {author} {\bibfnamefont {T.~A.}\ \bibnamefont {Darden}},
  \bibinfo {author} {\bibfnamefont {I.}~\bibnamefont {Cheatham}}, \bibinfo
  {author} {\bibfnamefont {C.~L.}\ \bibnamefont {Simmerling}}, \bibinfo
  {author} {\bibfnamefont {J.}~\bibnamefont {Wang}}, \bibinfo {author}
  {\bibfnamefont {R.~E.}\ \bibnamefont {Duke}}, \bibinfo {author}
  {\bibfnamefont {R.}~\bibnamefont {Luo}}, \bibinfo {author} {\bibfnamefont
  {R.~C.}\ \bibnamefont {Walker}}, \bibinfo {author} {\bibfnamefont
  {W.}~\bibnamefont {Zhang}}, \bibinfo {author} {\bibfnamefont {K.~M.}\
  \bibnamefont {Merz}}, \bibinfo {author} {\bibfnamefont {B.}~\bibnamefont
  {Roberts}}, \bibinfo {author} {\bibfnamefont {S.}~\bibnamefont {Hayik}},
  \bibinfo {author} {\bibfnamefont {A.}~\bibnamefont {Roitberg}}, \bibinfo
  {author} {\bibfnamefont {G.}~\bibnamefont {Seabra}}, \bibinfo {author}
  {\bibfnamefont {J.}~\bibnamefont {Swails}}, \bibinfo {author} {\bibfnamefont
  {A.~W.}\ \bibnamefont {G\"{o}tz}}, \bibinfo {author} {\bibfnamefont
  {I.}~\bibnamefont {Kolossvari}}, \bibinfo {author} {\bibfnamefont {K.~F.}\
  \bibnamefont {Wong}}, \bibinfo {author} {\bibfnamefont {F.}~\bibnamefont
  {Paesani}}, \bibinfo {author} {\bibfnamefont {J.}~\bibnamefont {Vanicek}},
  \bibinfo {author} {\bibfnamefont {R.~M.}\ \bibnamefont {Wolf}}, \bibinfo
  {author} {\bibfnamefont {J.}~\bibnamefont {Liu}}, \bibinfo {author}
  {\bibfnamefont {X.}~\bibnamefont {Wu}}, \bibinfo {author} {\bibfnamefont
  {S.~R.}\ \bibnamefont {Brozell}}, \bibinfo {author} {\bibfnamefont
  {T.}~\bibnamefont {Steinbrecher}}, \bibinfo {author} {\bibfnamefont
  {H.}~\bibnamefont {Gohlke}}, \bibinfo {author} {\bibfnamefont
  {Q.}~\bibnamefont {Cai}}, \bibinfo {author} {\bibfnamefont {X.}~\bibnamefont
  {Ye}}, \bibinfo {author} {\bibfnamefont {J.}~\bibnamefont {Wang}}, \bibinfo
  {author} {\bibfnamefont {M.-J.}\ \bibnamefont {Hsieh}}, \bibinfo {author}
  {\bibfnamefont {G.}~\bibnamefont {Cui}}, \bibinfo {author} {\bibfnamefont
  {D.~R.}\ \bibnamefont {Roe}}, \bibinfo {author} {\bibfnamefont {D.~H.}\
  \bibnamefont {Mathews}}, \bibinfo {author} {\bibfnamefont {M.~G.}\
  \bibnamefont {Seetin}}, \bibinfo {author} {\bibfnamefont {R.}~\bibnamefont
  {Salomon-Ferrer}}, \bibinfo {author} {\bibfnamefont {C.}~\bibnamefont
  {Sagui}}, \bibinfo {author} {\bibfnamefont {V.}~\bibnamefont {Babin}},
  \bibinfo {author} {\bibfnamefont {T.}~\bibnamefont {Luchko}}, \bibinfo
  {author} {\bibfnamefont {S.}~\bibnamefont {Gusarov}}, \bibinfo {author}
  {\bibfnamefont {A.}~\bibnamefont {Kovalenko}},\ and\ \bibinfo {author}
  {\bibfnamefont {P.~A.}\ \bibnamefont {Kollman}},\ }\href@noop {} {\enquote
  {\bibinfo {title} {{AMBER} 12},}\ } (\bibinfo {year} {2012}),\ \bibinfo
  {note} {{University of California: San Francisco, CA}}\BibitemShut {NoStop}%
\bibitem [{\citenamefont {Titov}\ \emph {et~al.}(2013)\citenamefont {Titov},
  \citenamefont {Ufimtsev}, \citenamefont {Luehr},\ and\ \citenamefont
  {Martinez}}]{Titov2013}%
  \BibitemOpen
  \bibfield  {author} {\bibinfo {author} {\bibfnamefont {A.~V.}\ \bibnamefont
  {Titov}}, \bibinfo {author} {\bibfnamefont {I.~S.}\ \bibnamefont {Ufimtsev}},
  \bibinfo {author} {\bibfnamefont {N.}~\bibnamefont {Luehr}},\ and\ \bibinfo
  {author} {\bibfnamefont {T.~J.}\ \bibnamefont {Martinez}},\ }\bibfield
  {title} {\enquote {\bibinfo {title} {{Generating Efficient Quantum Chemistry
  Codes for Novel Architectures}},}\ }\href {https://doi.org/10.1021/ct300321a}
  {\bibfield  {journal} {\bibinfo  {journal} {J. Chem. Theory Comput.}\
  }\textbf {\bibinfo {volume} {9}},\ \bibinfo {pages} {2213--221} (\bibinfo
  {year} {2013})}\BibitemShut {NoStop}%
\bibitem [{\citenamefont {Wang}\ \emph {et~al.}(2004)\citenamefont {Wang},
  \citenamefont {Wolf}, \citenamefont {Caldwell},\ and\ \citenamefont
  {Case}}]{GAFF}%
  \BibitemOpen
  \bibfield  {author} {\bibinfo {author} {\bibfnamefont {J.}~\bibnamefont
  {Wang}}, \bibinfo {author} {\bibfnamefont {R.~M.}\ \bibnamefont {Wolf}},
  \bibinfo {author} {\bibfnamefont {P.~A.}\ \bibnamefont {Caldwell},
  \bibfnamefont {J.~W. an~Kollman}},\ and\ \bibinfo {author} {\bibfnamefont
  {D.~A.}\ \bibnamefont {Case}},\ }\bibfield  {title} {\enquote {\bibinfo
  {title} {{Developing and testing of a general amber force field}},}\ }\href
  {https://doi.org/10.1002/jcc.20035} {\bibfield  {journal} {\bibinfo
  {journal} {J. Comput. Chem.}\ }\textbf {\bibinfo {volume} {25}},\ \bibinfo
  {pages} {1157--1174} (\bibinfo {year} {2004})}\BibitemShut {NoStop}%
\bibitem [{\citenamefont {Wang}\ \emph {et~al.}(2006)\citenamefont {Wang},
  \citenamefont {Wang}, \citenamefont {Kollman},\ and\ \citenamefont
  {Case}}]{Antechamber}%
  \BibitemOpen
  \bibfield  {author} {\bibinfo {author} {\bibfnamefont {J.}~\bibnamefont
  {Wang}}, \bibinfo {author} {\bibfnamefont {W.}~\bibnamefont {Wang}}, \bibinfo
  {author} {\bibfnamefont {P.~A.}\ \bibnamefont {Kollman}},\ and\ \bibinfo
  {author} {\bibfnamefont {D.~A.}\ \bibnamefont {Case}},\ }\bibfield  {title}
  {\enquote {\bibinfo {title} {Automatic atom type and bond type perception in
  molecular mechanical calculations},}\ }\href@noop {} {\bibfield  {journal}
  {\bibinfo  {journal} {J. Mol. Graph. Model.}\ }\textbf {\bibinfo {volume}
  {25}},\ \bibinfo {pages} {247--260} (\bibinfo {year} {2006})}\BibitemShut
  {NoStop}%
\bibitem [{\citenamefont {Jorgensen}, \citenamefont {Chandrasekhar},\ and\
  \citenamefont {Madura}(1983)}]{TIP3P}%
  \BibitemOpen
  \bibfield  {author} {\bibinfo {author} {\bibfnamefont {W.~L.}\ \bibnamefont
  {Jorgensen}}, \bibinfo {author} {\bibfnamefont {J.}~\bibnamefont
  {Chandrasekhar}},\ and\ \bibinfo {author} {\bibfnamefont {J.~D.}\
  \bibnamefont {Madura}},\ }\bibfield  {title} {\enquote {\bibinfo {title}
  {{Comparison of simple potential functions for simulating liquid water}},}\
  }\href {https://doi.org/10.1063/1.445869} {\bibfield  {journal} {\bibinfo
  {journal} {J. Chem. Phys.}\ }\textbf {\bibinfo {volume} {79}},\ \bibinfo
  {pages} {926} (\bibinfo {year} {1983})}\BibitemShut {NoStop}%
\bibitem [{\citenamefont {Medders}\ \emph {et~al.}(2017)\citenamefont
  {Medders}, \citenamefont {Alguire}, \citenamefont {Jain},\ and\ \citenamefont
  {Subotnik}}]{Medders2017}%
  \BibitemOpen
  \bibfield  {author} {\bibinfo {author} {\bibfnamefont {G.~R.}\ \bibnamefont
  {Medders}}, \bibinfo {author} {\bibfnamefont {E.~C.}\ \bibnamefont
  {Alguire}}, \bibinfo {author} {\bibfnamefont {A.}~\bibnamefont {Jain}},\ and\
  \bibinfo {author} {\bibfnamefont {J.~E.}\ \bibnamefont {Subotnik}},\
  }\bibfield  {title} {\enquote {\bibinfo {title} {Ultrafast electronic
  relaxation through a conical intersection: Nonadiabatic dynamics disentangled
  through an oscillator strength-based diabatization framework},}\ }\href@noop
  {} {\bibfield  {journal} {\bibinfo  {journal} {J. Phys. Chem. A}\ }\textbf
  {\bibinfo {volume} {121}},\ \bibinfo {pages} {1425--1434} (\bibinfo {year}
  {2017})}\BibitemShut {NoStop}%
\bibitem [{\citenamefont {Krasnoshchekov}, \citenamefont {Isayeva},\ and\
  \citenamefont {Stepanov}(2014)}]{Krasnoshchekov2014}%
  \BibitemOpen
  \bibfield  {author} {\bibinfo {author} {\bibfnamefont {S.~V.}\ \bibnamefont
  {Krasnoshchekov}}, \bibinfo {author} {\bibfnamefont {E.~V.}\ \bibnamefont
  {Isayeva}},\ and\ \bibinfo {author} {\bibfnamefont {N.~F.}\ \bibnamefont
  {Stepanov}},\ }\bibfield  {title} {\enquote {\bibinfo {title} {{Determination
  of the Eckart molecule-fixed frame by use of the apparatus of quaternion
  algebra}},}\ }\href {https://doi.org/10.1063/1.4870936} {\bibfield  {journal}
  {\bibinfo  {journal} {J. Chem. Phys.}\ }\textbf {\bibinfo {volume} {140}},\
  \bibinfo {pages} {154104} (\bibinfo {year} {2014})}\BibitemShut {NoStop}%
\bibitem [{\citenamefont {Wiethorn}\ \emph {et~al.}(2023)\citenamefont
  {Wiethorn}, \citenamefont {Hunter}, \citenamefont {Zuehlsdorff},\ and\
  \citenamefont {Montoya-Castillo}}]{Wiethorn2023}%
  \BibitemOpen
  \bibfield  {author} {\bibinfo {author} {\bibfnamefont {Z.~R.}\ \bibnamefont
  {Wiethorn}}, \bibinfo {author} {\bibfnamefont {K.~E.}\ \bibnamefont
  {Hunter}}, \bibinfo {author} {\bibfnamefont {T.~J.}\ \bibnamefont
  {Zuehlsdorff}},\ and\ \bibinfo {author} {\bibfnamefont {A.}~\bibnamefont
  {Montoya-Castillo}},\ }\bibfield  {title} {\enquote {\bibinfo {title}
  {{Beyond the Condon limit: Condensed phase optical spectra from atomistic
  simulations}},}\ }\href@noop {} {\bibfield  {journal} {\bibinfo  {journal}
  {J. Chem. Phys.}\ }\textbf {\bibinfo {volume} {159}},\ \bibinfo {pages}
  {244114} (\bibinfo {year} {2023})}\BibitemShut {NoStop}%
\bibitem [{\citenamefont {Chin}\ \emph {et~al.}(2010)\citenamefont {Chin},
  \citenamefont {Rivas}, \citenamefont {Huelga},\ and\ \citenamefont
  {Plenio}}]{Chin2010}%
  \BibitemOpen
  \bibfield  {author} {\bibinfo {author} {\bibfnamefont {A.~W.}\ \bibnamefont
  {Chin}}, \bibinfo {author} {\bibfnamefont {{\'{A}}.}~\bibnamefont {Rivas}},
  \bibinfo {author} {\bibfnamefont {S.~F.}\ \bibnamefont {Huelga}},\ and\
  \bibinfo {author} {\bibfnamefont {M.~B.}\ \bibnamefont {Plenio}},\ }\bibfield
   {title} {\enquote {\bibinfo {title} {{Exact mapping between system-reservoir
  quantum models and semi-infinite discrete chains using orthogonal
  polynomials}},}\ }\href@noop {} {\bibfield  {journal} {\bibinfo  {journal}
  {J. Math. Phys.}\ }\textbf {\bibinfo {volume} {51}} (\bibinfo {year}
  {2010})},\ \Eprint {https://arxiv.org/abs/1006.4507} {arXiv:1006.4507}
  \BibitemShut {NoStop}%
\bibitem [{\citenamefont {Dunnett}\ \emph {et~al.}(2021)\citenamefont
  {Dunnett}, \citenamefont {Gowland}, \citenamefont {Isborn}, \citenamefont
  {Chin},\ and\ \citenamefont {Zuehlsdorff}}]{Dunnett2021}%
  \BibitemOpen
  \bibfield  {author} {\bibinfo {author} {\bibfnamefont {A.~J.}\ \bibnamefont
  {Dunnett}}, \bibinfo {author} {\bibfnamefont {D.}~\bibnamefont {Gowland}},
  \bibinfo {author} {\bibfnamefont {C.~M.}\ \bibnamefont {Isborn}}, \bibinfo
  {author} {\bibfnamefont {A.~W.}\ \bibnamefont {Chin}},\ and\ \bibinfo
  {author} {\bibfnamefont {T.~J.}\ \bibnamefont {Zuehlsdorff}},\ }\bibfield
  {title} {\enquote {\bibinfo {title} {Influence of non-adiabatic effects on
  linear absorption spectra in the condensed phase: Methylene blue},}\ }\href
  {https://doi.org/10.1063/5.0062950} {\bibfield  {journal} {\bibinfo
  {journal} {J. Chem. Phys.}\ }\textbf {\bibinfo {volume} {155}},\ \bibinfo
  {pages} {144112} (\bibinfo {year} {2021})}\BibitemShut {NoStop}%
\bibitem [{\citenamefont {Hunter}\ \emph {et~al.}(2024)\citenamefont {Hunter},
  \citenamefont {Mao}, \citenamefont {Chin},\ and\ \citenamefont
  {Zuehlsdorff}}]{Hunter2024}%
  \BibitemOpen
  \bibfield  {author} {\bibinfo {author} {\bibfnamefont {K.~E.}\ \bibnamefont
  {Hunter}}, \bibinfo {author} {\bibfnamefont {Y.}~\bibnamefont {Mao}},
  \bibinfo {author} {\bibfnamefont {A.~W.}\ \bibnamefont {Chin}},\ and\
  \bibinfo {author} {\bibfnamefont {T.~J.}\ \bibnamefont {Zuehlsdorff}},\
  }\bibfield  {title} {\enquote {\bibinfo {title} {Environmentally driven
  symmetry breaking quenches dual fluorescence in proflavine},}\ }\href@noop {}
  {\bibfield  {journal} {\bibinfo  {journal} {J. Phys. Chem. Lett.}\ }\textbf
  {\bibinfo {volume} {15}},\ \bibinfo {pages} {4623--4632} (\bibinfo {year}
  {2024})}\BibitemShut {NoStop}%
\bibitem [{\citenamefont {Gautschi}(1994)}]{gautschi_algorithm_1994}%
  \BibitemOpen
  \bibfield  {author} {\bibinfo {author} {\bibfnamefont {W.}~\bibnamefont
  {Gautschi}},\ }\bibfield  {title} {\enquote {\bibinfo {title} {Algorithm 726:
  {ORTHPOL}–a package of routines for generating orthogonal polynomials and
  gauss-type quadrature rules},}\ }\href
  {https://doi.org/10.1145/174603.174605} {\bibfield  {journal} {\bibinfo
  {journal} {ACM Trans. Math. Softw.}\ }\textbf {\bibinfo {volume} {20}},\
  \bibinfo {pages} {21--62} (\bibinfo {year} {1994})}\BibitemShut {NoStop}%
\bibitem [{\citenamefont {Raab}\ \emph {et~al.}(1999)\citenamefont {Raab},
  \citenamefont {Worth}, \citenamefont {Meyer},\ and\ \citenamefont
  {Cederbaum}}]{Raab1999}%
  \BibitemOpen
  \bibfield  {author} {\bibinfo {author} {\bibfnamefont {A.}~\bibnamefont
  {Raab}}, \bibinfo {author} {\bibfnamefont {G.~A.}\ \bibnamefont {Worth}},
  \bibinfo {author} {\bibfnamefont {H.-D.}\ \bibnamefont {Meyer}},\ and\
  \bibinfo {author} {\bibfnamefont {L.~S.}\ \bibnamefont {Cederbaum}},\
  }\bibfield  {title} {\enquote {\bibinfo {title} {{Molecular dynamics of
  pyrazine after excitation to the S2 electronic state using a realistic
  24-mode model Hamiltonian}},}\ }\href@noop {} {\bibfield  {journal} {\bibinfo
   {journal} {J. Chem. Phys.}\ }\textbf {\bibinfo {volume} {110}},\ \bibinfo
  {pages} {936--946} (\bibinfo {year} {1999})}\BibitemShut {NoStop}%
\bibitem [{\citenamefont {Frisch}\ \emph {et~al.}(2016)\citenamefont {Frisch},
  \citenamefont {Trucks}, \citenamefont {Schlegel}, \citenamefont {Scuseria},
  \citenamefont {Robb}, \citenamefont {Cheeseman}, \citenamefont {Scalmani},
  \citenamefont {Barone}, \citenamefont {Petersson}, \citenamefont {Nakatsuji},
  \citenamefont {Li}, \citenamefont {Caricato}, \citenamefont {Marenich},
  \citenamefont {Bloino}, \citenamefont {Janesko}, \citenamefont {Gomperts},
  \citenamefont {Mennucci}, \citenamefont {Hratchian}, \citenamefont {Ortiz},
  \citenamefont {Izmaylov}, \citenamefont {Sonnenberg}, \citenamefont
  {Williams-Young}, \citenamefont {Ding}, \citenamefont {Lipparini},
  \citenamefont {Egidi}, \citenamefont {Goings}, \citenamefont {Peng},
  \citenamefont {Petrone}, \citenamefont {Henderson}, \citenamefont
  {Ranasinghe}, \citenamefont {Zakrzewski}, \citenamefont {Gao}, \citenamefont
  {Rega}, \citenamefont {Zheng}, \citenamefont {Liang}, \citenamefont {Hada},
  \citenamefont {Ehara}, \citenamefont {Toyota}, \citenamefont {Fukuda},
  \citenamefont {Hasegawa}, \citenamefont {Ishida}, \citenamefont {Nakajima},
  \citenamefont {Honda}, \citenamefont {Kitao}, \citenamefont {Nakai},
  \citenamefont {Vreven}, \citenamefont {Throssell}, \citenamefont
  {Montgomery}, \citenamefont {Peralta}, \citenamefont {Ogliaro}, \citenamefont
  {Bearpark}, \citenamefont {Heyd}, \citenamefont {Brothers}, \citenamefont
  {Kudin}, \citenamefont {Staroverov}, \citenamefont {Keith}, \citenamefont
  {Kobayashi}, \citenamefont {Normand}, \citenamefont {Raghavachari},
  \citenamefont {Rendell}, \citenamefont {Burant}, \citenamefont {Iyengar},
  \citenamefont {Tomasi}, \citenamefont {Cossi}, \citenamefont {Millam},
  \citenamefont {Klene}, \citenamefont {Adamo}, \citenamefont {Cammi},
  \citenamefont {Ochterski}, \citenamefont {Martin}, \citenamefont {Morokuma},
  \citenamefont {Farkas}, \citenamefont {Foresman},\ and\ \citenamefont
  {Fox}}]{gdv}%
  \BibitemOpen
  \bibfield  {author} {\bibinfo {author} {\bibfnamefont {M.~J.}\ \bibnamefont
  {Frisch}}, \bibinfo {author} {\bibfnamefont {G.~W.}\ \bibnamefont {Trucks}},
  \bibinfo {author} {\bibfnamefont {H.~B.}\ \bibnamefont {Schlegel}}, \bibinfo
  {author} {\bibfnamefont {G.~E.}\ \bibnamefont {Scuseria}}, \bibinfo {author}
  {\bibfnamefont {M.~A.}\ \bibnamefont {Robb}}, \bibinfo {author}
  {\bibfnamefont {J.~R.}\ \bibnamefont {Cheeseman}}, \bibinfo {author}
  {\bibfnamefont {G.}~\bibnamefont {Scalmani}}, \bibinfo {author}
  {\bibfnamefont {V.}~\bibnamefont {Barone}}, \bibinfo {author} {\bibfnamefont
  {G.~A.}\ \bibnamefont {Petersson}}, \bibinfo {author} {\bibfnamefont
  {H.}~\bibnamefont {Nakatsuji}}, \bibinfo {author} {\bibfnamefont
  {X.}~\bibnamefont {Li}}, \bibinfo {author} {\bibfnamefont {M.}~\bibnamefont
  {Caricato}}, \bibinfo {author} {\bibfnamefont {A.~V.}\ \bibnamefont
  {Marenich}}, \bibinfo {author} {\bibfnamefont {J.}~\bibnamefont {Bloino}},
  \bibinfo {author} {\bibfnamefont {B.~G.}\ \bibnamefont {Janesko}}, \bibinfo
  {author} {\bibfnamefont {R.}~\bibnamefont {Gomperts}}, \bibinfo {author}
  {\bibfnamefont {B.}~\bibnamefont {Mennucci}}, \bibinfo {author}
  {\bibfnamefont {H.~P.}\ \bibnamefont {Hratchian}}, \bibinfo {author}
  {\bibfnamefont {J.~V.}\ \bibnamefont {Ortiz}}, \bibinfo {author}
  {\bibfnamefont {A.~F.}\ \bibnamefont {Izmaylov}}, \bibinfo {author}
  {\bibfnamefont {J.~L.}\ \bibnamefont {Sonnenberg}}, \bibinfo {author}
  {\bibfnamefont {D.}~\bibnamefont {Williams-Young}}, \bibinfo {author}
  {\bibfnamefont {F.}~\bibnamefont {Ding}}, \bibinfo {author} {\bibfnamefont
  {F.}~\bibnamefont {Lipparini}}, \bibinfo {author} {\bibfnamefont
  {F.}~\bibnamefont {Egidi}}, \bibinfo {author} {\bibfnamefont
  {J.}~\bibnamefont {Goings}}, \bibinfo {author} {\bibfnamefont
  {B.}~\bibnamefont {Peng}}, \bibinfo {author} {\bibfnamefont {A.}~\bibnamefont
  {Petrone}}, \bibinfo {author} {\bibfnamefont {T.}~\bibnamefont {Henderson}},
  \bibinfo {author} {\bibfnamefont {D.}~\bibnamefont {Ranasinghe}}, \bibinfo
  {author} {\bibfnamefont {V.~G.}\ \bibnamefont {Zakrzewski}}, \bibinfo
  {author} {\bibfnamefont {J.}~\bibnamefont {Gao}}, \bibinfo {author}
  {\bibfnamefont {N.}~\bibnamefont {Rega}}, \bibinfo {author} {\bibfnamefont
  {G.}~\bibnamefont {Zheng}}, \bibinfo {author} {\bibfnamefont
  {W.}~\bibnamefont {Liang}}, \bibinfo {author} {\bibfnamefont
  {M.}~\bibnamefont {Hada}}, \bibinfo {author} {\bibfnamefont {M.}~\bibnamefont
  {Ehara}}, \bibinfo {author} {\bibfnamefont {K.}~\bibnamefont {Toyota}},
  \bibinfo {author} {\bibfnamefont {R.}~\bibnamefont {Fukuda}}, \bibinfo
  {author} {\bibfnamefont {J.}~\bibnamefont {Hasegawa}}, \bibinfo {author}
  {\bibfnamefont {M.}~\bibnamefont {Ishida}}, \bibinfo {author} {\bibfnamefont
  {T.}~\bibnamefont {Nakajima}}, \bibinfo {author} {\bibfnamefont
  {Y.}~\bibnamefont {Honda}}, \bibinfo {author} {\bibfnamefont
  {O.}~\bibnamefont {Kitao}}, \bibinfo {author} {\bibfnamefont
  {H.}~\bibnamefont {Nakai}}, \bibinfo {author} {\bibfnamefont
  {T.}~\bibnamefont {Vreven}}, \bibinfo {author} {\bibfnamefont
  {K.}~\bibnamefont {Throssell}}, \bibinfo {author} {\bibfnamefont {J.~A.}\
  \bibnamefont {Montgomery}, \bibfnamefont {{Jr.}}}, \bibinfo {author}
  {\bibfnamefont {J.~E.}\ \bibnamefont {Peralta}}, \bibinfo {author}
  {\bibfnamefont {F.}~\bibnamefont {Ogliaro}}, \bibinfo {author} {\bibfnamefont
  {M.~J.}\ \bibnamefont {Bearpark}}, \bibinfo {author} {\bibfnamefont {J.~J.}\
  \bibnamefont {Heyd}}, \bibinfo {author} {\bibfnamefont {E.~N.}\ \bibnamefont
  {Brothers}}, \bibinfo {author} {\bibfnamefont {K.~N.}\ \bibnamefont {Kudin}},
  \bibinfo {author} {\bibfnamefont {V.~N.}\ \bibnamefont {Staroverov}},
  \bibinfo {author} {\bibfnamefont {T.~A.}\ \bibnamefont {Keith}}, \bibinfo
  {author} {\bibfnamefont {R.}~\bibnamefont {Kobayashi}}, \bibinfo {author}
  {\bibfnamefont {J.}~\bibnamefont {Normand}}, \bibinfo {author} {\bibfnamefont
  {K.}~\bibnamefont {Raghavachari}}, \bibinfo {author} {\bibfnamefont {A.~P.}\
  \bibnamefont {Rendell}}, \bibinfo {author} {\bibfnamefont {J.~C.}\
  \bibnamefont {Burant}}, \bibinfo {author} {\bibfnamefont {S.~S.}\
  \bibnamefont {Iyengar}}, \bibinfo {author} {\bibfnamefont {J.}~\bibnamefont
  {Tomasi}}, \bibinfo {author} {\bibfnamefont {M.}~\bibnamefont {Cossi}},
  \bibinfo {author} {\bibfnamefont {J.~M.}\ \bibnamefont {Millam}}, \bibinfo
  {author} {\bibfnamefont {M.}~\bibnamefont {Klene}}, \bibinfo {author}
  {\bibfnamefont {C.}~\bibnamefont {Adamo}}, \bibinfo {author} {\bibfnamefont
  {R.}~\bibnamefont {Cammi}}, \bibinfo {author} {\bibfnamefont {J.~W.}\
  \bibnamefont {Ochterski}}, \bibinfo {author} {\bibfnamefont {R.~L.}\
  \bibnamefont {Martin}}, \bibinfo {author} {\bibfnamefont {K.}~\bibnamefont
  {Morokuma}}, \bibinfo {author} {\bibfnamefont {O.}~\bibnamefont {Farkas}},
  \bibinfo {author} {\bibfnamefont {J.~B.}\ \bibnamefont {Foresman}},\ and\
  \bibinfo {author} {\bibfnamefont {D.~J.}\ \bibnamefont {Fox}},\ }\href@noop
  {} {\enquote {\bibinfo {title} {Gaussian {D}evelopment {V}ersion, {R}evision
  {I}.09},}\ } (\bibinfo {year} {2016}),\ \bibinfo {note} {{G}aussian Inc.
  Wallingford CT}\BibitemShut {NoStop}%
\bibitem [{\citenamefont {Samir}\ \emph {et~al.}(2020)\citenamefont {Samir},
  \citenamefont {Kalalian}, \citenamefont {Roth}, \citenamefont {Salghi},\ and\
  \citenamefont {Chakir}}]{Samir2020_pyrazine_vac}%
  \BibitemOpen
  \bibfield  {author} {\bibinfo {author} {\bibfnamefont {B.}~\bibnamefont
  {Samir}}, \bibinfo {author} {\bibfnamefont {C.}~\bibnamefont {Kalalian}},
  \bibinfo {author} {\bibfnamefont {E.}~\bibnamefont {Roth}}, \bibinfo {author}
  {\bibfnamefont {R.}~\bibnamefont {Salghi}},\ and\ \bibinfo {author}
  {\bibfnamefont {A.}~\bibnamefont {Chakir}},\ }\bibfield  {title} {\enquote
  {\bibinfo {title} {Gas-phase {UV} absorption spectra of pyrazine, pyrimidine
  and pyridazine},}\ }\href@noop {} {\bibfield  {journal} {\bibinfo  {journal}
  {Chem. Phys. Lett.}\ }\textbf {\bibinfo {volume} {751}},\ \bibinfo {pages}
  {137469} (\bibinfo {year} {2020})}\BibitemShut {NoStop}%
\bibitem [{\citenamefont {Sobolewski}, \citenamefont {Woywod},\ and\
  \citenamefont {Domcke}(1993)}]{Sobolewski1993}%
  \BibitemOpen
  \bibfield  {author} {\bibinfo {author} {\bibfnamefont {A.~L.}\ \bibnamefont
  {Sobolewski}}, \bibinfo {author} {\bibfnamefont {C.}~\bibnamefont {Woywod}},\
  and\ \bibinfo {author} {\bibfnamefont {W.}~\bibnamefont {Domcke}},\
  }\bibfield  {title} {\enquote {\bibinfo {title} {{Ab initio investigation of
  potential‐energy surfaces involved in the photophysics of benzene and
  pyrazine}},}\ }\href {https://doi.org/10.1063/1.464907} {\bibfield  {journal}
  {\bibinfo  {journal} {J. Chem. Phys.}\ }\textbf {\bibinfo {volume} {98}},\
  \bibinfo {pages} {5627--5641} (\bibinfo {year} {1993})}\BibitemShut {NoStop}%
\bibitem [{\citenamefont {Krempl}\ \emph {et~al.}(1994)\citenamefont {Krempl},
  \citenamefont {Winterstetter}, \citenamefont {Plöhn},\ and\ \citenamefont
  {Domcke}}]{Krempl1994}%
  \BibitemOpen
  \bibfield  {author} {\bibinfo {author} {\bibfnamefont {S.}~\bibnamefont
  {Krempl}}, \bibinfo {author} {\bibfnamefont {M.}~\bibnamefont
  {Winterstetter}}, \bibinfo {author} {\bibfnamefont {H.}~\bibnamefont
  {Plöhn}},\ and\ \bibinfo {author} {\bibfnamefont {W.}~\bibnamefont
  {Domcke}},\ }\bibfield  {title} {\enquote {\bibinfo {title} {{Path‐integral
  treatment of multi‐mode vibronic coupling}},}\ }\href
  {https://doi.org/10.1063/1.467253} {\bibfield  {journal} {\bibinfo  {journal}
  {J. Chem. Phys.}\ }\textbf {\bibinfo {volume} {100}},\ \bibinfo {pages}
  {926--937} (\bibinfo {year} {1994})}\BibitemShut {NoStop}%
\bibitem [{\citenamefont {Schneider}\ and\ \citenamefont
  {Domcke}(1988)}]{SCHNEIDER1988}%
  \BibitemOpen
  \bibfield  {author} {\bibinfo {author} {\bibfnamefont {R.}~\bibnamefont
  {Schneider}}\ and\ \bibinfo {author} {\bibfnamefont {W.}~\bibnamefont
  {Domcke}},\ }\bibfield  {title} {\enquote {\bibinfo {title} {{S1-S2 Conical
  intersection and ultrafast S2→S1 Internal conversion in pyrazine}},}\
  }\href {https://doi.org/https://doi.org/10.1016/0009-2614(88)80034-4}
  {\bibfield  {journal} {\bibinfo  {journal} {Chem. Phys. Lett.}\ }\textbf
  {\bibinfo {volume} {150}},\ \bibinfo {pages} {235--242} (\bibinfo {year}
  {1988})}\BibitemShut {NoStop}%
\bibitem [{\citenamefont {Schneider}, \citenamefont {Domcke},\ and\
  \citenamefont {Köppel}(1990)}]{Schneider1990}%
  \BibitemOpen
  \bibfield  {author} {\bibinfo {author} {\bibfnamefont {R.}~\bibnamefont
  {Schneider}}, \bibinfo {author} {\bibfnamefont {W.}~\bibnamefont {Domcke}},\
  and\ \bibinfo {author} {\bibfnamefont {H.}~\bibnamefont {Köppel}},\
  }\bibfield  {title} {\enquote {\bibinfo {title} {{Aspects of dissipative
  electronic and vibrational dynamics of strongly vibronically coupled
  systems}},}\ }\href {https://doi.org/10.1063/1.458167} {\bibfield  {journal}
  {\bibinfo  {journal} {J. Chem. Phys.}\ }\textbf {\bibinfo {volume} {92}},\
  \bibinfo {pages} {1045--1061} (\bibinfo {year} {1990})}\BibitemShut {NoStop}%
\bibitem [{\citenamefont {Stock}\ and\ \citenamefont
  {Domcke}(1990{\natexlab{a}})}]{Stock1990}%
  \BibitemOpen
  \bibfield  {author} {\bibinfo {author} {\bibfnamefont {G.}~\bibnamefont
  {Stock}}\ and\ \bibinfo {author} {\bibfnamefont {W.}~\bibnamefont {Domcke}},\
  }\bibfield  {title} {\enquote {\bibinfo {title} {Theory of femtosecond
  pump--probe spectroscopy of ultrafast internal conversion processes in
  polyatomic molecules},}\ }\href@noop {} {\bibfield  {journal} {\bibinfo
  {journal} {J. Opt. Soc. Am. B}\ }\textbf {\bibinfo {volume} {7}},\ \bibinfo
  {pages} {1970--1980} (\bibinfo {year} {1990}{\natexlab{a}})}\BibitemShut
  {NoStop}%
\bibitem [{\citenamefont {Stock}\ and\ \citenamefont
  {Domcke}(1990{\natexlab{b}})}]{Stock1990_2}%
  \BibitemOpen
  \bibfield  {author} {\bibinfo {author} {\bibfnamefont {G.}~\bibnamefont
  {Stock}}\ and\ \bibinfo {author} {\bibfnamefont {W.}~\bibnamefont {Domcke}},\
  }\bibfield  {title} {\enquote {\bibinfo {title} {{Theory of resonance Raman
  scattering and fluorescence from strongly vibronically coupled excited states
  of polyatomic molecules}},}\ }\href {https://doi.org/10.1063/1.459619}
  {\bibfield  {journal} {\bibinfo  {journal} {J. Chem. Phys.}\ }\textbf
  {\bibinfo {volume} {93}},\ \bibinfo {pages} {5496--5509} (\bibinfo {year}
  {1990}{\natexlab{b}})}\BibitemShut {NoStop}%
\bibitem [{\citenamefont {Worth}, \citenamefont {Meyer},\ and\ \citenamefont
  {Cederbaum}(1996)}]{Worth1996}%
  \BibitemOpen
  \bibfield  {author} {\bibinfo {author} {\bibfnamefont {G.~A.}\ \bibnamefont
  {Worth}}, \bibinfo {author} {\bibfnamefont {H.~D.}\ \bibnamefont {Meyer}},\
  and\ \bibinfo {author} {\bibfnamefont {L.~S.}\ \bibnamefont {Cederbaum}},\
  }\bibfield  {title} {\enquote {\bibinfo {title} {{The effect of a model
  environment on the S2 absorption spectrum of pyrazine: A wave packet study
  treating all 24 vibrational modes}},}\ }\href
  {https://doi.org/10.1063/1.472327} {\bibfield  {journal} {\bibinfo  {journal}
  {J. Chem. Phys.}\ }\textbf {\bibinfo {volume} {105}},\ \bibinfo {pages}
  {4412--4426} (\bibinfo {year} {1996})}\BibitemShut {NoStop}%
\end{thebibliography}%

\end{document}


\title[]{Supplementary material for ``Computing linear optical spectra in the presence of nonadiabatic effects on Graphics Processing Units using molecular dynamics and tensor-network approaches''}
\author{Evan Lambertson}
 \affiliation{Department of Chemistry, Oregon State University, Corvallis, Oregon 97331, USA}

\author{Dayana Bashirova}
 \affiliation{Department of Chemistry, Oregon State University, Corvallis, Oregon 97331, USA}

\author{Kye E. Hunter}
 \affiliation{Department of Chemistry, Oregon State University, Corvallis, Oregon 97331, USA}
  \affiliation{Present address: CRG (Barcelona Collaboratorium for Modelling and Predictive Biology),  Dr. Aiguader 88, Barcelona 08003, Spain}

 \author{Benhardt Hansen}
 \affiliation{Department of Chemistry, Oregon State University, Corvallis, Oregon 97331, USA}

\author{Tim J. Zuehlsdorff}
\email{tim.zuehlsdorff@oregonstate.edu}
 \affiliation{Department of Chemistry, Oregon State University, Corvallis, Oregon 97331, USA}

\date{\today}

\maketitle

\section{Computational Details of the MD}
\subsection{Pyrazine in vacuum}
For pyrazine in vacuum, an \emph{ab-initio} MD trajectory of 22~ps in length was generated using the \texttt{TeraChem} package\cite{Ufimtsev2009}. The molecule was initialized in its ground state structure, and all simulations were performed at the 6-31+G*/CAM-B3LYP\cite{Dunning1990, Yanai2004} level of theory. The temperature was kept at 300~K using a Langevin thermostat with a collision frequency of 1~ps$^{-1}$ and a time-step of 0.5~fs was used. The first 2~ps of the trajectory were discarded to allow for equilibration, resulting in 20~ps of usable trajectory. Sampling the trajectory at a rate of 2~ps then yielded 10,000 snapshots to construct correlation functions and spectral densities. 
\subsection{Pyrazine in cyclohexane}
\label{cyclohexane}
A cubic box with 353 cyclohexane molecules and 25~\AA\ side length was generated using \texttt{LEaP}\cite{amber}. The energy of the box was minimized, followed by heating from 0 to 300~K using Langevin dynamics for 300~ps, followed by equilibration at constant temperature and pressure (NPT ensemble) for 400~ps. Using the equilibrated solvent box, a mixed QM/MM trajectory of 22~ps length was generated using the \texttt{TeraChem}\cite{Ufimtsev2009, Titov2013} (QM) and \texttt{AMBER}\cite{amber} (MM) interface for pyrazine in cyclohexane. 

Force-field parameters for both cyclohexane and pyrazine were generated using the general AMBER force field (GAFF)\cite{GAFF} and \texttt{Antechamber}\cite{Antechamber}. The ground state structure of pyrazine was solvated in a 28~\AA\ cyclohexane sphere, containing 332 cyclohexane molecules, constructed from the pre-equilibrated solvent box. The system was heated to 300~K using a Langevin thermostat with a collision frequency of 1~ps$^{-1}$ and a time-step of 0.5~fs, followed by a 50~ps MM equilibration under open boundary conditions. Subsequently, a 22~ps QM/MM simulation was performed with the temperature kept at 300~K using a Langevin thermostat with a collision frequency of 1~ps$^{-1}$ and a time-step of 0.5~fs. Pyrazine was treated at the QM level with 6-31+G*/CAM-B3LYP\cite{Dunning1990, Yanai2004} level of theory, the solvent environment was treated through the \texttt{AMBER} force field. The first 2~ps of the QM/MM trajectory were discarded to allow for additional equilibration after switching pyrazine from the MM to the QM Hamiltonian, resulting in 20~ps of usable trajectory. The trajectory was sampled every 2~fs, yielding 10.000 snapshots to construct correlation functions and spectral densities. For pyrazine in cyclohexane, it was found that the last 2,000 snapshots of the 20~ps trajectory showed anomalous energy gap fluctuations compared to the rest of the trajectory, the vacuum and the solvent results. As a consequence, only the first 8,000 snapshots were processed for computing tensor-network dynamics. 

\subsection{Pyrazine in water}
 A 22~ps QM/MM trajectory was generated for pyrazine in water. The ground state structure of the molecule was solvated in a pre-equilibrated solvent sphere of 28~\AA, containing 2855 TIP3P\cite{TIP3P} water molecules. Force-field parameters for pyrazine were again generated using the General AMBER Force Field (GAFF)\cite{GAFF} and \texttt{Antechamber}\cite{Antechamber}. Then, for pyrazine in water, the same MD simulation procedure was repeated as described for the system in cyclohexane (See SI Sec.~\ref{cyclohexane}).

\section{Transition dipole extraction and diabatization procedure in Pyrazine}

When computing excited state energies for pyrazine within TDDFT at the 6-31+G*/CAM-B3LYP level of theory, the two low lying excited states of interest do not correspond to the two lowest excitations in the system. Specifically, the bright $\pi\pi^*$ state referred to as S$_2$ in the main text actually corresponds to S$_3$ as computed at the ground-state optimized geometry, and along most of the MD trajectory. Additionally, crossings of the $\pi\pi^*$ state with a higher-lying state (S$_4$ at the ground state optimized geometry) occur periodically. For this reason, some care has to be taken to select the correct two \emph{adiabatic} excited states with $n\pi^*$ and $\pi\pi^*$ character for both the Gaussian non-Condon theory (GNCT) calculations and the diabatization procedure. 

For the purpose of this work, the computation of adiabatic dipole fluctuations for GNCT calculations and diabatizations are carried out using the following procedure (a \texttt{Python} script performing these steps is provided on GitHub and in a Zenodo repository, see data availability statement):
\begin{enumerate}
    \item 
    The ground state optimized geometry $\textbf{R}^\textrm{GS}_0$ for Pyrazine in vacuum is computed using 6-31+G*/CAM-B3LYP. Excited states at the reference geometry are computed using TDDFT and the states with $n\pi^*$ and $\pi\pi^*$ character are identified to obtain the reference geometry and transition dipole moments $\left\{ \textbf{R}^\textrm{GS}_0, \boldsymbol{\mu}_{n\pi^*}\left(\textbf{R}^\textrm{GS}_0\right), \boldsymbol{\mu}_{\pi\pi^*}\left(\textbf{R}^\textrm{GS}_0\right)\right\}$.
    \item 
    For a given MD snapshot $\textbf{R}_N$, excitation energies of the four lowest states are computed using TDDFT. The molecule and the transition dipole moments of all excited states are rotated into the Eckart frame defined by $\textbf{R}^\textrm{GS}_0$ (see description below, and Sec.~IIA.1 of the main text). 
    \item 
    \emph{Adiabatic} states $n\pi^*$ and $\pi\pi^*$ are defined by selecting the two states from the excited state manifold of $\textbf{R}_N$ whose Eckart-rotated transition dipoles have the largest overlaps with $\boldsymbol{\mu}_{n\pi^*}\left(\textbf{R}^\textrm{GS}_0\right)$ and $\boldsymbol{\mu}_{\pi\pi^*}\left(\textbf{R}^\textrm{GS}_0\right)$ respectively. Additionally, since the signs of the transition dipole moments are arbitrary, signs of Eckart rotated states are flipped if $\boldsymbol{\mu}_{n\pi^*}\left(\textbf{R}_N\right)\cdot \boldsymbol{\mu}_{n\pi^*}\left(\textbf{R}^\textrm{GS}_0\right)<0$ or $\boldsymbol{\mu}_{\pi\pi^*}\left(\textbf{R}_N\right)\cdot \boldsymbol{\mu}_{\pi\pi^*}\left(\textbf{R}^\textrm{GS}_0\right)<0$. The resulting \emph{adiabatic} states can be directly used in the GNCT formalism. 
    \item 
    To define \emph{diabatic} states, the resulting GNCT \emph{adiabatic} states, after an Eckart rotation and alignment with reference dipoles, are taken as input for the dipole-based diabatization approach\cite{Medders2017} (outlined in Sec.~IIB.2 in the main text). The approach yields $\left\{ E_{n\pi^*}\left(\textbf{R}_{N}\right), E_{\pi\pi^*}\left(\textbf{R}_{N}\right), \delta_{12}\left(\textbf{R}_N\right)\right\}$, the \emph{diabatic} energies for the $n\pi^*$ and $\pi\pi^*$ transition, as well as their coupling. 
    \item 
    Repeating steps 2. and 3. for all $N_\textrm{frames}$ MD-snapshots yields adiabatic states with correctly aligned dipoles for the GNCT approach, and diabatic energies and couplings with correctly chosen signs for tensor-network based calculations. 
\end{enumerate}

While the above procedure is specific for the chosen molecule, we expect it to work for a large variety of molecules where the mixing between adiabatic states along the trajectory remains sufficiently small such that states can be identified by the overlap of their respective transition dipoles with a set of reference dipole moments computed for a single reference geometry. The \texttt{Python} script performing both the rotation and selection of states can be easily adapted to other systems. 

Transforming the transition dipole moments along the trajectory into the Eckart frame reduces to applying a 3D rotation matrix $\textbf{U}$ to the transition dipole moment computed for a given frame $\textbf{R}_N$. In defining the Eckart rotation matrix, we follow the algorithm outlined by Krasnoshcheckov and coworkers\cite{Krasnoshchekov2014}. First, both the reference geometry $\textbf{R}^\textrm{GS}_0$ and the geometry of a given MD snapshot $\textbf{R}_N$ are transformed into the center-of-mass frame with respect to the chromophore coordinates only (ignoring any potential solvent molecules or other condensed phase environment). The 3D rotation matrix ${\textrm{\textbf{U}}}$ rotating the transition dipole moments into the reference frame is then given by 
\begin{equation}
    \textrm{\textbf{U}}=
    \begin{pmatrix}
    \left(q_0^2+q^2_1-q^2_2-q^2_3\right) & 2\left(q_1q_2+q_0q_3\right)& 2\left(q_1q_3-q_0q_2\right)\\
    2\left(q_1q_2-q_0q_3\right)& \left(q_0^2-q^2_1+q^2_2-q^2_3\right)& 2\left(q_2q_3+q_0q_1\right)\\
    2\left(q_1q_3+q_0q_2\right)&2\left(q_2q_3-q_0q_1\right) &\left(q_0^2-q^2_1-q^2_2+q^2_3\right)
    \end{pmatrix}.
\end{equation}
Here, $\textbf{\textrm{q}}=\left[q_0,q_1,q_2,q_3\right]$ is the eigenvector with the lowest eigenvalue of the 4x4 symmetric matrix $\textbf{\textrm{C}}$, whose elements are
\begin{eqnarray} \nonumber
C_{11}=\sum^N_a m_a\left(x_{-a}^2+y_{-a}^2+z_{-a}^2 \right);\,C_{12}=\sum^N_a m_a\left(y_{+a}z_{-a}-y_{-a}z_{+a} \right) \\ \nonumber
C_{13}=\sum^N_a m_a\left(x_{-a}z_{+a}-x_{+a}z_{-a} \right);\,C_{14}=\sum^N_a m_a\left(x_{+a}y_{-a}-x_{-a}y_{+a} \right) \\ 
C_{22}=\sum^N_a m_a\left(x_{-a}^2+y_{+a}^2+z_{+a}^2 \right);\, C_{23}=\sum^N_a m_a\left(x_{-a}y_{-a}-x_{+a}y_{+a} \right) \\ \nonumber
C_{24}=\sum^N_a m_a\left(x_{-a}z_{-a}-x_{+a}z_{+a} \right);\,C_{33}=\sum^N_a m_a\left(x_{+a}^2+y_{-a}^2+z_{+a}^2 \right) \\ \nonumber
C_{34}=\sum^N_a m_a\left(y_{-a}z_{-a}-y_{+a}z_{+a}\right); \,C_{44}=\sum^N_a m_a\left(x_{+a}^2+y_{+a}^2+z_{-a}^2 \right)
\end{eqnarray}
$N$ denotes the number of atoms and $m_a$ is the mass of atom $a$. We have introduced the notation of $x_{+a}=x_{a,\textrm{ref}}+x_{a}$ and $x_{-a}=x_{a,\textrm{ref}}-x_{a}$, where $x_{a,\textrm{ref}}$ denotes the $x$-coordinate of the $a$-th atom of the reference geometry $\textbf{R}^\textrm{GS}_0$ in the center-of-mass frame, and $x_{a}$ denotes the $x$-coordinate of the $a$-th atom of the given MD geometry $\textbf{R}_N$ that is to be transformed into the Eckart frame, again in the center-of-mass frame. Constructing the rotation matrix for a given MD snapshot thus reduces to calculating the matrix elements of $\textbf{\textrm{C}}$ and computing its lowest eigenvalue and corresponding eigenvector. 

\section{Full mathematical expression for the GNCT}
\label{si_sec:GNCT}
As stated in the main text, the response function for the absorption lineshape of a molecule in the GNCT formalism\cite{Wiethorn2023} can be expressed as
\begin{equation}
\chi^{01}_\textrm{GNCT}(t)=A_{\delta \mu \delta U}\left[\mathcal{J}^{01}_{\delta \mu},\boldsymbol{\mathcal{J}}^{01}_{\delta \mu\delta U}\right](t)e^{-\textrm{i}\omega_{01}^\textrm{av}t-g_2\left[\mathcal{J}_{01}\right](t)}.
\label{eqn:GNCT_expression}
\end{equation}
where $\boldsymbol{\mathcal{J}}^{01}_{\delta \mu\delta U}(\omega)$ and $\mathcal{J}^{01}_{\delta \mu\delta \mu}(\omega)$ are mixed dipole-energy and dipole-dipole spectral densities. A full derivation of the response function can be found in Ref.~\onlinecite{Wiethorn2023}. Here, we just present the numerical expression for the $A_{\delta \mu \delta U}(t)$ term. The $A_{\delta \mu \delta U}(t)$ term can be written as
\begin{eqnarray} \nonumber
A_{\delta \mu \delta U}(t)&=&|\boldsymbol{\mu}^\textrm{av}_{01}|^2-2\boldsymbol{\mu}^\textrm{av}_{01}\cdot \boldsymbol{\mathcal{A}}\left[\boldsymbol{\mathcal{J}}^{01}_{\delta \mu\delta U}(\omega)\right](t)+\boldsymbol{\mathcal{A}}\left[\boldsymbol{\mathcal{J}}^{01}_{\delta \mu\delta U}(\omega)\right](t) \cdot \boldsymbol{\mathcal{A}}\left[\boldsymbol{\mathcal{J}}^{01}_{\delta \mu\delta U}(\omega)\right](t) \\
&&+ \mathcal{B}\left[\mathcal{J}^{01}_{\delta \mu\delta \mu}(\omega) \right](t) \\
\boldsymbol{\mathcal{A}}\left[\boldsymbol{\mathcal{J}}^{01}_{\delta \mu\delta U}(\omega)\right](t)&=& \frac{1}{\pi}\int \textrm{d}\omega \, \frac{\boldsymbol{\mathcal{J}}^{01}_{\delta \mu\delta U}(\omega)}{\omega}\left[\left(\cos(\omega t)-1\right)-\textrm{i}\,\textrm{coth}\left(\frac{\beta \omega}{2} \right)\sin(\omega t)\right]\\
\mathcal{B}\left[\mathcal{J}^{01}_{\delta \mu\delta \mu}(\omega) \right](t)&=&\frac{1}{\pi}\int \textrm{d}\omega \, \mathcal{J}^{01}_{\delta \mu\delta \mu}(\omega) \left[\textrm{coth}\left(\frac{\beta \omega}{2}\right)\cos(\omega t)-\textrm{i}\sin(\omega t) \right]
\end{eqnarray}
Evaluating the necessary terms to compute the prefactor $A_{\delta \mu \delta U}(t)$ thus comes at a negligible additional computational cost with respect to computing a cumulant spectrum within the Condon approximation. 

\section{Chain mapping}
To allow for an efficient representation of the many-body wavefunction in terms of a matrix product state (MPS), we apply a unitary transformation to the Hamiltonian resulting in a 1D or quasi-1D (tree)-like topology, known as a chain mapping. This chain mapping can be naturally formulated in terms of orthonormal polynomials\cite{Chin2010}. For a Hamiltonian with linear vibronic coupling to the $|S_1\rangle$ state described by a spectral density $\mathcal{J}(\omega)$, the energy level interaction Hamiltonian can be expressed as:
\begin{equation}
\hat{H}^{S_1}_\textrm{I}=\int \textrm{d}\omega \, \sqrt{\mathcal{J}(\omega)}\left(a^\dagger (\omega)+a(\omega) \right)|S_1\rangle\langle S_1 |.
\end{equation}

To bring the Hamiltonian into chain form, we start by finding the set of polynomials $\{\tilde{p}_{n} \in \mathbb{P}_{n}, n=0,1,2,...\}$ which are orthonormal with respect to the measure $\mathcal{J}(\omega)\text{d}\omega$, such that:
\begin{equation}
    \int \textrm{d}\omega \mathcal{J}(\omega) \tilde{p}_n(\omega) \tilde{p}_m (\omega)=\delta_{nm}.
\end{equation}

This allows us to define a unitary transformation $U_n(\omega)=\sqrt{\mathcal{J}(\omega)}\tilde{p}_n(\omega)$ that maps the infinite bath of bosonic operators described by the spectral density to a discrete set of chain modes. Due to special properties of the orthogonal polynomials, namely their orthogonality and the fact that they satisfy a three term recurrence relation, the resulting chain modes couple only to their nearest neighbours and only the first chain mode ($n$=0) couples to the system. Introducing the new set of chain modes as $b^\dagger_n$, the bath modes $\alpha(\omega)$ and semi-infinite chain modes are related through the following transformations: 
\begin{eqnarray}
    b^\dagger_n&=&\int \textrm{d}\omega\, U_n(\omega) a^\dagger (\omega)=\int \textrm{d}\omega \sqrt{\mathcal{J}(\omega)}\tilde{p}_n(\omega)  a^\dagger (\omega) \\
    a^\dagger(\omega)&=&\sum_n^\infty U_{n}(\omega) b^\dagger_n=\sum_n^\infty \sqrt{\mathcal{J}(\omega)}\tilde{p}_n(\omega)b^\dagger_n
\end{eqnarray}
and the interaction Hamiltonian can be rewritten as 
\begin{equation}
    H_{S_1}^\text{EL}=\sum_{n=0}^{\infty}\left(\underbrace{\int \text{d}\omega\,\mathcal{J}(\omega)\tilde{p}_{n}(\omega)}_{=\delta_{n,0}/\tilde{p}_{0}}|{S_{1}}\rangle\langle{S_{1}}| \right) \left(b_{n}^{\dagger}+b_{n}\right)
    \label{si_eqn:chain_mapped_ham}
\end{equation}
where, thanks to the orthogonality of the polynomials, the coupling reduces to a local interaction of between the system energy level and the first ($n=0$) site on the chain. In practice, the semi-infinite chain is truncated to a finite, user-defined length $N$. However, this approximation to the unitary transformation does not influence the system dynamics as long as the chain is chosen sufficiently long such that the last chain element does not get significantly populated during propagation.   

To map the full three level Hamiltonian to a tree-like topology, a unitary transformation with a separate set of orthogonal polynomials needs to be constructed for the S$_1$, S$_2$ and the coupling spectral density. For the S$_1$ and S$_2$ tuning modes, we additionally account for cross correlations (see SI Sec.~\ref{si_sec:bath_corr}), which, in practice, leads to an extra set of long-range couplings in Eqn.~\ref{si_eqn:chain_mapped_ham} to $|S_2\rangle$ (see Refs.~\onlinecite{Dunnett2021, Hunter2024}). 

The task to perform the chain mapping thus reduces to finding sets of orthogonal polynomials using $\mathcal{J}_{01}(\omega)$, $\mathcal{J}_{02}(\omega)$, and $\mathcal{J}_{12}(\omega)$ as a measure. The site energies and hopping strengths for the chain components are directly determined from the recurrence coefficients of the polynomials. Whereas for certain model spectral densities, these coefficients can be determined analytically, for the arbitrary spectral densities resulting from MD, this is instead done numerically. Here, we use a Python implementation of the ORTHOPOL package\cite{gautschi_algorithm_1994} (see script \texttt{spectral\_dens\_to\_chain} in Data Availability). Care is taken to guarantee the orthogonality of the polynomials with respect to the measure, by avoiding numerical artifacts due to insufficient frequency resolution of the spectral density. This is achieved through both a cubic spline interpolation of MD energy gap fluctuations, and a Fourier interpolation by padding the classical time-correlation function with zeros (see \texttt{Python} script \texttt{compute\_diabatic\_sds.py} for details). 

\section{Fluorescence spectra from tensor networks}
In a matrix product state (MPS) format, the relaxed state $\left|\Psi_\textrm{abs}(t_\textrm{abs}^\textrm{max})\right\rangle$ at the end of an absorption calculation must take the following form:
\begin{equation}
    \left|\Psi_\textrm{abs}(t_\textrm{abs}^\textrm{max})\right\rangle= \sum_{i_E} \left(A^0_S A^{i_E}_E|S_0\rangle |E_{i_E}\rangle+A^1_S A^{i_E}_E|S_1\rangle |E_{i_E}\rangle+A^2_S A^{i_E}_E|S_2\rangle |E_{i_E}\rangle\right)
\end{equation}
where $A^i_S$ is the tensor corresponding to the $i$th electronic state $|S_i\rangle$, and $A^{i_E}_E$ is a product of all tensors corresponding to environmental state configurations. We note that $|S_0\rangle$ is completely decoupled from the excited state manifold in our model, such that the tensor $A^0_S$ does not evolve during the absorption calculation. Since the dipole operator is independent of nuclear degrees of freedom for the diabatic states defining the Hamiltonian, it does not act on environmental states. Thus $|\phi_R\rangle=\mu^{-}\left|\Psi_\textrm{abs}(t_\textrm{abs}^\textrm{max})\right\rangle$ can be expressed as
\begin{eqnarray}
|\phi_R\rangle&=&\sum_{i_E}\left(\mu_1 A^1_S A^{i_E}_E |S_0\rangle|E_{i_E}\rangle + \mu_2 A^2_S A^{i_E}_E |S_0\rangle|E_{i_E}\rangle\right) \\
&=&\tilde{A}^0_S |S_0\rangle |E_{i_E}\rangle ,
\end{eqnarray}
where $\tilde{A}^0_S=\mu_1 A^1_S+\mu_2 A^2_S $ is a new tensor that can be expressed in terms of tensors $A^1_S $ and $A^2_S $ that make up the final state of the absorption calculation $\left|\Psi_\textrm{abs}(t_\textrm{abs}^\textrm{max})\right\rangle$. 

For the calculation of emission spectra, we need to construct the state $|\chi_R (0)\rangle $, formed from $|\phi_R\rangle$ and the components of $\left|\Psi_\textrm{abs}(t_\textrm{abs}^\textrm{max})\right\rangle$ involving the excited electronic states only. In other words, $|\chi_R (0)\rangle$ can be expressed in MPS format as
\begin{equation}
|\chi_R (0)\rangle=\sum_{i_E} \left(\tilde{A}^0_S A^{i_E}_E|S_0\rangle |E_{i_E}\rangle+A^1_S A^{i_E}_E|S_1\rangle |E_{i_E}\rangle+A^2_S A^{i_E}_E|S_2\rangle |E_{i_E}\rangle\right).
\end{equation}
It follows that to construct the required initial state for an emission calculation, all that is needed is to take $\left|\Psi_\textrm{abs}(t_\textrm{abs}^\textrm{max})\right\rangle$ and overwrite the tensor $A^0_S$ with $\tilde{A}^0_S=\mu_1 A^1_S+\mu_2 A^2_S $, with no manipulation of the (generally very large) environment object $A^{i_E}_E$ required.  

\section{Bath correlations}
\label{si_sec:bath_corr}

\begin{figure*}
    \begin{center}
        \includegraphics[width=0.98\textwidth]{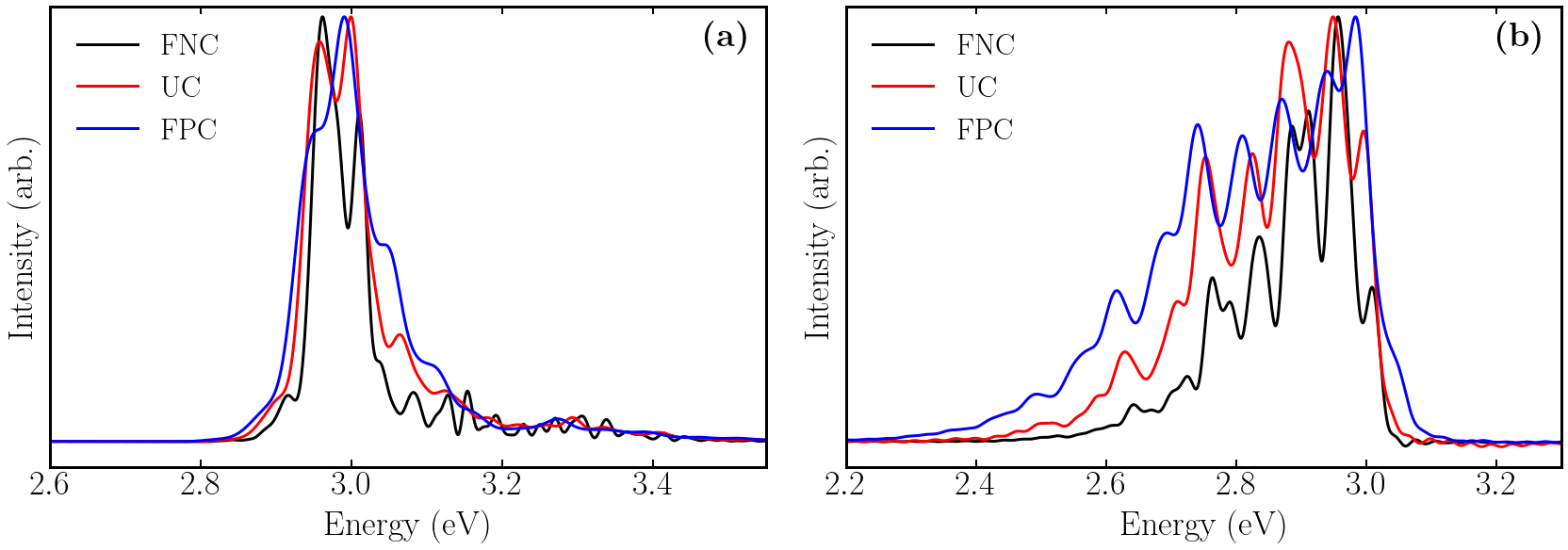}
    \end{center}
    \caption{Comparison of the MSA200 (a) absorption and (b) emission spectra in the cases of FNC, uncorrelated (UC) and FPC cross-correlation spectral densities.}
    \label{fig:si_model_sys_correlations}
\end{figure*}

The spectral densities of the tuning modes, $\mathcal{J}_{01}(\omega)$ and $\mathcal{J}_{02}(\omega)$, are positive semi-definite functions and thus can only encode strength of the couplings to S$_1$ and S$_2$ energy gap fluctuations, but not the relative sign. The degree to which fluctuations in S$_1$ are \emph{correlated} or \emph{anti-correlated} with fluctuations in S$_2$ is therefore introduced into the outlined formalism through long-range couplings in the baths describing the two spectral densities\cite{Dunnett2021,Hunter2024}. In an MD-based formalism, these long-range couplings can be directly computed from the cross correlation spectral density $\mathcal{J}_\textrm{cross}(\omega)$ and its normalized counterpart $\widetilde{\mathcal{J}}_\text{cross}\left(\omega\right) = \mathcal{J}_\text{cross}\left(\omega\right)/\sqrt{\mathcal{J}_{01}\left(\omega\right)\mathcal{J}_{02}\left(\omega\right)}$. Here, $\widetilde{\mathcal{J}}_\text{cross}\left(\omega\right)=1$ indicates fully correlated S$_1$ and S$_2$ fluctuations, whereas $\widetilde{\mathcal{J}}_\text{cross}\left(\omega\right)=-1$ indicates anti-correlated fluctuations. In the MD sampling procedure, where all system fluctuations are mapped to an infinite bath of harmonic oscillators, $\widetilde{\mathcal{J}}_\text{cross}\left(\omega\right)$ can take any value between $-1$ and 1 for a given frequency, whereas for a system with a finite number of modes, individual modes can only be correlated, anti-correlated, or uncorrelated $\left(\widetilde{\mathcal{J}}_\text{cross}\left(\omega\right)=0\right)$

In all model system calculations presented in the main manuscript, we assume fully positively correlated $\left(\mathrm{FPC, }\,\,\widetilde{\mathcal{J}}_\text{cross}\left(\omega\right)=1\right)$ tuning modes. However, previous results obtained for realistic systems\cite{Dunnett2021, Hunter2024} have demonstrated that in general a variety of degrees of cross correlations between fluctuations exist in real molecules, and these correlations can have a strong impact on the resulting quantum dynamics and spectroscopic observables. To confirm this impact for our two-mode model system of a conical intersection, we perform additional calculations on model system parameterization MSA, assuming fully negatively correlated $\left(\mathrm{FNC, }\,\,\widetilde{\mathcal{J}}_\text{cross}\left(\omega\right)=-1\right)$ and uncorrelated $\left(\widetilde{\mathcal{J}}_\text{cross}\left(\omega\right)=0\right)$ fluctuations. The results are presented in Fig.~\ref{fig:si_model_sys_correlations}.

We note that bath correlations also have a strong influence on the simple two-mode model system. Specifically, FPC S$_1$ and S$_2$ fluctuations are found to strongly enhance nonadiabatic effects, resulting in a very broad emission lineshape, with FNC fluctuations yielding a significantly reduced low energy tail of the spectrum. Interestingly, for the absorption lineshape, the FNC spectrum, while narrower than the FPC is found to gain additionally vibronic fine-structure not present in the fully positively correlated lineshape. 

This phenomenon is likely driven by the fact that for FNC fluctuations, the effective S$_1$-S$_2$ gap in the Condon region periodically increases and decreases, whereas it stays more constant for the FPC parameterization. Since the S$_1$-S$_2$ gap at the Condon point is 0.2~eV in the model system, for FPC fluctuations, the impact of nonadiabatic effects on the absorption spectrum is minimal, but the periodically reduced gap for FNC fluctuations is sufficient to induce some nonadiabatic coupling effects in the lineshape. For the emission spectrum on the other hand, the strong nonadiabatic effects are driven by the relaxed S$_1$-S$_2$ minima being close in energy to allow for strong intensity borrowing, and this intensity borrowing is most effective for FPC baths. 

\section{Convergence of the quantum dynamics}

\begin{figure*}
    \begin{center}
        \includegraphics[width=0.98\textwidth]{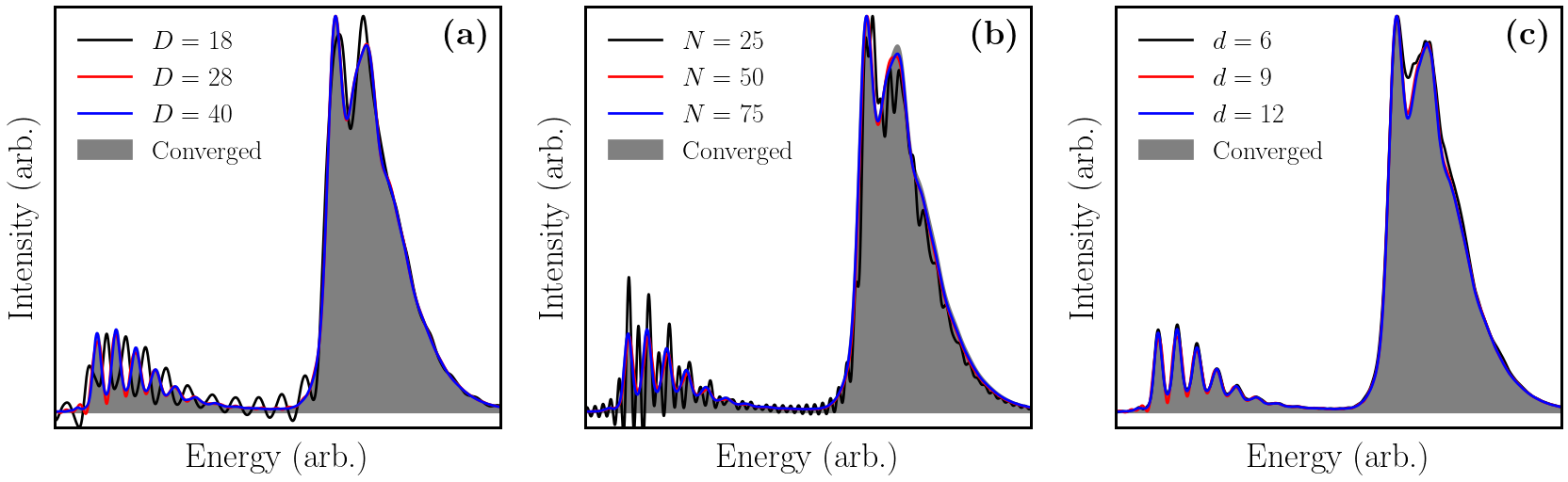}
    \end{center}
    \caption{Convergence of the absorption spectrum for pyrazine in vacuum with respect to (a) bond dimension $D$, (b) chain length $N$ and (c) local Fock space dimension $d$. In all cases, the fully converged reference spectrum is obtained for $D=50$, $N=150$, and $d=30$.}
    \label{fig:si_pyrazine_convergences}
\end{figure*}

The convergence of the quantum dynamics and resulting optical spectra in the T-TEDOPA formalism is controlled by three parameters: the bond dimension $D$ controlling the expressiveness of the tensor network in representing the many-body wavefunction, the chain length $N$ defining the number of effective modes involved in the system dynamics after the chain-mapping procedure, and the size of the local Fock space $d$ controlling the physical dimensions of the tensor network. 

Convergence tests involving all three quantities can be found in SI Fig.~\ref{fig:si_pyrazine_convergences}. The results show a systematic convergence of the absorption lineshape, with $D=40$, $N=75$, and $d=12$ yielding results indistinguishable from the fully converged benchmark results that were calculated for $D=50$, $N=150$, and $d=30$. All results presented for pyrazine in the main text are computed at the benchmark convergence settings. 

For the 2D model systems, similar convergence tests reveal that absorption and fluorescence lineshapes are well converged for $D=35$, $N=125$, and $d=18$, and these parameters are used throughout for results presented in the main text. 

\section{Model system parameters}

For each of the model systems presented in the main text, vertical S$_1$ and S$_2$ energy gaps were set at 3.0~eV and 3.2~eV with coresponding dipole moments of 2.54~a.u. and 0.0~a.u. The central frequencies of the Gaussian portion of the tuning and coupling spectral densities were 1000~cm$^{-1}$ and 500~cm$^{-1}$ with standard deviations of 8.493~cm$^{-1}$. The peak heights of the S$_1$, S$_2$ and coupling spectral density lineshapes were then fixed using the reorganization energies specified in Table~\ref{tab:model_system_params}. The total reorganization energy in the S$_2$ and coupling spectral densities of each model system was set to 124.0~meV. However, the A and B model systems differ in that MSA has an additional 99.2~meV of reorganization energy in the S$_1$ tuning spectral density. The frequency $\omega_\text{env}$ characterizing the environmental, Debye contributions to the spectral densities were set to 30~cm$^{-1}$ in MSA30 and MSB30, and 200~cm$^{-1}$ in MSA200 and MSB200. Tuning modes were FPC in each model system.

\begin{table}[h!]
\centering
\renewcommand{\arraystretch}{1.5}
\setlength{\tabcolsep}{3pt}
\newcolumntype{C}[1]{>{\centering\arraybackslash}p{#1}}
\begin{tabular}{C{1.9cm}|C{2.15cm}|C{2.15cm}|C{2.15cm}|C{2.15cm}|C{2.15cm}|C{2.15cm}} 
 & $\lambda_{t,1}$ (meV) & $\lambda_{\mathrm{env},1}$ (meV) & $\lambda_{t,2}$ (meV) & $\lambda_{\mathrm{env},2}$ (meV) & $\lambda_c$ (meV) & $\lambda_{\mathrm{env},c}$ (meV) \\ 
\hline
MSA & 12.4 & 12.4 & 117.8 & 6.2 & 111.6 & 12.4 \\ 
\hline
MSB & 111.6 & 12.4 & 62.0 & 62.0 & 111.6 & 12.4
\end{tabular}
\caption{Reorganization energies used for the ``MSA" and ``MSB" model systems introduced in the main text. The same cutoff frequency $\omega_\mathrm{env}$ for the Debye spectral densities was used in the tuning and coupling spectral densities of each model system.}
\label{tab:model_system_params}
\end{table}

\section{Additional results: System parameters and comparison of spectral densities for pyrazine} 
\label{si_sec:pyrazine_sds}

Beyond the chain coefficients, the main parameters entering the tensor network simulations are the average diabatic energies of the excited states, computed directly from MD, and the average magnitudes of the transition dipole moments. Table~\ref{tab:system_params} shows the computed parameters for the pyrazine molecule in vacuum, cyclohexane and water. Additionally, the total reorganization energies $\lambda^\textrm{reorg}$ contained in the $n\pi^*$, the $\pi\pi^*$ and the coupling spectral densities are also displayed, where the reorganization energy is defined as:
\begin{equation}
\lambda^\textrm{reorg}\left[\mathcal{J}\right]=\frac{1}{\pi}\int_0^\infty \textrm{d}\omega\, \frac{\mathcal{J}(\omega)}{\omega}. 
\end{equation}

The CAM-B3LYP functional overestimates the average energy of the $\pi\pi^*$ transition relative to the $n\pi^*$ transition, leading to a poor agreement of the computed absorption lineshape with experiment when left unaltered. Instead, we adjust the value such that the minimum of the diabatic $\pi\pi^*$ PES is 0.84~eV higher in energy than the minimum of the $n\pi^*$ PES in vacuum, in agreement with earlier model calculations based on MCTDH\cite{Raab1999}. Here, the minimum of the PES is defined as $E^\textrm{av}$, the average diabatic energy in the Condon region, minus the reorganization energy $\lambda^\textrm{reorg}$ for each electronic state. The same relative shift obtained for vacuum is applied to the $\pi\pi^*$ transitions in cyclohexane and water, resulting in the $E_{\pi\pi^*}^\textrm{av}$(adjusted) values specified in Table~\ref{tab:system_params} that are used throughout the main text. We note that these are the only adjustments made to the quantum dynamics Hamiltonian and all other system parameters are directly obtained from the MD sampling. 

\begin{center}
\begin{table}
\begin{tabular}{ c|c|c|c } 
  & Vacuum & Cyclohexane & Water \\  \hline
$E_{n\pi^*}^\textrm{av}$  (eV)  & 4.241 & 4.247  & 4.484 \\ 
$E_{\pi\pi^*}^\textrm{av}$(DFT) (eV)  & 5.598 & 5.619 & 5.573 \\ 
$E_{\pi\pi^*}^\textrm{av}$(adjusted) (eV)  & 5.129 & 5.150 & 5.104 \\ \hline
$|\boldsymbol{\mu}^\textrm{av}_{n\pi^*}|$ (D) & 0.727 & 0.727 & 0.717 \\
$|\boldsymbol{\mu}^\textrm{av}_{\pi\pi^*}|$ (D) & 2.336 & 2.339 & 2.556 \\ \hline
$\lambda_{n\pi^*}^\textrm{reorg}$ (eV)  & 0.138  & 0.147 & 0.209 \\ 
$\lambda_{\pi\pi^*}^\textrm{reorg}$ (eV)  & 0.186 & 0.187 & 0.250 \\ 
$\lambda_{\textrm{coupling}}^\textrm{reorg}$ (eV) & 0.206 & 0.206 & 0.214\\ 
\end{tabular}
\caption{System parameters for pyrazine in vacuum as computed from the MD sampling. Average diabatic energies and reorganization energies are given in eV, average transition dipole moments are given in Debye. For the diabatic energy $E_{\pi\pi^*}^\textrm{av}$, two values are quoted: The average obtained directly from the (TD)DFT calculations, and an adjusted value used in calculations in the main manuscript.}
\label{tab:system_params}
\end{table}
\end{center}

\begin{figure*}
    \begin{center}
        \includegraphics[width=0.98\textwidth]{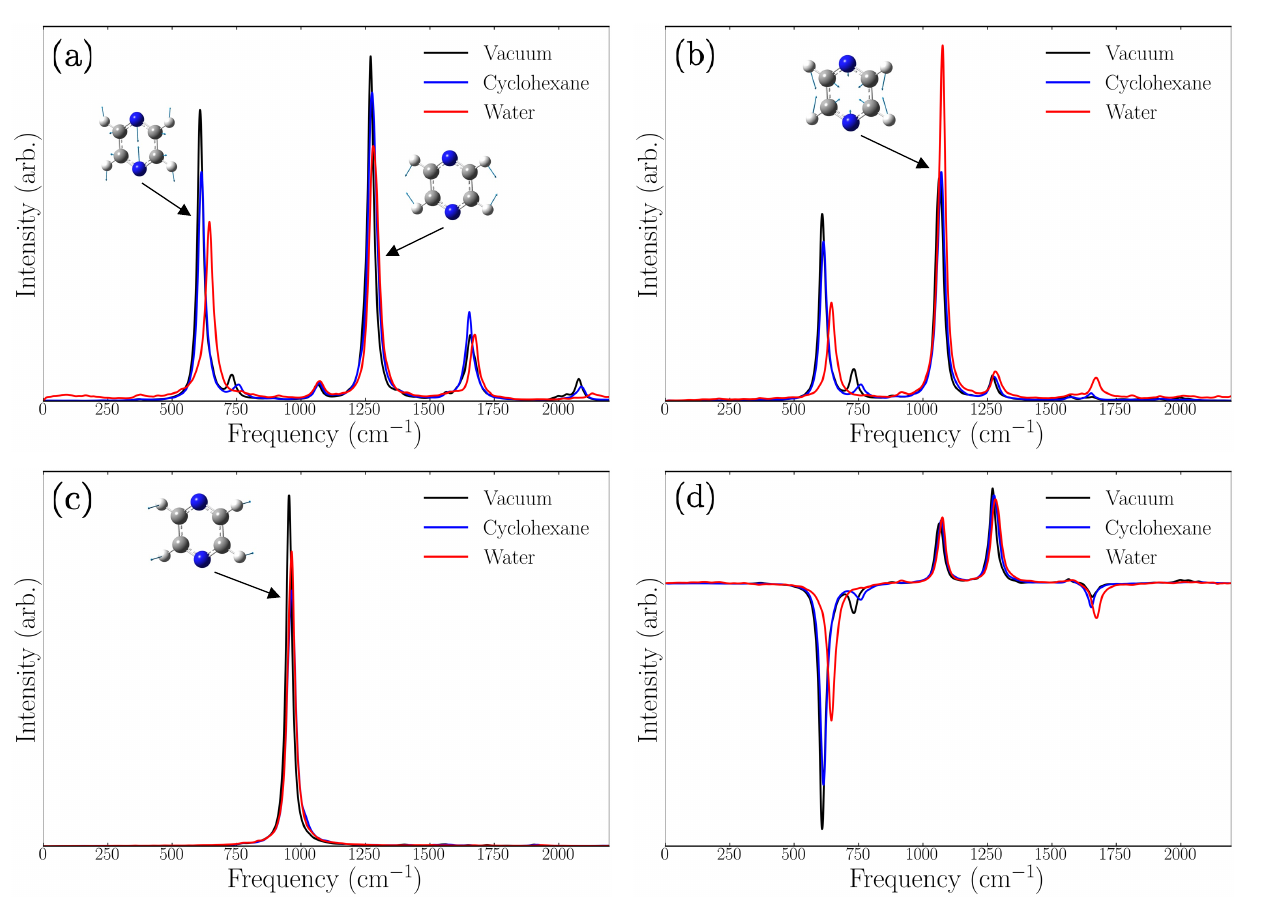}
    \end{center}
    \caption{Comparison of pyrazine spectral densities computed based on the 20~ps AIMD trajectory in vacuum (orange) and 20~ps QM/MM trajectories in cyclohexane (green) and water (blue). (a) and (b) represent the diabatic $n\pi^*$ and $\pi\pi^*$ spectral densities, respectively. (c) illustrates the coupling spectral densities. (d) is the cross correlation spectral density. Prominent peaks are assigned to vibrational frequencies computed via a normal mode analysis of the molecule in vacuum performed in Gaussian\cite{gdv}. }
    \label{fig:pyrazine_sd}
\end{figure*}

The results in Table~\ref{tab:system_params} reveal that system parameters undergo minor changes between vacuum and cyclohexane, as would be expected in a non-polar solvent environment. However, in water, the average energy of the $n\pi^*$ transition undergoes a strong hypsochromic shift of around 0.24~eV, whereas the $\pi\pi^*$ transition undergoes a much weaker bathochromic shift of around 50~meV. Additionally, the reorganization energies of the $\pi\pi^*$ and $n\pi^*$ transition increase significantly due to the contribution of low frequency solvent coupling. The coupling spectral density mixing the two excited states on the other hand is significantly less impacted by solvent interactions.

Fig.~\ref{fig:pyrazine_sd} shows computed $n\pi^*$, $\pi\pi^*$, coupling and cross-correlation spectral densities for pyrazine in vacuum, cyclohexane and water. We note that the $n\pi^*$ and $\pi\pi^*$ are very similar in cyclohexane and vacuum, both in terms of peak position and intensity. This is expected for a small, rigid molecule in a non-polar solvent. For pyrazine in water, some peaks undergo significant frequency shifts and intensity changes, showing that hydrogen bonding interactions and polarization effects influence the energy gap fluctuations. Additionally, the $n\pi^*$ spectral density has a very significant low frequency ($<250$~cm$^{-1}$) contribution, that can be directly ascribed to solvent fluctuations and polarization effects, as the lowest vibrational frequency of the molecule itself is approximately 370~cm$^{-1}$. For the cross-correlation spectral density, we note that the strong mid-frequency vibration at approximately 610~cm$^{-1}$ is \emph{anti-correlated} between $n\pi^*$ and $\pi\pi^*$, whereas higher energy vibrations are mostly correlated. 

The coupling spectral density has only a single clearly identifiable peak at approximately 960~cm$^{-1}$. Contrary to findings in recent work of some of the authors\cite{Hunter2024} on the proflavine molecule, the coupling spectral density for pyrazine is relatively insensitive to coupling to solvent environment. However, given the significant changes to the spectral densities describing the tuning modes modulating the fluctuations of the $n\pi^*$ and $\pi\pi^*$ transitions, the quantum dynamics are still expected to be significantly impacted by condensed phase environment interactions, and the approach described here is ideally suited to modeling these interactions from first principles. 

We note that several features of the pyrazine PES encoded in our spectral densities match the model system parameterization obtained by Raab and coworkers\cite{Raab1999}. Specifically, the low frequency mode at 610~cm$^{-1}$ (experimentally observed at approximately 600~cm$^{-1}$) causes anti-correlated fluctuations in the $n\pi^*$ and $\pi\pi^*$ transition energies, with higher frequency vibrations mostly correlated, and only a single mode is found to strongly couple the two electronic states. 

\section{Additional results: GNCT dipole-dipole spectral densities}

\begin{figure}
    \begin{center}
        \includegraphics[width=0.8\textwidth]{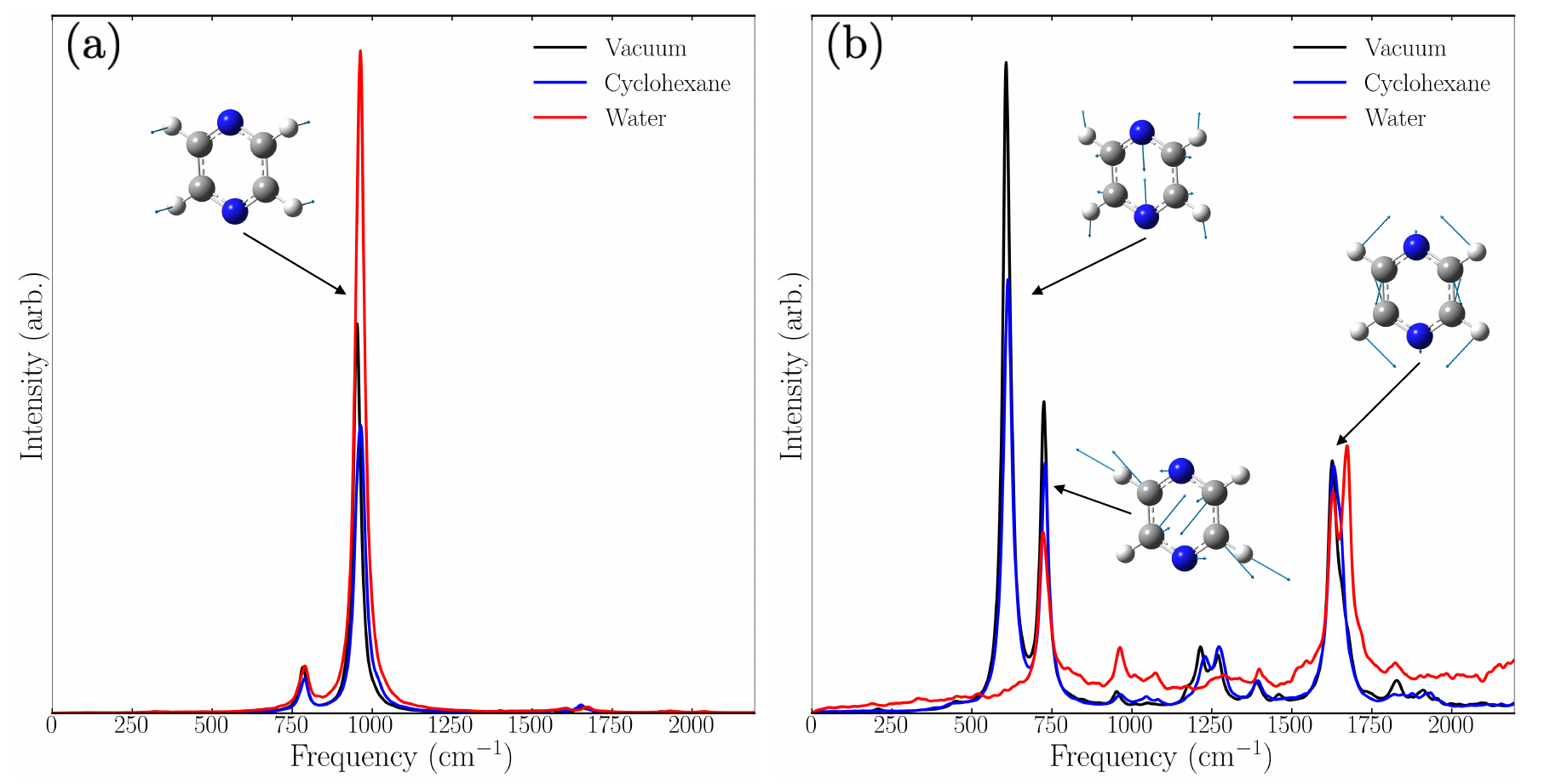}
    \end{center}
    \caption{The $n\pi^*$ (a) and the $\pi\pi^*$ (b) dipole-dipole spectral densities for the pyrazine molecule, as used in the GNCT. Prominent peaks are assigned to vibrational frequencies computed via a normal mode analysis of the molecule in vacuum.}
    \label{fig:pyrazine_dd_sd}
\end{figure}

Fig.~\ref{fig:pyrazine_dd_sd} shows the dipole-dipole spectral density $\mathcal{J}_{\delta \mu}(\omega)$ for the $n\pi^*$ and the $\pi\pi^*$ state of pyrazine in vacuum and in solution, as computed in the GNCT approach (see SI Sec.~\ref{si_sec:GNCT}). We note that for the $n\pi^*$ state, non-Condon effects in the lineshape as captured by the GNCT are driven by the same out-of-plane mode that strongly couples the diabatic $n\pi^*$ and $\pi\pi^*$ states in the tensor network formalism. Solvent effects have a significant impact on the intensity of the dipole fluctuations driven by this mode, with water leading to a significantly stronger coupling than cyclohexane or vacuum. For the adiabatic $\pi\pi^*$ dipole-dipole spectral densities, we note an even stronger influence of solvent interactions, with water fully quenching coupling to a low frequency symmetric stretching mode of the pyrazine ring. This is likely caused by direct solute-solvent interactions in form of hydrogen bonding that can couple strongly to the dipole moment of the $\pi\pi^*$ transition. The results highlight the importance of treating environmental interactions on the same footing with nuclear vibrations of the chromophore, as is done in the MD-based approaches outlined in this work. 

\section{Additional results: Tensor network simulations with reduced diabatic coupling} 

\begin{figure}
    \centering
    \includegraphics[width=0.9\textwidth]{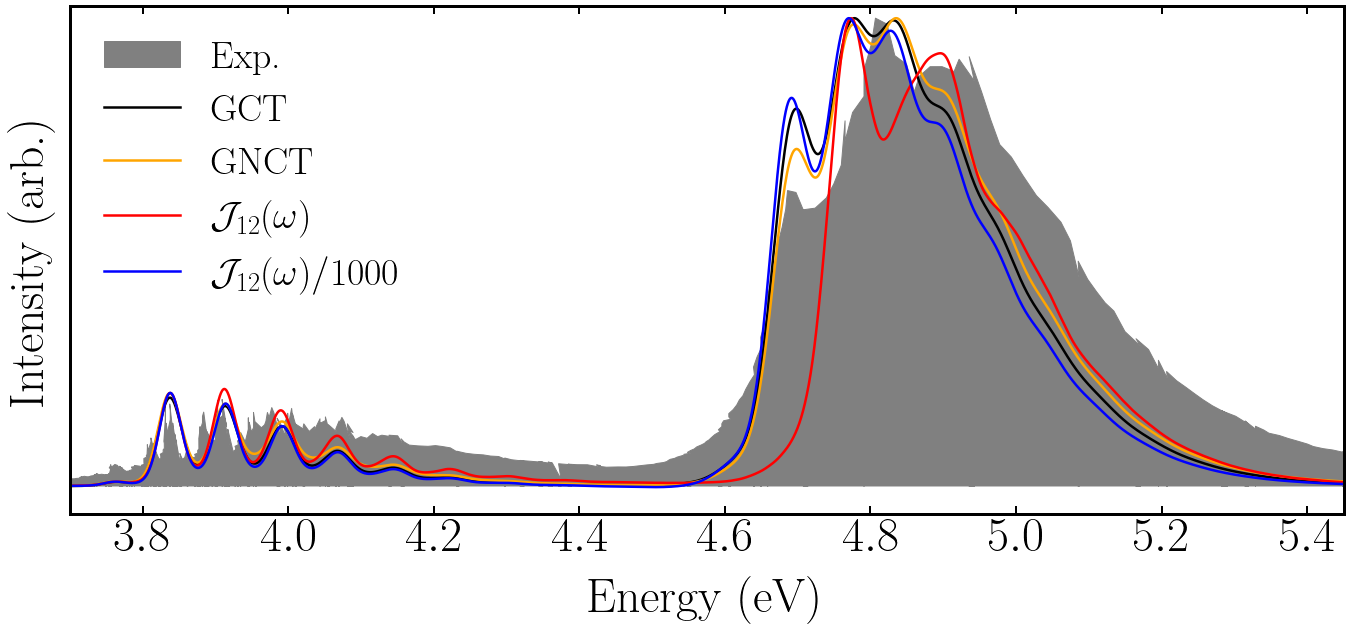}
    \caption{Comparison between GCT, GNCT and T-TEDOPA absorption lineshapes for pyrazine in vacuum, in comparison with the experimental spectrum\cite{Samir2020_pyrazine_vac}. For T-TEDOPA, we include the predicted lineshapes from the unaltered coupling spectral density $\mathcal{J}_{12}\left(\omega\right)$ and a modified coupling spectral density, uniformly scaled by $1/1000$.}
    \label{fig:pyrazine_reduced_coupling}
\end{figure}

Fig.~\ref{fig:pyrazine_reduced_coupling} shows the spectrum of pyrazine in vacuum, as computed with the GCT and GNCT approaches based on decoupled \emph{adiabatic} states and the T-TEDOPA approach based on non-perturbative quantum dynamics of fully coupled \emph{diabatic} states. To provide a cross-check of the diabatization procedure used, it is interesting to confirm that the two approaches begin to resemble each other in certain limits. Specifically, we assess the influence of reducing the $n\pi^*$-$\pi\pi^*$ coupling in the T-TEDOPA formalism, by scaling the coupling spectral density $\mathcal{J}_{12}(\omega)$ by a constant factor. As can be seen in Fig.~\ref{fig:pyrazine_reduced_coupling}, reducing the coupling by a factor of 1000 in the T-TEDOPA formalism yields an absorption lineshape in close agreement with that of the GCT. The results indicate that the diabatic states largely retain the character of their adiabatic counterparts, producing very similar energy gap fluctuations and spectral densities for the tuning modes. Differences in the predicted lineshapes between GNCT and T-TEDOPA can thus be directly ascribed to the different treatment of non-Condon effects, through accounting for Gaussian fluctuations in the transition dipole moment for GNCT, and explicit non-perturbative quantum dynamics of coupled diabatic states in T-TEDOPA.

\section{Additional results: Tensor network simulations with varying cross-correlation} 

\begin{figure}
    \centering
    \includegraphics[width=0.9\textwidth]{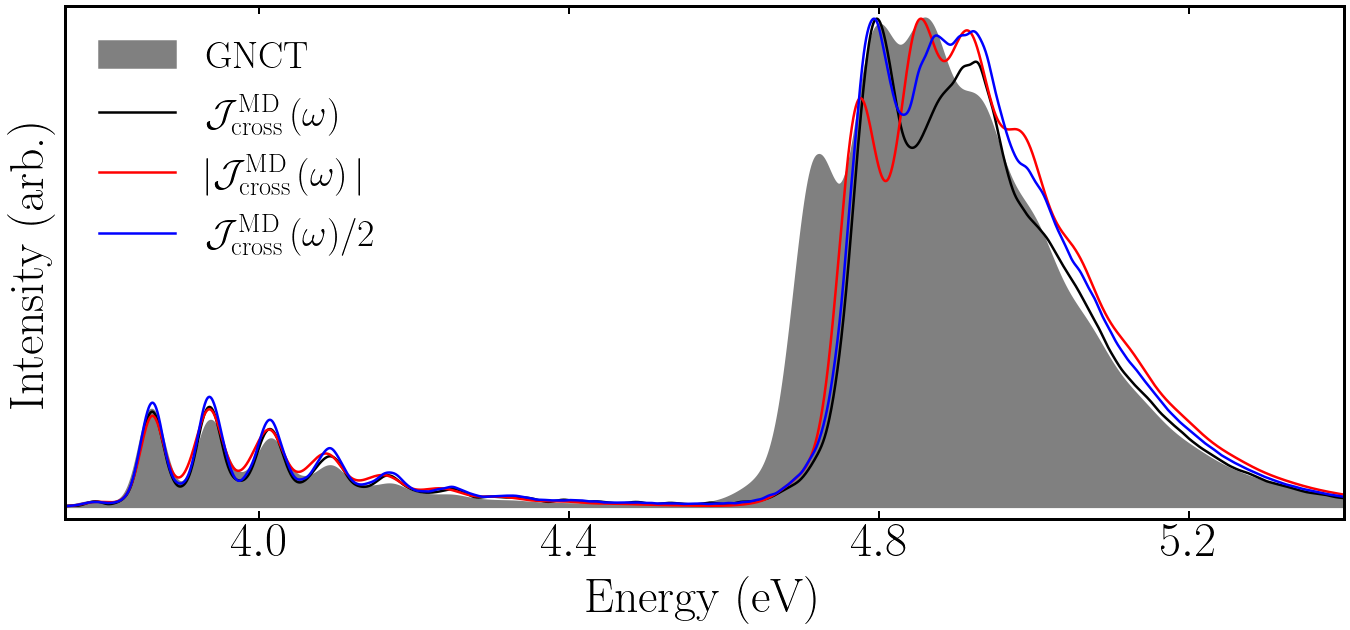}
    \caption{GNCT and T-TEDOPA absorption lineshapes for pyrazine in vacuum. We compare the influence on the predicted T-TEDOPA lineshape when the MD coupling spectral density is made strictly positive ($\vert \mathcal{J}_\mathrm{cross}^\mathrm{MD}\left(\omega\right) \vert$) and reduced by a factor of 2 ($\mathcal{J}_\mathrm{cross}^\mathrm{MD}\left(\omega\right)/2$). }
    \label{fig:pyrazine_vac_correlations}
\end{figure}

Fig.~\ref{fig:pyrazine_vac_correlations} shows the predicted absorption spectra for pyrazine in vacuum using GNCT and T-TEDOPA with varied cross-correlation spectral densities. When the cross-correlation spectral densities are made strictly positive, the T-TEDOPA lineshape becomes very similar to that of GNCT (up to an arbitrary shift between the S$_1$ and S$_2$ spectral features in GNCT). We note that, as suggested by SI Fig.~\ref{fig:pyrazine_sd}(d), making the vacuum pyrazine cross-correlation spectral density strictly positive is essentially equivalent to the FPC limit, since correlated and anti-correlated tuning modes are well-separated in frequency. Fig.~\ref{fig:pyrazine_vac_correlations} also contains a lineshape computed where the cross-correlation spectral density has been scaled down by a factor of 2, reducing the cross correlations between the S$_1$ and S$_2$ baths over the entire frequency range (thus pushing the system towards the uncorrelated bath limit). We find that the spectral lineshape of the $\pi\pi^*$ transition for the reduced cross correlations shows some features more closely aligned with the spectrum generated for MD-derived cross-correlations (like the lowest energy peak in the $\pi\pi^*$ transition showing the highest intensity), whereas others more closely resemble the GNCT (such as an additional peak appearing in between the most prominent peaks of the standard MD-based cross-correlated spectrum).

The results reveal that bath correlations are crucial in determining the spectral lineshape of pyrazine, and in particular the fact that $n\pi^*$ and $\pi\pi^*$ fluctuations are strongly anti-correlated with respect to the dominant tuning mode at 610~cm$^{-1}$ (see SI Sec.~\ref{si_sec:pyrazine_sds}). 

\section{Additional results: Pyrazine spectral densities with reparameterized couplings}

An apparent shortcoming of the T-TEDOPA method is the absence of the low-energy vibronic band seen in the experimental S$_2$ spectrum of pyrazine (c.f. Fig.~6 of the main text). In order to determine whether this absence is due to the T-TEDOPA method itself, or the spectral densities used to perform the T-TEDOPA dynamics, we investigate the case where specific peaks in the spectral densities are rescaled to be consistent with couplings previously obtained through extensive multireference configuration interaction (MRCI) calculations of Pyrazine in vacuum\cite{Sobolewski1993}. This is done by fitting Lorentzian lineshapes to each mode in the spectral density and scaling their height to reproduce the correct coupling strength/reorganization energy over the appropriate interval.

The MRCI couplings have previously shown to produce very accurate absorption spectra from dynamics simulated through path integral and MCTDH methods\cite{Krempl1994,Raab1999}. Moreover, 20 of the 24 modes previously considered in pyrazine are having a relatively minor impact on the overall absorption lineshape\cite{SCHNEIDER1988,Schneider1990,Stock1990,Stock1990_2}, so we only modify the four most relevant ones. These four modes, one coupling ($\nu_\mathrm{10a}$) and three tuning ($\nu_\mathrm{6a}$, $\nu_1$, $\nu_\mathrm{9a}$), are indicated with image insets in Fig.~\ref{fig:pyrazine_sd}. 

Table \ref{tab:pyrazine_couplings} shows the comparison of linear couplings obtained from both MD-based TDDFT/CAM-B3LYP sampling (this work) and static MRCI calculations. Note that the couplings appearing in the S$_1$ and S$_2$ rows of the table are the coefficient of the linear displacement terms appearing in Eqn.~10 and 11 of the main text, but scaled by an additional factor $1/\sqrt{\omega_j}$. On the other hand, the couplings listed in the $\nu_\mathrm{10a}$ coupling mode correspond to the vibronic couplings ($\Lambda_\mathrm{10a}$) appearing in main text Eqn.~8. While the frequencies of the three tuning modes obtained via TDDFT/CAM-B3LYP were consistent with those obtained through MRCI to within 40~cm$^{-1}$, the frequency of the coupling mode $\nu_\mathrm{10a}$ shows a considerable difference of nearly 205~cm$^{-1}$. Hence, in order to maintain consistency with the parameterizations used in Refs.~\onlinecite{Krempl1994,Worth1996}, we have shifted the central frequency of the peak in the coupling spectral density down from 960~cm$^{-1}$ to 755~cm$^{-1}$. The central frequency of all other modes was left unaltered.

\begin{figure}
    \centering
    \includegraphics[width=0.9\textwidth]{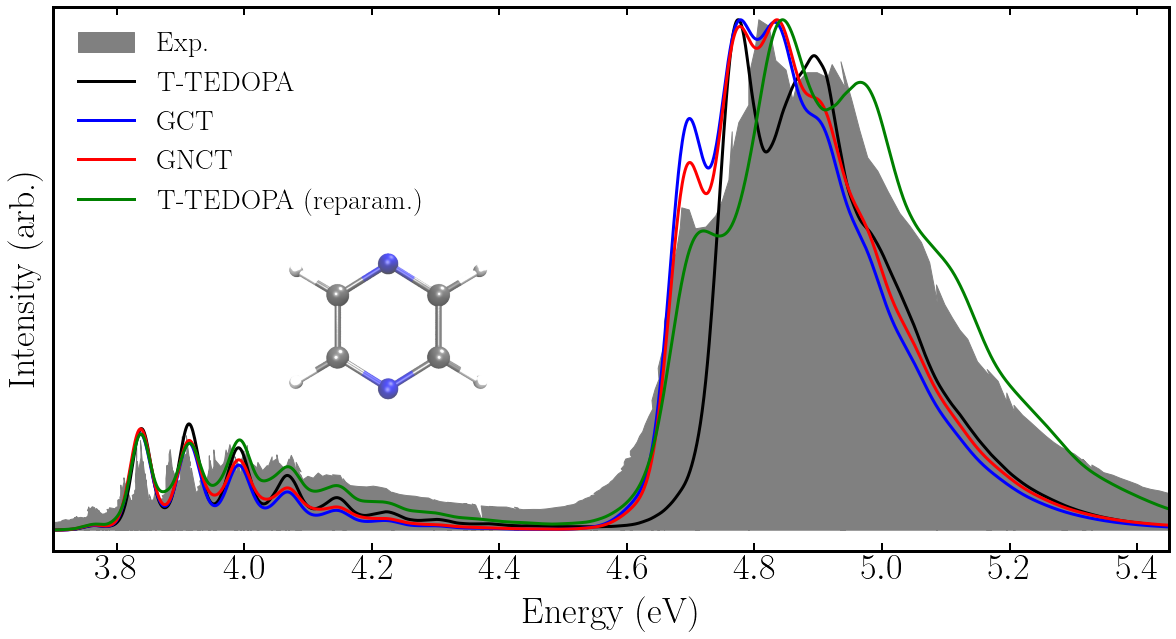}
    \caption{Reproduced Fig.~6(a) from the main text but with reparameterized spectral densities according to the MRCI couplings listed in Table~\ref{tab:pyrazine_couplings}.}
    \label{fig:pyrazine_sd_reparam}
\end{figure}

The predicted pyrazine absorption lineshape computed from T-TEDOPA dynamics with altered spectral densities is shown in Fig.~\ref{fig:pyrazine_sd_reparam}. It is clear that this lineshape is in much better agreement with experiment than the other three considered and serves to demonstrate that the accuracy of the calculation is mainly limited by the accuracy of the TDDFT method used to parameterize the spectral densities.

\begin{table}[h!]
\centering
\renewcommand{\arraystretch}{1.5}
\setlength{\tabcolsep}{5pt}
\newcolumntype{L}[1]{>{\raggedright\arraybackslash}p{#1}}
\newcolumntype{C}[1]{>{\centering\arraybackslash}p{#1}}

\begin{tabular}{ C{1.6cm}|C{1.6cm}|C{1.6cm}|C{1.6cm}|C{1.6cm}|C{1.6cm}|C{1.6cm}|C{1.6cm}|C{1.6cm}|}
\multirow{2}{*}{} & \multicolumn{2}{c|}{$\nu_\mathrm{6a}$ (610~cm$^{-1}$)} & \multicolumn{2}{c|}{$\nu_1$ (1065~cm$^{-1}$)} & \multicolumn{2}{c|}{$\nu_\mathrm{9a}$ (1271~cm$^{-1}$)} & \multicolumn{2}{c}{$\nu_\mathrm{10a}$ (755~cm$^{-1}$)} \\ \cline{2-9}
 & \multicolumn{1}{c|}{TDDFT} & \multicolumn{1}{c|}{MRCI} & \multicolumn{1}{c|}{TDDFT} & \multicolumn{1}{c|}{MRCI} & \multicolumn{1}{c|}{TDDFT} & \multicolumn{1}{c|}{MRCI} & \multicolumn{1}{c|}{TDDFT} & \multicolumn{1}{c}{MRCI} \\ \cline{1-9}
S$_1$ & \multicolumn{1}{c|}{0.095} & \multicolumn{1}{c|}{0.096} & \multicolumn{1}{c|}{0.032} & \multicolumn{1}{c|}{0.047} & \multicolumn{1}{c|}{0.113} & \multicolumn{1}{c|}{0.159} & \multicolumn{1}{c|}{-} & \multicolumn{1}{c}{-} \\ 
S$_2$ & \multicolumn{1}{c|}{0.102} & \multicolumn{1}{c|}{0.119} & \multicolumn{1}{c|}{0.128} & \multicolumn{1}{c|}{0.201} & \multicolumn{1}{c|}{0.053} & \multicolumn{1}{c|}{0.048} & \multicolumn{1}{c|}{-} & \multicolumn{1}{c}{-} \\ 
Coupling & \multicolumn{1}{c|}{-} & \multicolumn{1}{c|}{-} & \multicolumn{1}{c|}{-} & \multicolumn{1}{c|}{-} & \multicolumn{1}{c|}{-} & \multicolumn{1}{c|}{-} & \multicolumn{1}{c|}{0.215} & \multicolumn{1}{c}{0.183} \\
\end{tabular}
\caption{Linear couplings in eV for the four-mode model of pyrazine as computed from TDDFT/CAM-B3LYP and MRCI\cite{Sobolewski1993} techniques. The central frequencies used in the altered spectral densities are indicated in the parentheses of each header column.}
\label{tab:pyrazine_couplings}
\end{table}

\section{Additional results: Model systems with dark S$_1$ and bright S$_2$}

\begin{figure}
    \centering
    \includegraphics[width=0.9\textwidth]{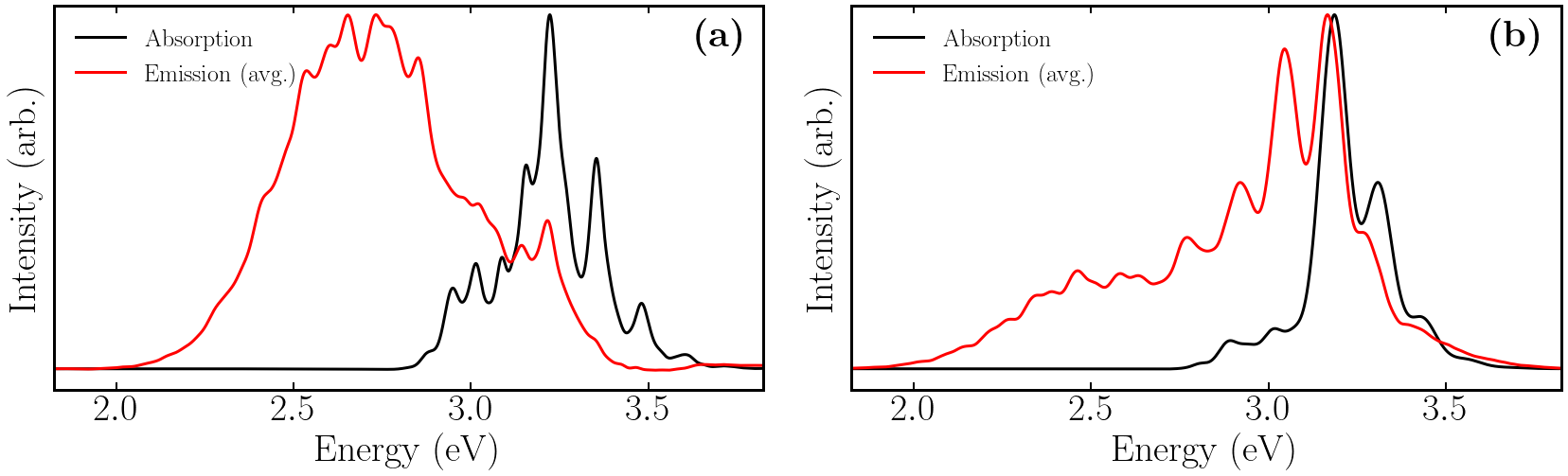}
    \caption{Absorption and averaged emission spectra of (a) MSA30-S and (b) MSB30-S. The emission spectra in each case are formed from the average of 30 uniformly-spaced samples of the trajectory between 116.1 and 203.2~fs.}
    \label{fig:si_tdm_swap_spectra}
\end{figure}

\begin{figure}
    \centering
    \includegraphics[width=0.9\textwidth]{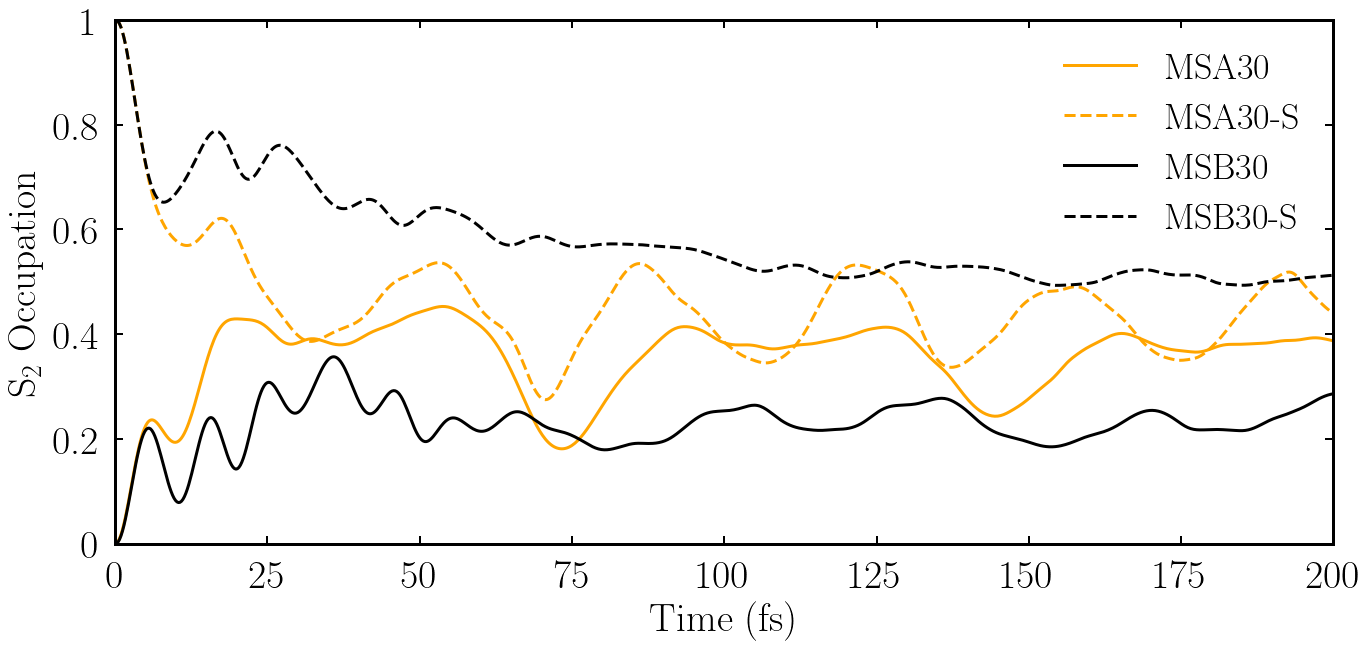}
    \caption{Comparison of the S$_2$ population dynamics between model systems MSA30, MSA30-S, MSB30, and MSB30-S.}
    \label{fig:si_tdm_swap_pop}
\end{figure}

Throughout this work, the model systems we have considered possess respective diabatic S$_1$ and S$_2$ transition dipole moments of magnitude 2.54 and 0 a.u. In this section, we consider modified versions of the MSA30 and MSB30 model systems, labeled MSA30-S and MSB30-S, in which transition dipole moments are swapped so that instead S$_2$ is bright with a moment of 2.54~a.u. and S$_1$ is completely dark with no net moment. 

Fig.~\ref{fig:si_tdm_swap_spectra} shows the absorption and averaged emission spectra for the MSA30-S and MSB30-S model systems computed using the T-TEDOPA approach described in the main text. Note that this figure is identical to Fig.~3 in the main text but with diabatic S$_1$ and S$_2$ states in each model system having swapped transition dipole moments. These model systems exhibit many of the same spectral features seen in Fig.~3 of the main text, such as broken mirror symmetry between absorption and emission, intensity borrowing between bright and dark states in the emission of both systems, and the distortion of regularly-spaced vibronic progressions into more diffuse, irregular profiles. 

Fig.~\ref{fig:si_tdm_swap_pop} outlines the S$_2$ population of the swapped model systems during the initial excited state relaxation, in comparison with the standard model system population dynamics described in the main text. We note that both system parameterizations evolve under the same underlying Hamiltonian, and the only difference lies in the nature of the initial state (either 100\%\,population in S$_1$, for the model system described in the main text, or 100\%\, population in S$_2$, for the model systems with swapped intensity). The resulting population dynamics however do not reach the same final relaxed excited state, which is particularly apparent for MSB, where the swapped dipole parameterization retains a significantly larger population in the S$_2$ state after 200~fs of relaxation. 

There is evidence that neither MSB30 or MSB30-S population dynamics are fully converged at the end of the 200~fs propagation time, with MSB30-S in particular undergoing a continued slow population shift from S$_2$ to S$_1$ between 75~fs and 200~fs. Similarly, for MSA30-S, the population dynamics remain highly oscillatory up to 200~fs of dynamics. Insufficient propagation time to reach a steady state is potentially the reason for the large overlap between absorption and emission lineshapes in Fig.~\ref{fig:si_tdm_swap_spectra}, suggesting that the initial state used in the emission spectrum is not fully relaxed. We note that this issue is exacerbated in our model system calculations due to the small number of modes coupling to the system, the small amount of reorganization energy confined to low frequency ``solvent" modes, and the slow solvent bath. In realistic condensed phase systems, population dynamics often quickly reach a steady state\cite{Dunnett2021, Hunter2024}, such that the relatively short propagation times easily accessible through T-TEDOPA are sufficient for obtaining converged steady state emission lineshapes. 

We also note a small negative intensity in the emission spectrum of the MSA30-S model system in Fig.~\ref{fig:si_tdm_swap_spectra}. This is likely due to a slight underconvergence in the averaging over relaxed states, although other factors (see below in Sec.~\ref{si_sec:converging_spectra}) may also contribute.

\section{Additional information on obtaining converged spectra}
\label{si_sec:converging_spectra}

\begin{figure}
    \centering
    \includegraphics[width=0.9\textwidth]{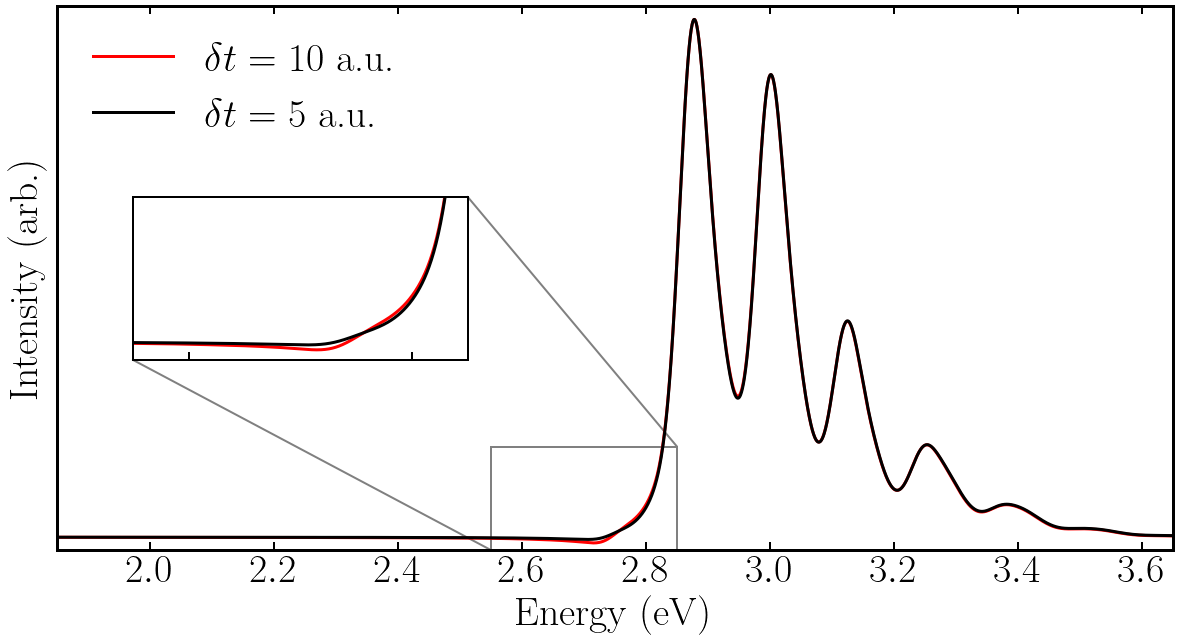}
    \caption{Convergence of the absorption spectrum of MSB30 with respect to the time-step used in the 1TDVP propagation.}
    \label{fig:si_timestep_conv}
\end{figure}

While a time-step of 10~a.u. in the 1TDVP propagation was found to be sufficient for the convergence of most optical spectra computed using the T-TEDOPA formalism, we also found that this time-step led to a slight underconvergence in the absorption spectrum of the MSB30 and MSB200 model systems. Fig.~\ref{fig:si_timestep_conv} shows that while the main features of the MSB30 absorption lineshape are converged, there exists a small, negative intensity region at the onset of the spectrum at $\approx2.7$~eV, which is systematically removed by running the dynamics with a reduced time-step of 5~a.u.

\bibliography{bibliography}